\RequirePackage{fix-cm}
\documentclass[numbook]{svjour3}                     
\smartqed  
 \journalname{J. Elast.}
\usepackage{graphicx}
\usepackage{color}
\usepackage{verbatim}
\usepackage{bm}
\usepackage{ntheorem}
\usepackage{float}

\def\salta#1{{}}
\newcommand{\EGV}[1]{{$\ro \clubsuit$}\marginpar{\scriptsize{\ro{EGV: #1}}}}

\newcommand{\ro}{\color{red}}

\newcommand{\blu}{\color{blue}}

\newcommand{\n}{\mathbf{n}}
\newcommand{\mol}{\bm{\ell}}
\newcommand{\e}{\mathbf{e}}
\newcommand{\pos}{\mathbf{r}}
\newcommand{\pot}{\Phi}
\newcommand{\cyl}{\mathsf{C}}
\newcommand{\bulk}{\mathcal{B}}
\newcommand{\disk}{\mathcal{D}}
\newcommand{\axis}{\mathcal{A}}
\newcommand{\plane}{\mathcal{P}}
\newcommand{\centre}{\mathcal{C}}
\newcommand{\tetra}{\mathcal{T}}
\newcommand{\dd}{\mathrm{d}}
\newcommand{\Frame}{(\e_1,\e_2,\e_3)}
\newcommand{\hypol}{\bm{\chi}^{(2)}}
\newcommand{\hycomp}[1]{\chi^{(2)}_{#1}}
\newtheorem{prop}{Proposition}
\theorembodyfont{\upshape}
\newtheorem{rem}{Remark}

\newlength{\irrl}
\newlength{\irrw}
\newcommand{\irr}[1]{
	\settowidth{\irrl}{\mbox{$\displaystyle #1$}}
	\setlength{\irrw}{0.12ex}
	\mbox{$\hspace{0.2em}
		\stackrel{
			\mbox{$\vphantom{\rule[-5\irrw]{\irrw}{6\irrw}}
				\rule[-4\irrw]{\irrw}{5\irrw}\hspace{-\irrw}
				\rule{\irrl}{\irrw}\hspace{-\irrw}
				\rule[-4\irrw]{\irrw}{5\irrw}$}}
		{\mbox{$\displaystyle #1$}}\hspace{0.2em}$}
}

\newcommand{\avir}[1]{\left\langle\irr{{#1}}\right\rangle}
\newcommand{\ave}[1]{\left\langle{#1}\right\rangle}
\newcommand{\avirro}[1]{\left\langle\irr{{#1}}\right\rangle_{\!\!\!\varrho}}

\def\giorno{\today}

\def\a{\alpha}
\def\b{\beta}
\def\ga{\gamma}

\def\vth{\vartheta}
\def\la{\lambda}
\def\vtheta{\vartheta}
\def\vphi{\varphi}
\def\s{\sigma}

\def\eps{\varepsilon}

\def\pa{\partial}

\def\xb{{\mathbf{x}}}
\def\pb{{\bf p}}

\def\R{{\bf R}}

\def\A{{\bf A}}

\def\ss{\subset}

\def\EOR{\hfill $\odot$}
\def\({\left(}
\def\){\right)}
\def\[{\left[}
\def\]{\right]}

\def\beq{\begin{equation}}
\def\beql#1{\begin{equation} \label{#1}}
\def\eeq{\end{equation}}
\def\eqref#1{(\ref{#1})}

\def\^#1{\widehat{#1}}

\def\w1f{250pt}
\def\w2f{150pt}
\def\w3f{120pt}

\begin{document}
	
	\title{The symmetries of octupolar tensors
	}
	
	
	\author{Giuseppe Gaeta       \and Epifanio G. Virga}
	
	
	\institute{Giuseppe Gaeta \at
		Dipartimento di Matematica,
		Universit\`a degli Studi di Milano, via Saldini 50, I-20133 Milano
		(Italy)\\
		\email{giuseppe.gaeta@unimi.it}           
		\and
		Epifanio G.
		Virga \at
	Dipartimento di Matematica, Universit\`a di Pavia,
	via Ferrata 5, I-27100 Pavia (Italy)\\
	\email{ eg.virga@unipv.it}  
	}
	
	\date{Received: \giorno / Accepted: date}

	\maketitle
	
	\begin{abstract}
		Octupolar tensors are third order, completely symmetric and traceless tensors. Whereas in 2D an octupolar tensor has the same symmetries as an equilateral triangle and can ultimately be identified with a vector in the plane, the symmetries that it enjoys in 3D are quite different, and only exceptionally reduce to those of a regular tetrahedron. By use of the octupolar potential that is, the cubic form  associated    on the unit sphere with an octupolar tensor, we shall classify all inequivalent octupolar symmetries. This is a mathematical study which also reviews and incorporates some previous, less systematic attempts.
	\end{abstract}



\section{Introduction}\label{sec:intro}
It is well known that the Landau theory of phase transitions
\cite{landau:theory,landau:theory_I,LL,Peliti,Toledano} describes the states of matter in
the vicinity of a critical point in terms of an \emph{order
parameter}; in the simplest cases this is a scalar quantity, but
it can be a vector, or more generally a tensor of any order.

In fact, in the case of liquid crystals it is rather common to
describe their state in terms of a second-order tensor
\cite{PdG,GLJ,VirgaBook}. More recently, it became apparent that
certain materials displaying \emph{tetrahedral} nematic phases
\cite{Fel1,Fel2} are better described in terms of a \emph{third}
order tensor $\mathbf{A}$ -- more precisely, a fully symmetric and completely
traceless one (see below for definitions). We stress that it is
conceivable, even probable, that order parameters described by
still higher order tensors will be needed in considering
generalized liquid crystals \cite{liu:classification,liu:generalized,liu:generic}.

This is precisely what is meant in this paper by an \emph{octupolar tensor}: a third order, fully symmetric and completely traceless tensor. Although, as also shown below, octupolar tensors feature in many branches of physics, we shall systematically use the paradigm of the Landau theory of phase transitions to illustrate the physical significance of our study. The reader is advised from the start that this is but one of many incarnations of our mathematical theory, which is concerned with the symmetries of the most general octupolar tensor in three space dimensions. 

One of us provided a complete description of the physics of a
material represented by an octupolr tensor in the
\emph{two dimensional} case \cite{Virga2D}. In this case, one
obtains a remarkably simple description, and the physical state is
basically identified by the orientation of an equilateral triangle
in the order parameter space.

Unfortunately, such a simple description is peculiar to the 2D
case, and as soon as we pass to consider a
\emph{three-dimensional} situation things become much more
involved. In a recent contribution \cite{GV2016} (see also
\cite{CQV}), we have studied this situation, providing a
description of the physics described by an octupolar tensor in three dimensions; this study also
showed an unexpected feature awaiting experimental confirmation,
i.e. the existence -- together with higher symmetric special
phases -- of \emph{two} different generic phases; the interface
between these has been investigated in detail in a subsequent work
\cite{CQV}.

In our efforts to classify all symmetries of an octupolar tensor $\mathbf{A}$ a prominent role is played by the appropriate notion of eigenvalues and eigenvectors applicable for $\A$. In multilinear algebra, such a notion is not as univocally defined as one might naively think. For real-valued tensors that bear a physical meaning, as the ones we are interested in, the issue arises as to whether complex eigenvalues, which would perfectly be allowed according to certain definitions, should be admitted or not. The definition of eigenvalues (and associated eigenvectors) that we adopt in this paper is essentially the one  put forward in \cite{qi:eigenvalues,qi:rank,qi:eigenvalues_invariants} (see also the recent book \cite{qi:tensor}, especially Chap.~7, which is specifically concerned with octupolar tensors and their mechanical applications). However, we give an equivalent characterization of this notion in terms of the critical points of a cubic polynomial defined over the unit sphere $S^2$. This is indeed the natural extension of what one learns from the lucid (and now rare) book of Noll \cite{noll:finite-dimensional}. In Sect.~84, the maximum and minimum of the spectrum of a symmetric, second order tensor $\mathbf{S}$ in a $n$-dimensional space are characterized as the corresponding extrema of the quadratic form associated with $\mathbf{S}$ on the unit sphere $S^{n-1}$. Noll's book was published in 1987, but its contents were available many years back.\footnote{In the Introduction to \cite{noll:finite-dimensional} (p.~IX), we read: 
\begin{quote}
	About 25 years ago I started to write notes for a course for seniors and beginning graduate students at Carnegie Institute of Technology (renamed Carnegie-Mellon University in 1968). At first, the course was entitled ``Tensor Analysis''. [\dots] The notes were rewritten several times. They were widely distributed and they served as the basis for appendices to the books \cite{coleman:viscometric} and \cite{truesdell:first}.
\end{quote}
}

In our previous work \cite{GV2016} we focused on the physics of
the problem, and in particular of its generic phases -- discovering an unexpected phenomenon, i.e. the existence of two generic (octupolar) phases, hence the possibility of an intra-octupolar phase transition (see also \cite{CQV} for details on this) -- providing little mathematical detail; the purpose of the present paper is twofold:
\begin{enumerate}
	\item[(a)]
	 On the one hand, we want to give a full account of the mathematical details needed for the study of such a problem. We trust that  -- beside the interest \emph{per se}  -- this will also be relevant  to systems  described by higher order tensors \cite{liu:classification,liu:generalized,liu:generic}.
\item[(b)]
On the other hand, also thanks to this higher mathematical detail provided here, we want to discuss in more detail the non-generic phases (and the transitions  between these),  thus completing the physical description provided in previous work \cite{GV2016,CQV}.
\end{enumerate}

The plan of the paper is as follows. In 
Sect.~\ref{sec:physical} we present the physical motivation of our work;
in Sect.~\ref{sec:freeenergy} we discuss the general features of
a prototypical  Landau potential, which will be referred to as the \emph{octupolar potential}, for short. As this potential is based on octupolar tensors in three spatial dimensions, the
subsequent Sect.~\ref{sec:symmtensors} is devoted to study these
objects and their eigenvectors. We can then pass, in 
Sect.~\ref{sec:potential}, to study the general octupolar potential; this
depends a priori on seven parameters, but by a suitable choice of
reference frame and of the potential scale -- as discussed in 
Sect.~\ref{sec:potential} --  we can reduce to study a problem depending
only on three parameters; the allowed parameters are described by
a cylinder in parameter space. It turns out that this potential and its
critical points bear some relation to the tetrahedral group;
Sect.~\ref{sec:tetra} is thus devoted to recalling some basic
facts about this. We can then pass to study the extremals of the
Landau potential; these depend of course on the parameters and
different regions in the allowed cylinder in parameter space
correspond actually to different symmetries (phases) for the
potential and its critical set, as discussed in detail in 
Sect.~\ref{sec:symmetries} and its subsections, each devoted to one of
these phases.  Finally in Sect.~\ref{sec:discu} we review and
summarize our findings, with special emphasis on the distribution in parameter space of the critical points of the octupolar potential. In the final Sect.~\ref{sec:conclu}
we draw our conclusions. The paper is completed by an appendix  with details on the tetrahedral
group beyond the brief discussion of Sect.~\ref{sec:tetra}.

Whenever indices are needed, summation over repeated pairs will be routinely understood,
unless explicitly stated otherwise.

\section{Physical motivation}\label{sec:physical}
Octupolar tensors have many manifestations in physics. In this section, we shall review a few of them, ranging from the most classical to the most innovative ones. 

Among the former is Buckingham's formula \cite{Buck} for 
the probability density $\varrho$ of the the distribution of a molecular director $\mol$
over the unit sphere $S^2$, which can be written as
\beq \varrho(\mol) =\frac{1}{4 \pi}  \( 1 \
+  \sum_{k=1}^\infty \frac{(2k+1)!!}{k!}   \avirro{\mol^{\otimes k}} \cdot\mol^{\otimes k} \),
\eeq
where, for a generic vector $\xb$, ${\bf x}^{\otimes k}$ is the $k$th-order tensor defined as in \cite{turzi:cartesian} by
\beq {\bf x}^{\otimes k} ={\bf x} \otimes ... \otimes {\bf x} \eeq
(with $k$ factors, of course), $\cdot$ denotes tensor contraction, and $\avirro{\bf P}$ is the multipole average,
\beq \avirro{\bf P}
:=  \frac{1}{4 \pi}  \int_{S^2} \avir{\bf P} 
\varrho ({\mol})  \dd \mol  , \eeq
of the symmetric, traceless part $\irr{\bf P} $ of a
tensor ${\bf P}$.

With this notation, our octupolar order tensor $\A$ is identified with
\beq \A  =
\avirro{\mol\otimes\mol\otimes\mol} .
\eeq
Passing to spherical coordinates, and writing $\mol$ as
\beq\label{eq:parperp} {\mol} =\sin \vartheta  \cos
\varphi\,  \e_1  +  \sin \vartheta  \sin \varphi\,  \e_2  +  \cos \vartheta\,\e_3  , \eeq
we easily express the scalar parameters that represent $\A$ in the Cartesian frame $\Frame$
in terms of multipole averages, which also reveal the bounds they are subject to \cite{GV2016}.

In nonlinear optics (see, for example Sect.~1.5 of \cite{boyd:nonlinear}, and also \cite{zyss:nonlinear} and \cite{kanis:design}), higher order susceptibilities tensors, often called  \emph{hypersusceptibilities} (and also \emph{hyperpolarizabilities}), are introduced to decompose the electromagnetic energy density in multipoles. In particular, the cubic term has the general form
\begin{equation}\label{eq:hypersusceptibility}
U^{(2)}=\hycomp{ijk}F_iF_jF_k,
\end{equation} 
where $F_i$ are the components of an external field and $\hycomp{ijk}$ are the components of the first hypersusceptibility tensor $\hypol$, which is a third order tensor.\footnote{The superscript $^{(2)}$ reminds us that this tensor expresses the field induced by polarization as a \emph{quadratic} function of the external field, whereas the ordinary susceptibility establishes a linear relationship between the two fields.}

When the frequencies of the applied fields are much smaller than the resonance frequency, $\hypol$ can safely be assumed to be independent of frequencies and fully symmetric in all its indices. Though this symmetry, which is often called Kleinman's symmetry after the name of the author who first introduced it \cite{kleinman:nonlinear}, has been widely criticized \cite{dailey:general} and also found in disagreement with some computational schemes \cite{wergifosse:evaluation}, it is still often accepted as an approximation for its simplicity. Assuming Kleinman's symmetry to be valid, we can extract an octupolar tensor $\A=\irr{\hypol}$ out of $\hypol$ and rewrite the latter in the equivalent form
\begin{equation}\label{eq:hypersusceptibility_equivalent}
\hycomp{ijk}=A_{ijk}+\frac15\(\hycomp{i}\delta_{jk}+\hycomp{j}\delta_{ki}+\hycomp{k}\delta_{ij}\),
\end{equation}
where $\hycomp{k}:=\hycomp{iik}=\hycomp{iki}=\hycomp{kii}$. Using \eqref{eq:hypersusceptibility_equivalent} in \eqref{eq:hypersusceptibility}, we also write
\begin{equation}\label{eq:hypersusceptibility_rewritten}
U^{(2)}=A_{ijk}F_iF_jF_k+\frac35F^2\hycomp{i}F_i,
\end{equation}
where $F$ is the strength of the applied field. Normalizing $F$ to unity, $U^{(2)}$ becomes the sum of an octupolar  and a dipolar potential, the latter of which contributes to a lower multipole.

A third order tensor very similar to $\hypol$ was introduced in \cite{lubensky:theory} to describe the ordering of bent-core molecules that possess liquid crystal phases. As also recalled in \cite{CQV}, the theory of bent-core liquid crystal phases features a mesoscopic third order tensor  derived from $\irr{\bm{\alpha}^{(3)}}$; here $\bm{\alpha}^{(3)}$ is the molecular structural tensor defined by
\begin{equation}
\bm{\alpha}^{(3)}:=\sum_{\mu=1}^Nm_\mu\pos_\mu\otimes\pos_\mu\otimes\pos_\mu,
\end{equation}
where the sum is extended to all the atoms in a constituent molecule, $m_\mu$ is the mass of each individual atom and $\pos_\mu$ is its position vector relative to the molecule's center of mass. The \emph{ensemble} average  $\A=\avir{\bm{\alpha}^{(3)}}$ is an octupolar tensor that plays an important role in classifying all possible new phases that bent-core liquid crystals are allowed to exhibit. They have been collectedly  called \emph{tetrahedral} by a symmetry that $\A$ can indeed enjoy, but our study has shown to be only too special, rather than generic.

Lately, octupolar tensors have also become popular with the classical elastic theory of nematic liquid crystals (both passive and active). The nematic director field $\n$ represents at the macroscopic scale the average orientation of the elongated molecules that constitute the medium. Elastic distortions of $\n$ are measured locally by its spatial gradient $\nabla\n$, which may become singular at certain points in space in response to external distorting stimuli. These are the \emph{defects} of $\n$, where molecular order is degraded; they can be classified into distinct topological classes, associated in 2D with the \emph{winding} number $m$ of $\n$ around a point defect ($m$, which is often referred to as the \emph{topological charge}, is half an integer; more information about defects in liquid crystals and their topological  charges can be obtained from the monographs \cite{stewart:static} and \cite{VirgaBook}). 

It has recently been proposed \cite{vromans:orientational} that a vector be associated with a point defect in 2D to represent the  average direction along which the field $\n$ is fluted away from the defect. It is believed that such a vector could play a role in describing the interaction of defects, as if they were \emph{oriented} particles interacting like \emph{vessels} in a viscous \emph{sea}. If this image is indeed suggestive for the charge $m=\frac12$, when the integral lines of $\n$ around the defect resemble a flame, associating a single direction with a defect with charge $m=-\frac12$ seems somehow troublesome at first sight, as in that case the integral lines of $\n$ escape along three  directions, ideally separating the plane in three equal sectors. To overcome this difficulty, it has been proposed in \cite{tang:orientation} to describe the orientation of a $-\frac12$ defect through the octupolar tensor
\begin{equation}\label{eq:defect_orientation}
\A=\avir{\nabla\n\otimes\n},
\end{equation}
where the average $\ave{\cdots}$ is now meant to be computed on a sufficiently small neighborhood surrounding the defect. As shown in \cite{Virga2D}, any octupolar tensor $\A$  in 2D   possesses the symmetries of an equilateral triangle, and its most general representation is given by
\begin{equation}\label{eq:2D_representation}
\A=\alpha\irr{\e\otimes\e\otimes\e},
\end{equation}
where $\alpha$ is a scalar and $\e$ is a unit vector in in the plane. Thus, in 2D, $\A$ effectively reduces to a vector and both approaches in \cite{vromans:orientational} and \cite{tang:orientation} are equivalent. However, as shown in \cite{GV2016}, in 3D $\A$ cannot in general be represented in terms of a single vector as in \eqref{eq:2D_representation}  and \eqref{eq:defect_orientation} becomes a more versatile tool to describe the directions along which the integral lines of $\n$ are fluted around a point defect. In particular, it might be interesting to compute $\A$ in \eqref{eq:defect_orientation} for the \emph{combed} defects described in \cite{saupe:disclinations} and their distortions possibly due to the interaction with other point defects  nearby (see also \cite{sonnet:reorientational}). The complete classification of the symmetries enjoyed by $\A$, which we give in this paper, may supplement the topological classification of defects by providing extra synthetic information on the qualitative features of the director field surrounding the defect. 

\section{Octupolar potential}
\label{sec:freeenergy}

We will work in three-dimensional space, with standard coordinates
$(x_1,x_2,x_3)$ for $\xb$ in a Cartesian frame $\Frame$.
In the theory we are interested in, the octupolar potential $\Phi (\xb)$
is described by a three-dimensional third
order tensor\footnote{More generally, we might consider potentials with contributions \emph{up to} third order; thus we would have the sum of a scalar
part, a vector one, another part described by a second order
tensor, and finally the one described by the third order one. Here
we focus on this last contribution, as the study of theories with
scalar, vector, o second-order tensor order parameters is sort of
standard (in principle; obviously concrete applications can
present endless complications).} $\A $ via
\beq
\label{eq:genPhi}
\Phi =A_{ijk}  x_i  x_j  x_k  ; \eeq
the physical states will be described by minima of this function.
In view of the homogeneity of  $\pot$, it is not restrictive to look for extrema of $\Phi$ constrained to the unit sphere
$S^2 \ss \R^3$.

The third order tensor $\A$ has some additional properties:
\begin{enumerate}
\item $\A$ is completely symmetric; in terms of the components of
$\A$, this means $A_{ijk} = A_{\pi (ijk)}$, with $\pi$ any
permutation; \item $\A$ is completely traceless, i.e. $A_{iik} =
A_{iki} = A_{kii} = 0$ for any $k$.
\end{enumerate}

Since $\Phi$ is homogeneous of odd degree, we always have
\beql{eq:Phiodd} \Phi ( - \xb) =-  \Phi (\xb)  . \eeq This also implies
that if $\xb_0$ is a minimum of $\Phi$, then $- \xb_0$ is a
maximum, and conversely; on the other hand, if $\xb_0$ is  saddle
point with $p$ unstable directions, then $- \xb_0$ is again a
saddle point, albeit with $p$ \emph{stable} directions (and hence
$\widetilde{p} = 3 - p$ unstable ones).

This means that we can equivalently describe the system in terms
of \emph{maxima} of $\Phi$ (this amounts to changing a global
sign); this is more convenient on graphical terms, and we will
thus adhere to this convention. 

\subsection{Extrema, eigenvectors, ray solutions}
\label{sec:extrema}

We thus have to maximize (or minimize) $\Phi (\xb)$ given by
\eqref{eq:genPhi} subject to the constraint $|\xb |=1$. This can
be obtained in two ways: \begin{itemize} \item[(a)] by augmenting
$\Phi$ to a function \beq \Phi_\la (\xb)  :=  \Phi (\xb)  - \
\frac12  \la  (|\xb|^2 - 1)  , \eeq where $\la$ is a Lagrange
multiplier; \item[(b)] or passing to spherical coordinates $(r,
\vphi, \vtheta)$ and setting $r=1$, thus obtaining a reduced
potential\footnote{As a general convention, we will denote the
potentials in Cartesian coordinates by $\Phi$ (with several
suffixes) and those in spherical coordinates -- which we always
consider only for $r=1$ -- by $\Psi$ (again with corresponding
suffixes).} $\Psi (\vphi, \vtheta) : S^2 \to \R$.
\end{itemize} We will mainly use the latter approach, but where
convenient we also employ the former.\footnote{It may be worth
mentioning that (in particular, if we are satisfied with studying
$\Phi$ on one hemisphere, which is justified by \eqref{eq:Phiodd})
a third option is present, i.e. setting
$z = \pm \sqrt{1 - x^2 - y^2}$ and considering $\Phi$ as a
function of $x$ and $y$; these take value in the unit disk. This
will be used in Sect.~\ref{sec:maximum}.}

The condition of constrained extremum results in requiring that
$(\nabla \Phi)(\xb)$ is collinear to $\xb$, i.e. \beq (\nabla
\Phi) (\xb) =k  \la  \xb  . \eeq We will look at this
problem in terms of \emph{eigenvectors} (and eigenvalues) of
a higher order tensor \cite{CS,qi:eigenvalues,qi:rank,qi:eigenvalues_invariants,ni:degree}, see Sect.~\ref{sec:evev}
below.

It should be noted that the same problem can be seen in a slightly
different way. That is, we can consider the associated dynamical
system \beq \frac{d \xb}{ d t } =(\nabla \Phi) (\xb)  , \eeq
and search for \emph{ray solutions}, i.e. solutions of the form
\beq \xb (t) =\a (t)  \xb (0)  . \eeq This problem has been
considered in the literature, and a number of results (in
particular, concerning the number of such solutions) are
available \cite{Rohrl,RohrlLong,Walcher}. These coincide with the results obtained in terms of eigenvectors of higher order tensors
\cite{CS,qi:eigenvalues,qi:rank,qi:eigenvalues_invariants,ni:degree}.

In the following Sects.~\ref{sec:ray} and \ref{sec:evev} we
recall both kind of results. We stress that we just want to report
results present in the literature, but these should not be seen in
terms of priority.\footnote{In fact, as pointed out by Walcher
\cite{WalGSD}, this kind of results follows ultimately from the
work of Bezout on intersection theory dating back to the $18$th century. See his forthcoming paper \cite{Wal18} for details.}

Unfortunately, as we will see, both approaches provide a complete
answer in terms of \emph{complex} numbers, while we need results
in the field of \emph{real} numbers. The theorems to be reported
below do not specify how many of these ray solutions or equivalently eigenvectors, and the associated eigenvalues, will be real
(more or less in the same way as the fundamental theorem of algebra
does not say how many of the roots of a polynomial are real).

\subsection{Ray solutions of homogeneous dynamical systems}
\label{sec:ray}

We start by reporting the results obtained by R\"{o}hrl \cite{Rohrl}
for ray solutions of dynamical systems in ${\bf C}^N$.
\begin{prop}\label{prop:1}
Consider the
dynamical system \beql{eq:rohrl} \dot{x}^i =B^i (x)  ,   \
 i = 1,...,q  , \eeq with $B^i$ homogeneous polynomials of
degree $p$ in $x$. If the coefficients of the polynomials $B^i$
are algebraically independent over the field of the rationals,
then \eqref{eq:rohrl} fails to have a critical point in the origin
and has precisely \beql{eq:rohrlN} N_R =p^q  -  1 \eeq ray
solutions.
\end{prop}

In the case of interest here we deal with
three-dimensional systems, thus $q=3$, and homogeneous
polynomials of degree two, as these result from considering the
gradient of $\Phi (\xb)$, thus $p=2$. This yields $2^3 -1 = 7$ ray
solutions. Each of these intersects the unit sphere in two
(antipodal) points, hence we get 14 critical points for our
constrained variational problem.

It should also be stressed that, strictly speaking, R\"{o}hrl's theorem
holds under the condition of algebraic independence of the
coefficients $B^i_{jk}$ in $ B^i = B^i_{jk} x^j x^k$; this is a
generic property for general polynomials, but fails for symmetric
ones. However one should note that only the symmetrized sums
$B^i_{jk} + B^i_{kj}$ will play a role, hence one should
understand the condition of algebraic independence refers to
these;\footnote{It should be noted that the ``disappearance'' of
real critical points -- w.r.t. the generic situation described by
R\"{o}hrl's theorem -- is related, at least in our model, to the
appearance of a ``monkey saddle'' \cite{GV2016}, i.e. of a
critical point with a non-generic index; see below for detail.}
moreover, one should note that algebraic independence is only a
sufficient -- but not necessary -- condition in R\"{o}hrl's theorem.

The reader is referred to the original contribution by R\"{o}hrl \cite{Rohrl}, or
to \cite{RohrlLong,Walcher}, for more detail.
\begin{rem}\label{rem:1}
Note also that the Equivariant Branching Lemma
\cite{CicEBL,VdB,CicNLEBL,GEBL} gives information about the
existence and stability of solutions along such critical rays when
we consider theories depending on external parameters. \EOR
\end{rem}

\subsection{Eigenvalues and eigenvectors for higher order tensors}
\label{sec:evev}

Determining eigenvalues and eigenvectors for a second order, symmetric tensor $\mathbf{S}$ in an $n$-dimensional vector space $V$ is a very classical problem in linear algebra
and amounts to solving the linear problem \beq \label{eq:evev} \mathbf{S}
{\bf v} =\la  {\bf v}  ; \eeq one can without loss of
generality restrict to unit vectors,\footnote{In fact, if ${\bf v}$
is an eigenvector of $M$ with eigenvalue $\la$, then for any
number $\a \not= 0$ also ${\bf w} = \a {\bf v}$ is an eigenvector
with the same eigenvalue $\la$.} i.e. complement the problem with
the side condition
\beq ( {\bf v} , {\bf v} ) =| {\bf v} |^2 =1. \eeq

The same problem is obtained if one looks for minimizers of a
function $\Phi : \R^n \to \R$ which is a quadratic form defined by
$\mathbf{S}$, i.e. $\Phi = S_{ij} x_i x_j$, with $S_{ij}=S_{ji}$, restricted to the unit sphere $S^{n-1}$.
In fact, in this case one introduces the Lagrange multiplier $\la$
and considers the extended (constrained) potential
\beq \Phi_\la =\Phi - \frac12 \la x^i x^i  , \eeq obtaining the same
equation \eqref{eq:evev} as the critical point equation for the
potential $\Phi_\la : \R^n \times \R \to \R$. It is well known (see, for example, Sect.~84 of \cite{noll:finite-dimensional}) that all eigenvalues of $\mathbf{S}$ are \emph{real} and bear a physical meaning.

It seems in many ways natural to consider the same problem in
\emph{multilinear algebra}, i.e. for higher order tensors. Quite
surprisingly, not only very little is known in this respect, but
moreover the available results are rather recent \cite{CS,qi:eigenvalues,qi:rank,qi:eigenvalues_invariants,ni:degree}
 -- albeit, as mentioned above, older results dealing with algebraic
analysis of differential equations and which can be interpreted in
this direction are available in the literature
\cite{Rohrl,RohrlLong,Walcher}.

In this respect, it turns out that the problem we are interested
in (that is, third order symmetric traceless tensors in three
dimensions) is precisely the simplest nontrivial and
non-degenerate class of tensors (the same problem in dimension two
turns out to display a rather special and in many ways degenerate
behavior \cite{Virga2D}), so we believe our results are also of
general interest, as they show the kind of -- rather
counterintuitive -- behavior one can meet in studying the
eigenvalues problem in multilinear algebra.

As well known, a tensor $T$ of order $k$ on $V$ is
a $k$-linear map, \beq \label{eq:tensors} T : V \times ... \times
V = V^k \to \R  ; \eeq it is also well known that the algebra of
completely symmetric tensors on $\R^n$ is isomorphic to the
algebra of polynomials in $\R^n$.

By duality \eqref{eq:tensors} also defines a $(k-1)$-linear map,
which we denote by $\^T$ (by a standard abuse of notation, in the
following we will also denote this by $T$),
\beq \^T : V^{k-1} \to V  ; \eeq for second order tensors this is just
the standard description of a matrix as a linear operator in $V$.
For third order tensors, this associates to $T$ a quadratic map
$\^T : V \times V \to V$.

We say that ${\bf v} \in V$ is an \emph{eigenvector} of $T$ (with
\emph{eigenvalue} $\la \in {\bf C}$) if \beq\label{eq:evtens} \^T
({\bf v}, ... , {\bf v}) =\la  {\bf v}  . \eeq For second
order tensors, this coincides with the standard definition of
eigenvectors and eigenvalues.
\begin{rem}\label{rem:2}
It should be stressed that for tensors of order $k
\not= 2$, eigenvectors come in linear spaces, but these do not
share a common eigenvalue -- that is, eigenvectors are well
defined but eigenvalues are not. In fact, if we consider multiples
of the eigenvector ${\bf v}$, i.e. ${\bf w} = \a {\bf v}$ (with
$\a \not= 0$), we have \begin{eqnarray} \^T ( {\bf w},... {\bf w}
) &=& \^T ( \a {\bf v}, ... , \a {\bf v} ) =\a^{k-1} 
\^T({\bf v} , ... , {\bf v} ) =\a^{k-1}  \la  {\bf v} \nonumber\\
&=&  \a^{k-2}  \la  \( \a  {\bf v} \) =\a^{k-2}  \la
 {\bf w}  . \end{eqnarray} The situation is of course
different if we require the eigenvectors to be of unit length, as
this requirement also uniquely determines the eigenvalue (up to a
sign in case $k$ is odd). \EOR
\end{rem}
We (obviously) reach the same equation \eqref{eq:evtens} if we
consider the minimization of the homogeneous function $\Phi_T$ of
degree $k$ associated to the tensor $T$, i.e. of
\beq \Phi_T  :=  T_{i_1 , ... , i_k}  x^{i_1} ... x^{i_k}  , \eeq
constrained to the unit sphere $|{\bf x}| = 1$. In fact, in this
case one introduces the Lagrange multiplier $\la$ and the
constraint term
$$ -  \frac{k}{2}  \la  x^i x^i  , $$ and minimization of
\beq \Psi_T =\Phi_T  -  \frac{k}{2}  \la  x^i x^i \eeq leads
precisely to \eqref{eq:evtens}. It should be noted that, again, in
this context one is only interested in \emph{real} solutions.

As well known, when one is properly taking into account
multiplicities, a square matrix in $\R^n$ always admits $n$
algebraic eigenvalues. This is generalized by Proposition~\ref{prop:2} below
\cite{qi:eigenvalues_invariants,CS} for arbitrary tensors of rank $m$ over $\mathbf{C}^n$. Note
that for an eigenpair $(\la , {\bf v} )$ we have an (equivalent)
eigenpair $(t^{m-2} \la , t {\bf v} )$ for any $t$ with $|t|=1$
(if we are in $\R^n$, only $t = \pm 1$ are admissible); thus we
should speak of equivalence classes of eigenpairs. The following result is given in Cartwright \& Sturmfels \cite{CS}; see also Qi \cite{qi:eigenvalues,qi:rank,qi:eigenvalues_invariants,ni:degree}.
\begin{prop}\label{prop:2}
If a tensor $A$ of rank $m \ge 3$ over $\mathbf{C}^n$ admits a finite number
of equivalence classes of eigenpairs, their number counted with
multiplicity is
\beq\label{eq:EVN1} E(m,n) =\frac{(m-1)^n -1}{m-2}
=\sum_{j=0}^{n-1} (m-1)^j  . \eeq
\end{prop}
For the case of interest here, i.e. $m=3$, formula
\eqref{eq:EVN1} provides \beq E(3,n) =2^n - 1  , \eeq in
agreement with R\"{o}hrl formula \eqref{eq:rohrlN}; in particular, we
get
\beq E(3,2) =3,\quad   E(3,3) =7   . \eeq

Note also that if we work in $\R^n$, the number of eigenpairs is
obtained simply by multiplying the number of equivalence classes
by two; on the other hand, some of the eigenpairs could be complex
rather than real, so Proposition~\ref{prop:2} only provides an upper bound on
the number of \emph{real} critical points. Thus the maximal number
of critical points for the constrained potential associated to a
tensor of order three in two dimensions is six, i.e. three pairs
(as in \cite{Virga2D}), in dimension three this number is fourteen
(as in \cite{GV2016}), and in dimension four it is thirty.

More detail about eigenvectors of higher order tensors is given in a paper by Walcher \cite{Wal18} (whom we thank for discussing a preliminary version of this with us).

\section{Octupolar tensors in dimension 3 and their eigenvalues}
\label{sec:symmtensors}

In the following we will be especially interested in
\emph{completely symmetric} tensors. In this case, it is convenient
to extract the trace terms. More precisely, one considers tensors
all of whose partial traces are zero.

We thus want to consider a fully symmetric and completely
traceless tensor
$\A$ with components $A_{ijk}$ (all partial traces being zero) . Taking into account the full
symmetry and the condition of zero partial traces, the only
independent components are \begin{eqnarray}
&&A_{123}= \a_0  ; \nonumber \\
&&A_{111} = \a_1  ,\   A_{222} =\a_2 ,\   A_{333}  = 
\a_3\,  ; \label{eq:Acomp} \\
&&A_{122} = \b_1 ,\   A_{233} =\b_2  ,\   A_{311}  = 
\b_3  . \nonumber \end{eqnarray} The partial trace condition
gives \beq A_{133} =- (\a_1 + \b_1)  ,   A_{211} =-
(\a_2 + \b_2)  ,   A_{322} =- (\a_3 + \b_3)  . \eeq The
other components are immediately recovered by the tensor symmetry.
The equations $A(\xb,\xb) = \la \xb$ for eigenvectors and
eigenvalues are then written as 
{\small\begin{eqnarray}
2 \a_0  x_2 x_3 + \a_1 (x_1^2 - x_3^2) - 2 \a_2  x_1
x_2 + \b_1  (x_2^2 - x_3^2)  
- 2 \b_2  x_1 x_2 + 2
\b_3  x_1 x_3&=& \la  x_1 ; \nonumber \\
2 \a_0  x_3 x_1 + \a_2 (x_2^2 - x_1^2) - 2 \a_3  x_2 x_3  
+ \b_2  (x_3^2 - x_1^2) - 2
\b_3  x_2 x_3 + 2 \b_1  x_1 x_2&=&\la  x_2 ; \label{eq:Axx} \\
2 \a_0  x_1 x_2 + \a_3 (x_3^2 - x_2^2) - 2 \a_1  x_3
x_1 + \b_3 (x_1^2 - x_2^2) 
 - 2 \b_1 
x_3 x_1 + 2 \b_2  x_3 x_2&=& \la  x_3 . \nonumber
\end{eqnarray}} 
\begin{rem}\label{rem:*1}
$\A$ on \emph{seven} parameters. A general completely symmetric tensor would depend on \emph{ten} parameters. Correspondingly, the potential we will consider in Sect.~\ref{sec:potential} will depend on seven parameters (these will then be reduced, see Sect.~\ref{sec:oriented}), while a generic potential homogeneous of order three in three spatial dimension would depend on ten parameters. \EOR
\end{rem}
\begin{rem}\label{rem:3}
By introducing the vectors \beq{\bf a}  = \
\pmatrix{\a_1 \cr \a_2 \cr \a_3 \cr}  ,\quad   {\bf b}  = \
\pmatrix{\b_1 \cr \b_2 \cr \b_3 \cr}  , \quad  {\bf u}  = \
\pmatrix{x_1 \cr x_2 \cr x_3 \cr}  , \quad  {\bf v} \pmatrix{x_2 x_3
\cr x_1 x_3 \cr x_1 x_2 \cr} \eeq and the matrices
\begin{eqnarray} L &=& \pmatrix{(x_3^2 - x_1^2) & 2 x_1 x_2 & 0
\cr 0 & (x_1^2 - x_2^2) & 2 x_2 x_3 \cr 2 x_1 x_3 & 0 & (x_2^2 -
x_3^2) \cr}  , \nonumber\\ 
\\
M &=& \pmatrix{(x_2^2 - x_3^2) & - 2
x_1 x_2 & 2 x_1 x_3 \cr 2 x_1 x_2 & (x_3^2 - x_1^2) & - 2 x_2 x_3
\cr - 2 x_1 x_3 & 2 x_2 x_3 & (x_1^2 - x_2^2) \cr}  ,\nonumber
\end{eqnarray}
equations \eqref{eq:Axx} are compactly rewritten
as \beq M  {\bf b} =L  {\bf a}  +  \la {\bf u}  -  2
 \a_0  {\bf v}  . \eeq Provided $\det (L) \not= 0$ and/or
$\det (M) \not= 0$ we can solve for ${\bf a}$ and/or for ${\bf
b}$; e.g., assuming $\det(M) \not= 0$ we have \beq {\bf b}  = \
M^{-1}  \left[ L  {\bf a}  +  \la {\bf u}  -  2  \a_0 
{\bf v} \right]  . \eeq
\end{rem}
It may be noted that there are special points on the unit sphere
where \emph{both} $L$ and $M$ have zero determinant; these are the
two poles $(0,0,\pm 1)$ together with $(\pm 1,0,0)$ and $(0,\pm 1
, 0)$; and the four points $(0,\pm 1/\sqrt{2} , \pm 1/\sqrt{2} )$
together with $(\pm 1/\sqrt{2} , 0 , \pm 1/\sqrt{2} )$ and $(\pm
1/\sqrt{2} , \pm 1/\sqrt{2} , 0)$.

This approach provides an expression for the value of the
parameters (belonging to one subset) which make a certain point
$\xb \in S^2$ a critical one for given values of the other
parameters; unfortunately, we are interested in the inverse -- and
natural -- problem, i.e. determining the critical points for a
given (full) set of parameters. These relations can however be
used to check the correctness of computations and results. \EOR

\section{Critical points}\label{sec:potential}

\subsection{General potential and the critical point equations}
\label{sec:genpot}

It follows from our discussion on the tensor ${\bf A}$ that the
general (unconstrained) potential \eqref{eq:genPhi} is thus
written, using coordinates $(x,y,z)$ rather than $(x_1,x_2,x_3)$
as we are from now on working in three dimensions, as
\begin{eqnarray}
\Phi &=& 6  {\a_0}  x y z  +   {\a_1}  x 
\left(x^2-3 z^2\right)  +  {\a_2}  y  \left(y^2-3 x^2
\right)  +  {\a_3}
 z   \left(z^2-3 y^2 \right)\nonumber \\
&+&  3  \left[ {\b_1}  x  \left(y^2-z^2\right)  + \
{\b_2}  y 
   \left(z^2-x^2\right)  +  {\b_3}  z  \left(x^2 - y^2
   \right) \right]  . \label{eq:Phi}
\end{eqnarray}
\begin{rem}\label{rem:4}
This is obviously covariant under inversion (this
implies that the potential can be constant only if it is
identically zero), i.e. \beql{eq:phicov} \Phi (-x,-y,-z) =- \
\Phi (x,y,z)  . \eeq Actually this is just the consequence of
having a homogeneous (of odd degree) potential. This potential
is also covariant under inversion of the parameters (which is a
consequence of its being homogeneous of degree one in these);
including the dependence on parameters in our notation we have
\beq \Phi ( - \xb ; \pb ) =\Phi ( \xb ; - \pb ) =- \Phi
( \xb ; \pb )  , \eeq where $\xb = (x,y,z)$ and $\pb =
(\a_0;\a_1,\a_2,\a_3;\b_1,\b_2,\b_3)$ is the vector of parameters.
\EOR
\end{rem}
\begin{rem}\label{rem:5}
It follows from \eqref{eq:phicov} that the average
of $\Phi$ over the unit sphere is zero. This also implies that
(unless the potential is identically zero) the absolute maximum --
or maxima in case of degenerate ones -- is necessarily positive.
\EOR
\end{rem}
\begin{rem}\label{rem:6}
The potential $\Phi$ is invariant under
simultaneous identical permutations in the triples $(x,y,z)$,
$(\a_1,\a_2,\a_3)$ and $(\b_1 , \b_2 , \b_3)$. This invariance
does \emph{not} extend to simultaneous general rotations in the
same three-dimensional spaces. \EOR
\end{rem}
We are actually interested in critical points for the potential
constrained to the unit sphere $S^2$ (this constraint does
not interfere with the symmetries mentioned in Remark~\ref{rem:4} above).
The constrained potential \beq \Phi_\la  :=  \Phi  +  \Phi_c \eeq
is obtained by adding to $\Phi$ the constraint term
\beq \Phi_c  :=  -  \frac32  \lambda (x^2 + y^2 + z^2 - 1)  ; \eeq
this breaks the covariance under spatial inversion, i.e. in general
\beq \Phi_\la (-x,-y,-z;\la)
 \not=  -  \Phi_\la (x,y,z;\la)  .
\eeq
On the other hand, it is clear that if we also take into account
the possibility of reversing $\la$, we get \beq\label{eq:Psirev}
\Phi_\la (-x,-y,-z; - \la) =-  \Phi_\la (x,y,z; \la)  .
\eeq
Since we shall only consider critical points of the
potential, any constant term can be omitted, and the constraint
term can be written simply as
\beq \Phi_c =\frac32  \la(x^2+y^2+z^2)  . \eeq

It is then immediate to obtain the equation for critical points of
the constrained potential $\Phi_\la$, which are
\begin{eqnarray}
2 \a_0 y z + \a_1 (x^2 - z^2) - 2 \a_2 x y + \b_1 (y^2 - z^2) - 2
\b_2 x y + 2 \b_3 x z &=&  \lambda x \nonumber \\
2 \a_0 z x + \a_2 (y^2 - x^2) - 2 \a_3 y z + \b_2 (z^2 - x^2) - 2
\b_3 y z + 2 \b_1 y x &=&   \lambda y \label{eq:eqcp} \\
2 \a_0 x y + \a_3 (z^2 - y^2) - 2 \a_1 z x + \b_3 (x^2 - y^2) - 2
\b_1 z x + 2 \b_2 z y &=&   \lambda z  . \nonumber
\end{eqnarray}
\begin{rem}\label{rem:7}
These equations are invariant -- as it should be in
view of \eqref{eq:Psirev} -- under the transformation
\begin{equation}\label{eq:phiconinv} (x,y,z;\la)  \to \
(-x,-y,-z;- \la)  . \end{equation} If we also consider inversion
in the parameters, denoting  the critical point equations \eqref{eq:eqcp} as $E (\xb
, \la , \pb) = 0$ (with the same notation used in Remark~\ref{rem:4}
above), we have
\begin{eqnarray}
E (- \xb; - \la ; \pb) &=& E (\xb; \la ; \pb),\nonumber \\
E (\xb; - \la ; - \pb) &=& E (\xb; \la ; \pb), \\
E (- \xb; - \la ; - \pb) &=& - E (\xb; \la ; \pb)  .\nonumber
\end{eqnarray}
(Obviously the map $E \to - E$ leaves the equations $E=0$
invariant as well.) Moreover they are still covariant under the
three simultaneous permutations mentioned in Remark~\ref{rem:6} above. \EOR
\end{rem}
The property \eqref{eq:phiconinv} together with \eqref{eq:phicov}
guarantee that to each critical point $\xb$ (with $k$ stable
directions) is associated another critical point $- \xb$ (with $k$
unstable directions, and hence $\widetilde{k} = n - k$ stable
ones, with $n$ the space dimensionality; in our case $n=3$). This
also means that it would suffice to consider one of the hemispheres.
\begin{rem}\label{rem:8}
As remarked above, the constrained potential reads
\beq \Phi_\la =\Phi  -  \frac32  \la  |{\bf x} |^2  ; \eeq
the equations for critical points are
\beq \nabla \Phi_\la  :=  \nabla \Phi  -  3  \la  {\bf x} =0  . \eeq
If we perform scalar product of (both members of) this equality with  ${\bf x}$, we get
\beq \( {\bf x} \cdot \nabla \Phi \) =3  \la  \( {\bf x} \cdot {\bf x} \)  . \eeq
Recalling now that $\Phi$ is homogeneous of degree three and that
critical points of $\Phi_\la$ are located, by construction, on the
unit sphere $|{\bf x} |^2 = 1$, we obtain in the end that at
critical points we always have
\beq \label{eq:laphi} \la =\Phi  . \eeq
That is, there is a simple relation -- actually, an identity --
between eigenvalues and the
value of the potential at the critical point identified by the
corresponding eigenvector. \EOR
\end{rem}

\subsection{Oriented potential}
\label{sec:oriented}
The potential $\Phi$ in \eqref{eq:Phi} depends on seven parameters; however,
it is possible to simplify it by choosing an adapted reference
frame in $\R^3$, as we are now going to discuss.

If the potential $\Phi$ is not constant on the sphere (the
potential is constant on $S^2$ only if all parameters are zero; we
are obviously not interested in this case), it will have at least
a critical point and actually at least a maximum -- and by
symmetry, a minimum.

We will then choose the $z$ axis so that one of the critical
points is in $(0,0,1)$; by trivial computations, this requires to
have
\beql{eq:oriented1}
\b_1 =-  \a_1  ,\quad   \b_2 =0
 . \eeq
\begin{rem}\label{rem:10}
We can actually choose $(0,0,1)$ to be
a maximum; this will set further constraints, as discussed in a
moment (in Sect.~\ref{sec:maximum}). \EOR
\end{rem}
Moreover, we can still choose an orientation in the $(x,y)$ plane;
in fact \eqref{eq:phicov} implies that there is at least a point
on the circle $S^1 \ss S^2$ identified by $z=0$ in which $\Phi$
vanishes. Thus we choose an orientation in the $(x,y)$ plane by
requiring that $\Phi (1,0,0) = 0$. Again by trivial computations
this implies, recalling also the expression for $\b_1$ in
\eqref{eq:oriented1}, \beql{eq:oriented2} \a_1 =0  ,\quad  \
\b_1 =0  . \eeq

In this way we obtain an ``oriented'' potential which depends only
on the four parameters
$\a_0  ,  \a_2  , \a_3  , \b_3$; 
this reads
\begin{equation} \Phi_{or} = 6  \a_0  x y z  + \
{\a_2}  (y^2-3 x^2)  y  +  \a_3  (z^2-3
y^2)  z 
 +  3  \b_3  ( x^2 - y^2 )  z  . \label{eq:Phior} \end{equation}
From now on we will only deal with this oriented potential.
\begin{rem}\label{rem:11}
By orienting the potential we have
disposed of the invariance under simultaneous permutations
discussed in Remark~\ref{rem:6} (see also below). On the other hand, the
inversion symmetries discussed in Remark~\ref{rem:7} are obviously inherited
by the oriented potential. \EOR
\end{rem}
\begin{rem}\label{rem:12}
When we have several critical points, each of
them can be chosen as the main orienting one; this trivial remark
may be used to simplify the discussion avoiding redundance in some
classification. In particular, in \cite{GV2016} (see in particular
Sect.~5) an alternative parametrization based on this
remark is used and allows one to reduce the parameter space from the
cylinder to a part of it. \EOR
\end{rem}

As we are working in the unit sphere, it is often convenient to
use angular coordinates (with $r=1$); we will set
\begin{eqnarray}  x &=& \cos \vth_1  \cos \vth_2  , \nonumber \\
y &=& \cos \vth_1  \sin \vth_2   , \label{eq:sphercoord} \\
z &=& \sin \vth_1  ; \nonumber \end{eqnarray}
with this choice of angular coordinates,
\beql{eq:vthlim} \vth_1 \in [- \pi/2,\pi/2]   ,\quad     \vth_2 \in [- \pi, \pi],    \eeq
and the
volume element  becomes \beq \dd x \wedge \dd y \wedge \dd z =r^2  \cos \vth_1 \
\dd r \wedge \dd \vth_2 \wedge \dd \vth_1.   \eeq
\begin{rem}\label{rem:13}
It should be noted that this potential has
generically (i.e. for generic\footnote{In this paper, the adjective ``generic'' is given the meaning common in algebraic geometry, that is, it designates a property valid away from the roots of a polynomial in parameter space \cite{CS}.} values of the parameters appearing
in it) no invariance under subgroups of $O(3)$; it retains the
covariance under inversion, which is described by
$$ \vth_1 \to - \vth_1  ,\quad   \vth_2 \to - \vth_2  $$
in polar coordinates. Special invariance properties are possible,
and will be studied, for special values of the parameters \EOR
\end{rem}

The critical point equations \eqref{eq:eqcp} (CPE) read now
\begin{eqnarray}
2  ( \a_0 y z  -  \a_2 x y  +  \b_3 x z ) &=& \lambda  x, \nonumber \\
 2\a_0 x z  -   \a_2 (x^2 - y^2)  - 2(\a_3 + \b_3) y z &=& \lambda  y,\label{eq:pc0} \\
2 \a_0 x y  -  \a_3 (y^2 - z^2)  +  \b_3 (x^2 - y^2) &=&
\lambda  z  . \nonumber 
\end{eqnarray}
In $(0,0,1)$ the first two are identically satisfied,
while the third one yields
\begin{equation} \lambda_N =\a_3  , \end{equation}
where $\la_N$ is the eigenvalue corresponding to the eigenvector
$(0,0,1)$ identifying the North Pole.\footnote{This means that
we can rule out the possibility to have $\a_3 = 0$. In fact, even
in the case this is a local maximum at height zero, we can always
-- see Remark~\ref{rem:5} -- choose the North Pole to be an absolute
maximum, and this is necessarily positive.} Thus normalizing the
potential by setting
\begin{equation}\label{eq:alpha_3=1}
\a_3 = 1,
\end{equation}
as we do in the following, is
equivalent to setting $\la_N = 1$, i.e. to normalizing the
``orienting'' eigenvalue. This will also correspond to normalizing
the potential in the maximum located at the North Pole.
The oriented potential is then written as
\begin{equation} \label{eq:psiorth}
\Psi_{or}=\sin^3 \vth_1   - 
\a_2  \sin 3 \vth_2   \cos ^3\vth_1
+\frac{3}{2}  \sin \vth_1  [(2 \b_3 + 1) \cos 2 \vth_2 
+  2 \a_0 \sin 2 \vth_2 -1 ]  \cos^2 \vth_1   .
\end{equation}
Needless to say, the CPE are also obtained by considering the
gradient of $\Psi_{or}$ w.r.t. the angular coordinates:\footnote{In
order to know the value for the corresponding $\la$, one needs to
express the solution in Cartesian coordinates and go back to
\eqref{eq:pc0}; this is due to the fact that our change of coordinates
was performed imposing $r=1$ and thus the constraint term, which
represents the dynamical origin of $\la$, is absent in the 
angular coordinates.} 
\begin{eqnarray}
\frac{3}{2} \cos \vth_1 \left[ (-1+(1+2 \b_3) \cos 2
\vth_2+2 \a_0 \sin 2 \vth_2) \cos^2\vth_1  \right.& &\nonumber \\
   \left. -  2 \sin^2\vth_1 (-2 +(1+2 \b_3) \cos 2
\vth_2+2 \a_0
\sin 2 \vth_2) + \a_2 \sin 2 \vth_1 \sin 3 \vth_2\right]&=& 0  ,\nonumber \\
\\
\frac{3}{2} \cos ^2\vth_1 \left[ \sin \vth_1 (4 \a_0 \cos
2 \vth_2 -2 (1+2 \b_3) \sin 2 \vth_2)-2 \a_2 \cos \vth_1
\cos 3 \vth_2 \right] &=&0  .\nonumber \end{eqnarray}
We obtain, in the end, a potential which depends on
\emph{three} parameters; this is still a substantial problem, but 
much simpler than the initial one, depending on \emph{seven}
parameters.

\subsection{The maximum condition}
\label{sec:maximum}
In the previous Sect.~\ref{sec:oriented} we have implemented
the requirement to have a critical point in the North Pole
$(0,0,1)$, setting some conditions on the potential parameters. Our
discussion, however, did not enter into the nature of this
critical point. This is what we will presently do.

To distinguish between maxima, minima and saddles, we
just have to compute the Hessian of the constrained potential at
its critical points. We need not consider the full
potential \eqref{eq:Phi}, but we can work directly on the oriented
potential \eqref{eq:Phior}; moreover, we are mainly interested in
the critical point at the North Pole.

If we work in the northern hemisphere, the constraint condition
can be implemented simply by setting
$z =\sqrt{1 - x^2 - y^2}$.
In this way the oriented potential is rewritten as \beq
\label{eq:psin} \Phi_N =[6 \a_0 x y + 3 \b_3 (x^2 - y^2) + (1
- x^2 - 4 y^2) ]  \sqrt{1 - x^2 - y^2}  +  \a_2  (y^2 - 3
x^2 )  y  . \eeq
The Hessian at the North Pole is simply computed as\footnote{To compare the expressions worked out in this paper for the Hessian matrix of the octupolar potential with those featuring in \cite{GV2016}, the reader should heed that these differ by a scaling factor: the Hessian matrix here is three times the Hessian matrix there.}
\beq H_{N} =\pmatrix{\pa^2 \Phi_N / \pa x^2 & \pa^2 \Phi_N / \pa
x\pa y \cr \pa^2 \Phi_N / \pa x \pa y & \pa^2 \Phi_N / \pa y^2
\cr}_{(0,0)}  ; \eeq with trivial computations we get
\beq\label{eq:HNP} H_{N} =3  \pmatrix{2 \b_3 - 1 & 2 \a_0
\cr 2 \a_0 & - (3  + 2 \b_3 ) \cr}  . \eeq
With the reparametrization\footnote{It should be noted that in
our previous work \cite{GV2016} we have used a slightly different
reparametrization, with $\rho$ instead of $\rho/2$. This accounts
for the differences in many of the forthcoming formulas.} 
\beq \label{eq:repar} \a_0 =\frac{\rho}{2}  \cos \chi  ,\quad
  \b_3 =- \frac12  +  \frac{\rho}{2} \sin \chi  ,  \quad
\a_2 = K  , \eeq
where
\begin{equation}\label{eq:first_bounds}
\rho \ge 0,\quad \chi \in [-\pi,\pi],\quad K \in {\bf R},
\end{equation}
 we have
\beq H_{N} =3  \left(
\begin{array}{cc}
 \rho  \sin \chi -2 & \rho  \cos \chi \\
 \rho  \cos \chi  & -\rho  \sin \chi -2
\end{array}
\right)  , \eeq and its eigenvalues are \beq \s_\pm =-  3 \
(2  \pm  \rho )  ; \eeq thus we have a maximum in the North
Pole if and only if \beq \label{eq:rho} 0 \le \rho \le 2  . \eeq
With \eqref{eq:repar}, the oriented (and
reparametrized) potential in \eqref{eq:psiorth} now reads as \beq\label{eq:potev}  \Psi_{or} \
=  -K \sin 3 \vth_{2} \cos^3\vth_{1}+\frac{3}{2} \sin
\vth_{1} \cos^2\vth_{1} [ \rho
    \sin (\chi +2 \vth_{2})-1 ]+\sin ^3\vth_{1}, \eeq
and its analogue in Cartesian coordinates as
\beq\label{eq:potential_53}
\Phi_{or} =z^3-3 y^2 z+3 \rho \cos \chi
x y z  +\frac{3}{2}(\rho  \sin \chi -1)
   \left(x^2-y^2\right)  z+K y
   \left(y^2-3 x^2\right).
   \eeq
The oriented octupolar potential, in either of its representations \eqref{eq:potev} or \eqref{eq:potential_53}, will be the founding stone of our future development.  
\begin{rem}\label{rem:15}
It is immediately apparent from \eqref{eq:potev}
that for $K=0$ the oriented potential is invariant under rotation
by an angle $\pi$ in $\vth_2$, i.e. for $\vth_2 \to \vth_2 \pm
\pi$; similarly, for $\rho = 0$ the oriented potential is
invariant under rotation by an angle $2 \pi /3$ in $\vth_2$, i.e.
for $\vth_2 \to \vth_2 \pm 2 \pi / 3$. More generally, the
potential is always invariant under a rotation by an angle $(2/3)
n \pi$ in $\vth_2$ accompanied by a rotation by an angle $ ( 2 
m  +  4  n/3 )  \pi $ in the parameter $\chi$. \EOR
\end{rem}

\subsection{Critical points and index}\label{sec:index}

The general results mentioned above (Propositions~\ref{prop:1} and \ref{prop:2}) show
that generically we have $14$ critical points (that is, $7$ pairs of
parity-conjugated ones); some of these could be complex, thus not
acceptable in the present context.

As we are mainly interested in maxima, it would be convenient to
have further information about the nature of these critical
points. Such information can be obtained through the use of
Poincar\'e-Hopf index. To discuss this, we pass to consider the
vector field \beq\label{eq:v_field} {\bf v} =\frac{\nabla \Phi}{|\nabla \Phi |}
 . \eeq  Obviously critical points of $\Phi$ correspond to
singularities for ${\bf v}$, and to these singular points we apply
Poincar\'e-Hopf theory \cite[pp.\,239--247]{stoker:differential}. In fact, each (isolated)
non-degenerate singular point $p_k$ has an index $\iota_k$, which
takes the value $\iota_k = +1$ if $p_k$ is a (local) maximum or
minimum, and $\iota_k = - 1$ if $p_k$ is a saddle. The sum of the
indices for all critical points must equal the Euler
characteristic of the two-sphere, i.e. \beq\label{eq:global_constraint} \sum_k \iota_k =2  .
\eeq

In other words, if all critical points are non-degenerate and $2N$
is their number, $2M$ is the number of maxima and minima
(together) and $2S$ is the number of saddles, we must have $M+S=N$
and $M-S = 1$.

The different combinations allowed by these constraints are
reported in the following Table~\ref{table:Ia}.
\begin{table}[H]
	\caption{Combinations of extrema and saddle points
			allowed by the Hopf-Poincar\'e index constraints for scalar
			potentials on a sphere $S^2 \ss \R^3$, allowing only
			non-degenerate critical points. The last column indicates if
			this is realized in our model; note that at bifurcations
			degenerate critical points can be present.}
	\centering
\begin{tabular}{||c|c|c|r|c||}
  \hline
  & $M$ & $S$ & $2N$ & \\
  \hline
  $(a)$ & 1 & 0 &  2 & no \\
  $(b)$ & 2 & 1 &  6 & no \\
  $(c)$ & 3 & 2 & 10 & yes \\
  $(d)$ & 4 & 3 & 14 & yes\\
  \hline
\end{tabular}

\label{table:Ia}
\end{table}

Thus, if all eigenvalues are real (hence $2N=14$) and all critical
points are isolated and non-degenerate (which is not always the
case, as we will see in our discussion), we will have 4 maxima, 4
minima, and 6 saddles; parity conjugation relates maxima to
minima, and saddles to saddles.

In the following, we will meet this ``expected'' situation; but we
will also meet the case where there are only ten real eigenvalues
(thus $2N=10$); in this case we will have 3 maxima, 3 minima and 4
saddles. Both these situations will be shown to be generic, but cases $(a)$ and $(b)$ will never be met.
\begin{rem}\label{rem:16}
It should be mentioned that the transition between 
the generic cases $(d)$ and  $(c)$ is made possible by the occurrence of
degenerate critical points (see also below); more
precisely, of ``monkey saddles'', see \cite{GV2016}; these have
index $\iota = -2$, to be compared with the index $\iota = -1$ for
ordinary saddles. Thus it appears possible that the non-appearance
of the ``more degenerate'' situations $(a)$ and $(b)$ is related
to the fact that, due to the small degree of the potential and to
its symmetries (which e.g. forbid $\Phi (x,y,z) = z^3$, leading to
case $(a)$ above), there is no possibility for the appearance of
``more degenerate'' critical points, with greater (in absolute
value) index, which would be needed for the associate bifurcation
to take place. \EOR
\end{rem}
In the above discussion, we have supposed that only non-degenerate
critical points are present. The index approach cannot give
results in the case of infinitely-degenerate critical points
(e.g. if we have a line of critical points, as we will find in
Sect.~\ref{sec:center}), but critical points with a finite
degeneration can also be present and can be taken into account. In
our present context, this concerns in particular the ``monkey
saddles'' we will meet in Sect.~\ref{sec:axis}, which have index
$\iota = - 2$. In principle, degenerate saddles with index $\iota = -
3$ could also be relevant to our classification. If we denote the
number of ordinary saddles by $2 S_1$, that of monkey saddles by
$2 S_2$, and the number of saddles with index $-3$ by $2 S_3$,
while still denoting by $2 N = 2 (M +S_1+S_2 + S_3)$ the number of
critical points, the different possible situations -- all having
total index $\iota = 2$ -- are summarized in Table~\ref{table:Ib}. It is perhaps  worth noting that
if we classify the different cases just by the number of maxima,
this gives the same classification as before, albeit the same
number of maxima can correspond to different numbers of saddles
and hence of critical points. 
\begin{table}[H]
	\caption{Combinations of extrema and saddle points
		allowed by the Hopf-Poincar\'e index constraints for scalar
		potentials on a sphere $S^2 \ss \R^3$, allowing regular critical
		points as well as finitely degenerate saddles of index
		$\iota=-2,-3$. The last column indicates if this is realized in our
		model; cases $(c_2)$ and $(d_3)$ will only be realized at bifurcation points. We shall indeed encounter instances where the octupolar potential possesses 12 critical points, but these do not fall under case $(d_2)$ (see Remark~\ref{rem:new_1}).}
	\label{table:Ib}
\centering
\begin{tabular}{||l|c|c|c|c|r|c||}
  \hline
  & $M$ & $S_1$ & $S_2$ & $S_3$ & $2N$ &  \\
  \hline
  $(a)$ & 1 & 0 & 0 & 0 & 2 & no \\
  $(b)$ & 2 & 1 & 0 & 0 & 6 & no \\
  \hline
  $(c_1)$ & 3 & 2 & 0 & 0 & 10 & yes \\
  $(c_2)$ & 3 & 0 & 1 & 0 &  8 & yes \\
  \hline
  $(d_1)$ & 4 & 3 & 0 & 0 & 14 & yes \\
  $(d_2)$ & 4 & 1 & 1 & 0 & 12 & no \\
  $(d_3)$ & 4 & 0 & 0 & 1 & 10 & yes \\
  \hline
\end{tabular}
\end{table}
\begin{rem}\label{rem:new_1}
Both Tables \ref{table:Ia} and \ref{table:Ib} describe possible scenarios with increasing, but moderate degrees of complexity. We have abstained from considering a type of critical points that may be added freely, as they do not affect the global constraint \eqref{eq:global_constraint}. These are critical points with index $\iota=0$. The vector field $\mathbf{v}$ in \eqref{eq:v_field} is still singular at these points, but it can be continuously altered all around each of them so as to be made locally equivalent to a uniform field. All such singularities of $\mathbf{v}$ then turn out to be \emph{removable}. As will be shown below (see Sect.~\ref{sec:refl}),  the octupolar potential can indeed exhibit critical points of this type, albeit in rather special circumstances, which are not included in either of the above tables.\EOR
\end{rem}

\subsection{Symmetry of the reduced potential}
\label{sec:symmred}

As already mentioned, orienting the potential destroys the
invariance under rotations (in the $\xb$ space and in the
parameter space). However, the high degree of symmetry of the
tensor $\A$ makes that even the oriented potential has some remnant
of this in the form of discrete symmetries. These are better
described by passing to the representation of $\Psi_{or}$ in polar
coordinates and with the reparametrization \eqref{eq:repar}, i.e.
in terms of $\Phi_{or}$, see \eqref{eq:potev}.

It is easily checked that this is invariant under several discrete
maps; as expected these do not involve the $\vth_1$ angle, which
is fixed by the requirement to have the North Pole as a maximum.
They do act on the other coordinate $\vth_2$ and on the
parameters\footnote{There are also maps acting on $\rho$ by
changing its sign and leaving the potential invariant; these are
not admitted as we have required $\rho \in [0,2]$.} $K, \chi $
(the action given below on the angles $\vth_2$ and $\chi$ are of
course to be meant \emph{modulo} $2 \pi$):
\begin{eqnarray}
\ga_1 & :  (\vth_2 ; K , \rho , \chi ) & \mapsto  (\vth_2 +
\pi , - K , \rho , \chi )  , \nonumber \\
\ga_2 & :  (\vth_2 ; K , \rho , \chi ) & \mapsto \
(- \vth_2 , - K , \rho , - \chi - \pi)  ,\label{eq:parsym1} \\
\ga_3 & : (\vth_2 ; K , \rho , \chi ) & \mapsto  (\vth_2 +
2\pi/3 , K , \rho , \chi + 2\pi/3 )  . \nonumber
\end{eqnarray}

The invariance under $\ga_1$ means that we can limit our study to the region with $K \ge 0$, which we will do. When taking into account the invariance under the three maps, we can limit our study to the parameter region \beq K \ge 0  ,\quad   \chi_0 \le \chi \le \chi_0  +  \frac{\pi}{3}, \eeq for any given $\chi_0$. This can be handy in the study of the more complex situations.

\section{The tetrahedral group}
\label{sec:tetra}

A special role in our discussion will be played by the tetrahedral
group. Some notions about it -- and explicit features of its
fundamental representation -- are collected in Appendix A. (The
reader is referred to classical books \cite{Ham,LLQM,Kit,Ash} for
further detail.)

In this section we collect the main facts needed for our following
discussion; in fact, the potential -- and in particular its maxima
-- are organized according to various subgroups of the tetrahedral
group, depending on the values assumed by the parameters.

We will denote by $(g \Phi)(\xb) = \Phi (g \xb)$ the value of the
potential computed at a transformed point; here $g$ is an element
of the group, and this is identified with its representation (we
always use the defining representation discussed in detail in
Appendix \ref{app:tetra}; this agrees with the orientation we have
considered and singles out accordingly a reference tetrahedron).

The full tetrahedron group consists of ternary rotations around
the main tetrahedral axes (identified by vertices of the
tetrahedron), of binary rotations around axes joining the middle
points of opposite tetrahedron edges, and by reflections through
the planes identified by two rotation axes.

Note also that we have considered an \emph{oriented} potential.
Thus we expect that in general, only the subgroup of $T_d$ leaving
the North Pole fixed (we will refer to these as \emph{oriented
subgroups}) will be relevant. This is given by ternary rotations
around the $z$ axis, and by reflections through the planes passing
through the $z$ axis; in the notations of Appendix
\ref{app:tetra}, these are the elements $\{ M_1 , M_2 , M_3 ,
M_{13} , M_{14} , M_{15} \}$, where $M_1$ is the identity $I$.

The nontrivial subgroups formed by these are the rotation group
$\{ M_1 , M_2 , M_3 \}$ (no reflections); and the reflection
groups $\{ M_1 , M_{13} \}$, $\{ M_1 , M_{14} \}$, $\{ M_1 ,
M_{15} \}$.

Actually, there are special values of the parameters such that $ g
\Phi = \Phi$ for all elements of the tetrahedron group. These are
$\rho = 0,K = \pm \sqrt{2}/2$.

It will turn out that for each of the oriented subgroups mentioned
above there are regions in the parameter space where the potential
admits these as symmetry groups; this will be discussed in Sect.~\ref{sec:symmetries} below.


\section{Classification of symmetries}
\label{sec:symmetries}

As mentioned above, we can always rescale the potential and adopt \eqref{eq:alpha_3=1}, thus
setting $\la = 1$ for the critical point at the North Pole. In such a way
the potential scale is set by the ``principal
eigenvalue'', the one which sets orientation, but  is not necessarily the largest.

\subsection{The cylinder in parameter space}
We have discussed above the limits to be set on the parameters in
order to accommodate our condition to have a maximum at the North
Pole.
These limits mean we have to investigate a limited subset of the
full parameter space. Having reduced the latter to $(\alpha_0,\alpha_2,\beta_3)$, by \eqref{eq:repar}, \eqref{eq:first_bounds}, and \eqref{eq:rho}, we readily see that we can effectively confine our study to  an infinite cylinder $\cyl$ of
unit radius and axis along the line $\alpha_0=0$, $\beta_3=-1/2$. In the equivalent parameters $(K,\rho,\chi)$, which we shall hereafter turn to, $\cyl$ is identified by the combination of \eqref{eq:first_bounds} and \eqref{eq:rho}, and its axis is the line $\rho = 0$. The discussion of
Sect.~\ref{sec:symmred} shows that we do not actually have to study
the full cylinder $\cyl$; in particular, it would suffice to
study the half with $K \ge 0$, which we denote as $\cyl_+$.

In Table~\ref{table:II} below we distinguish several subsets in the unit
cylinder $\cyl$, i.e. its center $\mathcal{C}$ (
$K=0$, $\rho=0$), its axis $\mathcal{A}$ ($\rho = 0$), the disk
$\mathcal{D}$ at $K=0$, and the special points $\mathcal{T}$ on
the axis at height $K =  1/\sqrt{2}$, together with generic points
in the bulk $\mathcal{B}$ of the cylinder. To each of these set
will correspond a given invariance group for the
potential.
Further splitting of these regions is also possible, as shown below. In particular, we
will find that there are special planes $\mathcal{P}$ in the bulk
$\mathcal{B}$ with special symmetry properties, and that
$\mathcal{B}$ has a richer structure than one could think of at
first.

We will now consider in detail the different cases, starting from
the more symmetric ones.
\begin{table}[H]
	\caption{Different subsets in the cylinder
		$\cyl$.} 
	\centering
\begin{tabular}{||l|l|l|l|l||}
  \hline
  & $G$ & parameters ($\a_0,\a_2,\b_3$) & parameters ($K,\rho$) & subset in $\cyl$ \\
  \hline
  (0) & $\{ e \}$ & & & bulk $\mathcal{B}$ \\
  (1) & $D_{\infty h}$ & $\a_0 = \a_2 = 0;  \b_3 = - 1/2$ & $K = \rho = 0$ & center ${\mathcal C}$ \\
  (2) & $D_{2 h}$ & $\a_2 = 0$ & $K = 0$ & disk ${\mathcal D}$ \\
  (3) & $D_{3 h}$ & $\a_0 = 0 ,  \b_3 = - 1/2$ & $\rho = 0 $ & axis ${\mathcal A}$ \\
  (4) & $T_d$ & $\a_0 = 0 ,  \b_3 = - 1/2;  \a_2 =  \pm1/\sqrt{2}$
  & $K =  \pm1/\sqrt{2} ,  \rho = 0$ & points $\mathcal{T} \in {\mathcal A}$ \\
  \hline
\end{tabular} 
\label{table:II}
\end{table}

\subsection{Tetrahedral symmetry: special points $\mathcal{T}$}
\label{sec:tetrasymm}
For the points $\mathcal{T}$, i.e. for $\rho = 0$ and $K = \pm
1/\sqrt{2}$, the potential in \eqref{eq:potential_53}
reduces to \beq \Phi_T^{(\pm )} =z^3  -  \frac32  (x^2 +
y^2)  z  \pm  \frac{1}{\sqrt{2}}  (y^2 - 3 x^2 )  y  .
\eeq This is invariant under the full tetrahedron group (referred
to the ``standard'' regular tetrahedron; by this we mean the one
complying with our orientation choices; see Appendix
\ref{app:tetra}); note that for generic points on $\mathcal{A}$
(see the next subsection) we do not have tetrahedral invariance,
but only that under the $D_{3h}$ group.

\begin{figure}
	\centering
		\includegraphics[width=100pt]{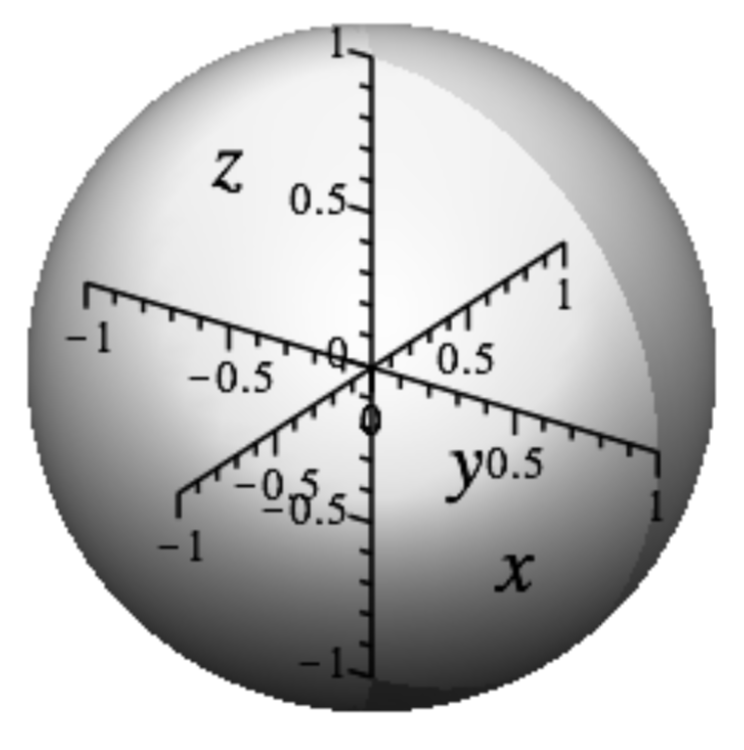}\\
	\includegraphics[width=150pt]{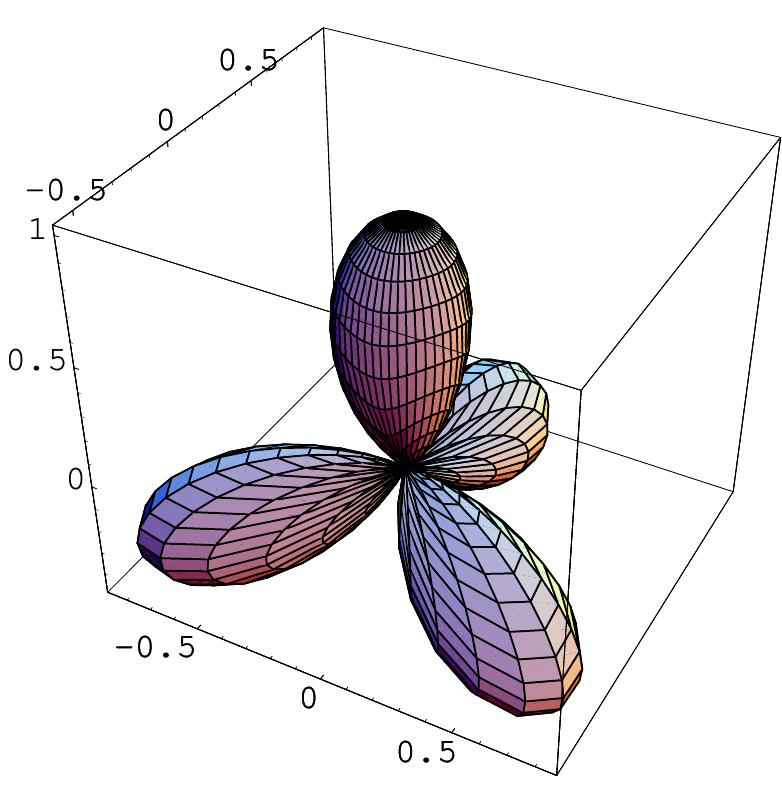} \hfill
		\includegraphics[width=150pt]{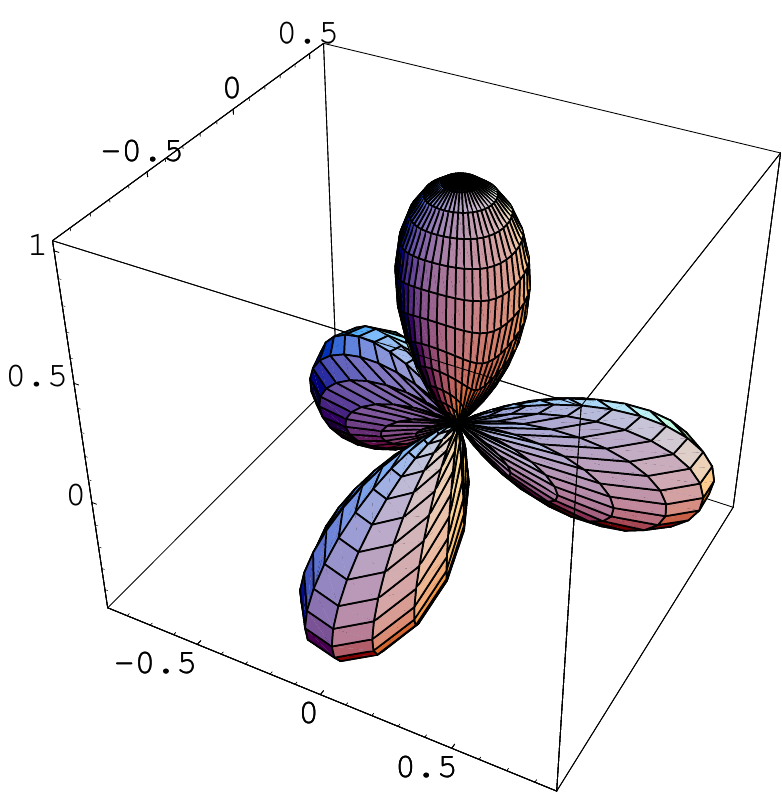} \\
	\caption{The potential $\Phi_T^{(\pm)}$ on the unit sphere.
		Left: $\Phi_T^{(+)}$ (i.e. $K = 1/\sqrt{2}$); right: $\Phi_T^{(-)}$ (i.e. $K = - 1 / \sqrt{2}$).
		The two cases are equivalent via a rotation by an angle $(2n +1) \pi/3$. The ``spherical compass'' on top of these panels shows the orientation of the Cartesian frame shared by the polar plots of the octupolar potential in all remaining panels below.}\label{fig:tetra}
\end{figure}

Now the maxima (and hence the minima) are all non-degenerate, take
the same value, and their position are just at the vertices of the
``standard'' regular tetrahedron. This is depicted in
Fig.~\ref{fig:tetra}.
All polar plots such as those in Fig.~\ref{fig:tetra} are three-dimensional renderings of the surface covered by the vector $\pot(\e_r)\e_r$ as $\e_r=\xb/|\xb|$ spans the unit sphere $S^2$. Since $\pot(\e_r)=-\pot(-\e_r)$, in all these plots the minima of $\pot$ are invaginated underneath its maxima.

\begin{figure}
	\includegraphics[width=150pt]{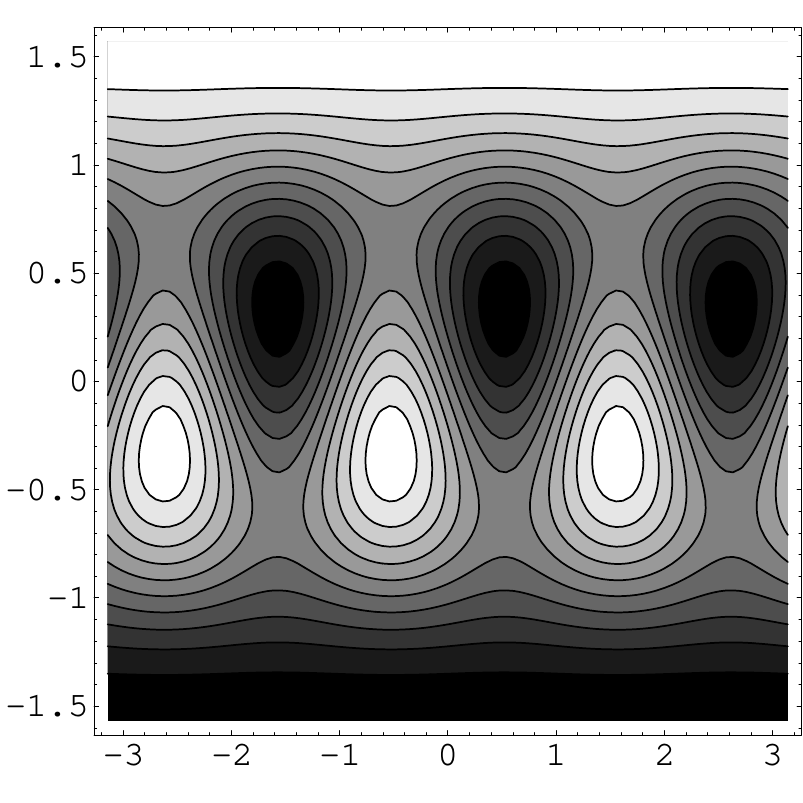}\hfill
	\includegraphics[width=150pt]{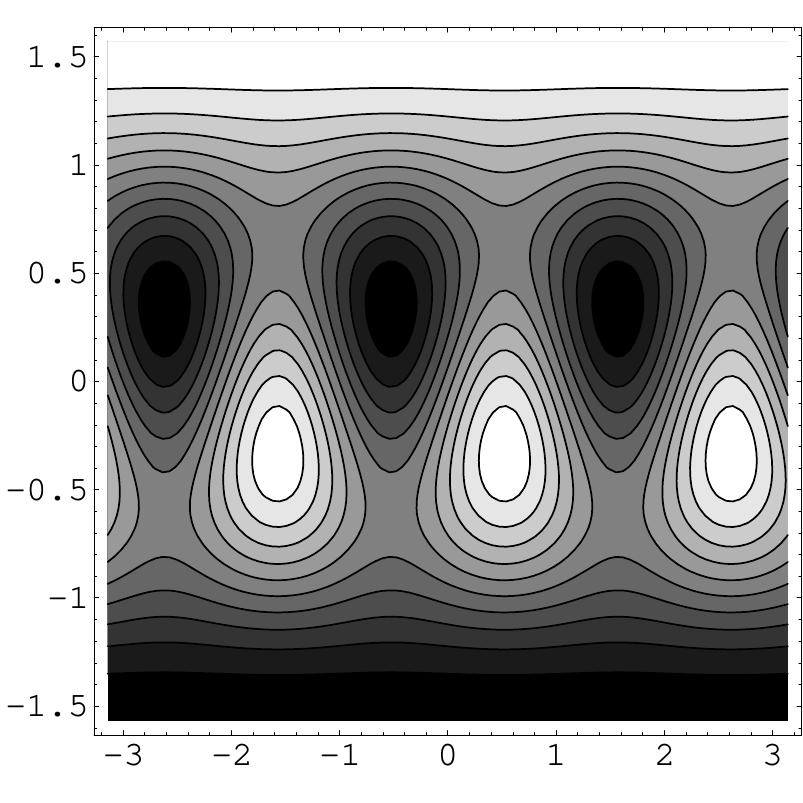}\\
	\caption{Contour plots of the potentials $\Psi_T^{(+)}$ (left) and $\Psi_T^{(-)}$ (right) on the plane $(\vth_2,\vth_1)$;
		see \eqref{eq:phitcc}. In these plots, as in the others that follow, the brightest points represent the maxima of the potential, whereas the darkest points represent the minima. One easily checks that there are indeed 14 critical points (two of them
		at the poles). The symmetry properties of $\Psi_T^{(\pm)}$ are evident
		from this picture. The upper and lower border, at $\vth_1 = \pm \pi/2$,
		correspond to the North and South Poles, respectively. The two plots are coincident up to a shift in $\vth_2$, in accord with the predicted invariance under $\ga_1$ in \eqref{eq:parsym1}.}\label{fig:phiT}
\end{figure}

In angular coordinates, from \eqref{eq:potev} we have
\begin{equation} \Psi_T^{(\pm)} = \sin^3 \vth_1  -  \frac32
 \sin \vth_1  \cos^2 \vth_1  \mp  \frac{1}{\sqrt{2}} 
\sin 3 \vth_1  \cos^3 \vth_1  . \label{eq:phitcc}
\end{equation}
See Fig.~\ref{fig:phiT} for a contour plot of this function. All contour plots such as those in Fig.~\ref{fig:phiT} are on the plane $(\vth_2,\vth_1)$, which develops the unit sphere $S^2$ so that the upper side $\vth_1={\pi}/{2}$ corresponds to the North Pole and the lower side $\vth_1=-{\pi}/{2}$ corresponds to the South Pole.

The potential is obviously invariant under shifts by
$(2/3) \pi$ in $\vth_2$, and that
parameter inversion $K \to - K$ is equivalent to an inversion or
to a shift by $\pi/3$ in $\vth_2$; i.e.
\begin{eqnarray}
\Psi_T^{(\pm)} (\vth_1 , \vth_2 +
2\pi/3 ) &=& \Psi_T^{(\pm)} (\vth_1, \vth_2)  , \\
\Psi_T^{(-)} (\vth_1, \vth_2 ) &=& \Psi_T^{(+)} (\vth_1 ,
- \vth_2 ) =\Psi_T^{(-)} (\vth_1 , \vth_2 + \pi/3 )  . \end{eqnarray}
These properties imply that it suffices to study one of the two
cases, say $\Psi_T^{(+)}$; from now on we will just consider this,
and refer to it simply as $\Psi_T$ (and correspondingly for
$\Phi^{(+)}_T$ and $\Phi_T$).

Eigenvalues  and the corresponding critical points (i.e. normalized eigenvectors) for $\Phi_T$ are listed in Table~\ref{table:III} below (recall that $\la = \Phi$ at critical points).
\begin{table}[H]
	\caption{Critical points for tetrahedral symmetry; here $\nu_m = \arcsin (1/3)$, $\nu_s = \arcsin (1/\sqrt{3})$.}
	\label{table:III}
	\centering
	\begin{tabular}{||r||l||c|c||c||l||}
		\hline
		$n$ & $\la$ & $\vth_1$ & $\vth_2$ &  $\Psi$ & type \\
		\hline
		1 &  -1 & $-\pi/2$ & -- &   $-1$ & min \\
		2 &  -1 & $- \nu_m$ & $- \pi/2$ &  $-1$ & min  \\
		3 &  -1 & $- \nu_m$ & $ 5\pi/6$ &   $-1$ & min \\
		4 &  -1 & $- \nu_m$ & $\pi/6$ &   $-1$ & min  \\
		\hline
		5 &  0 & $- \nu_s$ & $-\pi/2$ &  0  & saddle \\
		6 &  0 & $- \nu_s$ & $ {5\pi}/{6}$ &   0  & saddle \\
		7 &  0 & $- \nu_s$ & $\pi/6$ &   0  & saddle \\
		8 &  0 & $\nu_s$ & $\pi/2$ & 0  & saddle \\
		9 &  0 & $\nu_s$ & $-{5\pi}/{6} $ &  0 & saddle \\
		10 & 0 & $\nu_s$ & $- \pi/6$ &  0 & saddle \\
		\hline
		11 & 1 & $\nu_m$ & $\pi/2$ & 1 & max \\
		12 & 1 & $\nu_m$ & ${5\pi}/{6}$ & 1 & max \\
		13 & 1 & $\nu_m$ & $- \pi/6$ &  1 & max \\
		14 & 1 & $\pi/2$ & -- &  1 & max \\
		\hline
	\end{tabular} 
\end{table}

\subsection{Symmetry $D_{\infty h}$: center $\mathcal{C}$}
\label{sec:center}

At the center $\centre$ of the cylinder $\cyl$, i.e. for
$K=\rho=0$, the potential in \eqref{eq:potential_53} is just \beq \Phi_\infty =z^3  - \
\frac32  (x^2 + y^2)  z  ; \eeq correspondingly, its variant in spherical coordinates \eqref{eq:potev} is
\beq\label{eq:65} \Psi_\infty =\frac18  ( 3  \sin \vth_1  -  5  \sin 3 \vth_1 )  . \eeq

The symmetry under rotations (about the $z$ axis) is immediately
apparent, as well as the symmetry under reflection in any vertical
plane. We thus have a $D_{\infty h}$ symmetry. The potential is
also covariant under inversion in $z$,
\beq\Phi_\infty (x,y,-z) =-  \Phi_\infty (x,y,z)  . \eeq

The critical point equations \eqref{eq:eqcp} are now
\begin{eqnarray}
- 3  x  z &=& \la  x \nonumber \\
- 3  y  z &=& \la  y \\
3 z^2  -  \frac32 (x^2 + y^2) &=& \la z  , \nonumber
\end{eqnarray}
It follows immediately from the first two
equations that the critical points not lying at the poles of the
sphere have $\la = - 3 z$. More precisely, inserting this into the
third equation -- and recalling that $x^2 + y^2 = 1 - z^2$ -- it
turns out that for these we have \beq z =\pm  1 / \sqrt{5},\quad
   \la =\mp  3 / \sqrt{5}  . \eeq
   \emph{All}
points on this circle are (obviously, degenerate) critical ones;
the only  non-degenerate critical points are the maximum and the
minimum in the North and South Poles.

We get of course the same result working in angular coordinates. From \eqref{eq:65},
we have that the critical points are located in the North and
South Poles ($\vth_1 = \pm \pi/2$, value $\pm 1$); and on the
parallels with $ \vth_1 =  \pm  \arccos ( 2 / \sqrt{5} ) \doteq \pm
0.46$. In particular, we have a circle of degenerate minima
(value $- 1 / \sqrt{5}$) for $ \vth_1 = \arccos( 2 / \sqrt{5} )$, and a circle of degenerate maxima (value $1 /
\sqrt{5}$) for $ \vth_1 = - \arccos (2 / \sqrt{5} )$.

\begin{figure}
  \begin{tabular}{cc}
  \includegraphics[width=150pt]{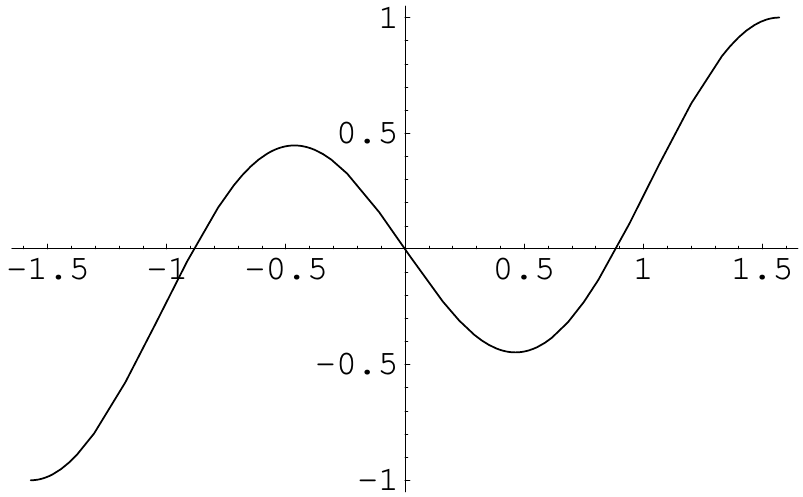} &
  \includegraphics[width=150pt]{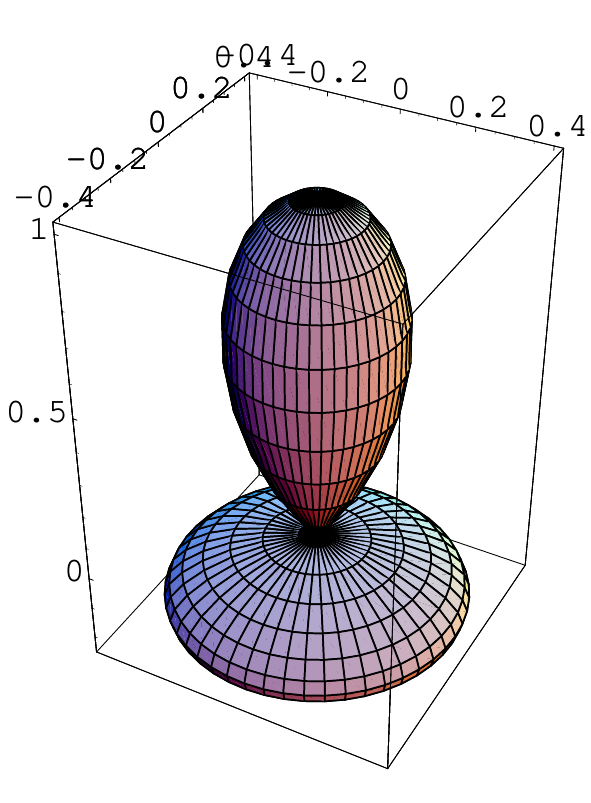}\\
  (a) & (b)
\end{tabular}
 \caption{The potential $\Psi_\infty$ depends only on $\vth_1$. $(a)$ Plot of $\Psi_\infty$ as a function of $\vth_1$: we observe the (circle of)
  maxima for $\vth_1 = \arcsin{(- 1 / \sqrt{5} )} \doteq - 0.46 $, and the
  (circle of) minima for $\vth_1 = \arcsin( 1 / \sqrt{5} ) $. $(b)$ The potential $\Phi_\infty$ as a ``spherical plot'': again we observe the circle of maxima at $z = - 1 / \sqrt{5}$, and the circle of minima for  $z = 1/\sqrt{5}$.}
\label{fig:dinf}
\end{figure}

The critical points are all non-degenerate \emph{modulo} the degeneration
enforced by the SO(2) symmetry (this also requires one of the
eigenvalues of the Hessian to be zero). Critical points at the
poles have an Hessian with nontrivial eigenvalue $\sigma_0 = \pm
6$; those on the circles have an Hessian with nontrivial
eigenvalue $\sigma_1 = \pm 12/\sqrt{5}$.
This case is depicted in Fig.\ref{fig:dinf}.
\subsection{Simmetry $D_{3 h}$: the axis $\mathcal{A}$}
\label{sec:axis}
\begin{figure}
	\begin{tabular}{cc}
		\includegraphics[width=150pt]{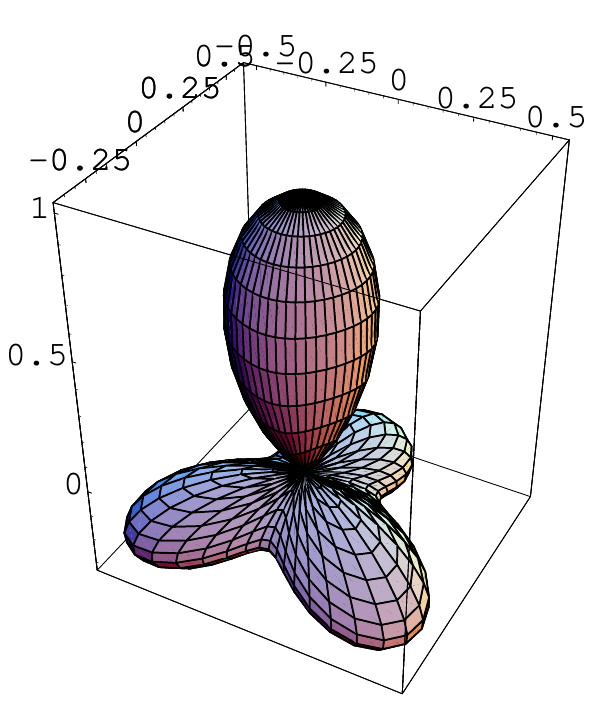} &
		\includegraphics[width=150pt]{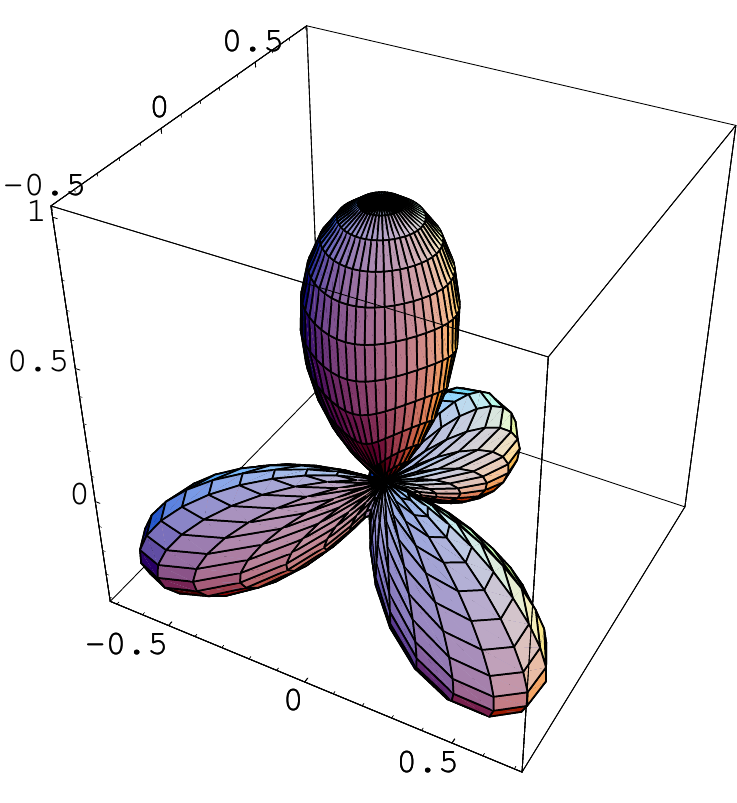} \\
		$K = 1/4$ & $K = 1/2$ \\
		\includegraphics[width=150pt]{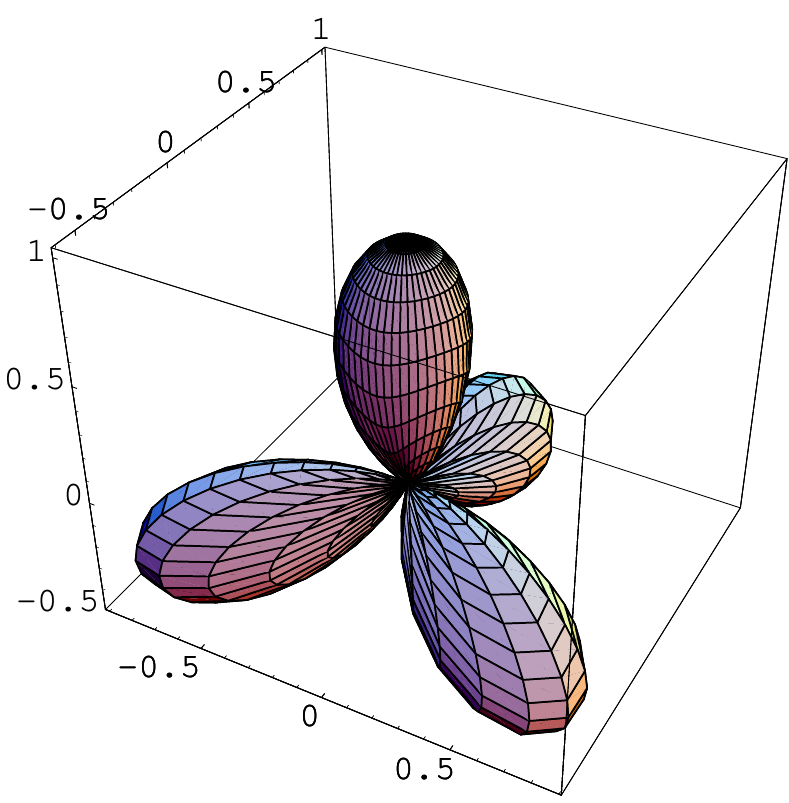} &
		\includegraphics[width=150pt]{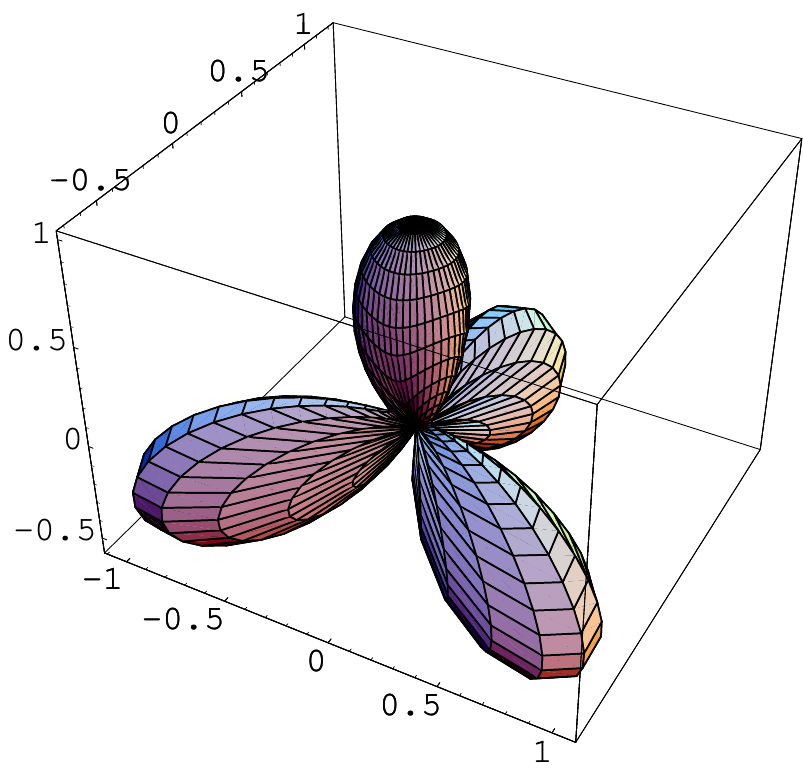} \\
		$K = 3/4$ & $K = 1$  \end{tabular} \\
	\caption{Polar plots of the potential $\Phi_3$ for different values of the
		parameter $K$. For low $K$ the maximum in the North Pole is the highest, while
		for higher $K$ it becomes smaller than the three symmetric maxima in the
		southern hemisphere.}\label{fig:d3hSP}
\end{figure}

\begin{figure}
	\begin{tabular}{cc}
		\includegraphics[width=150pt]{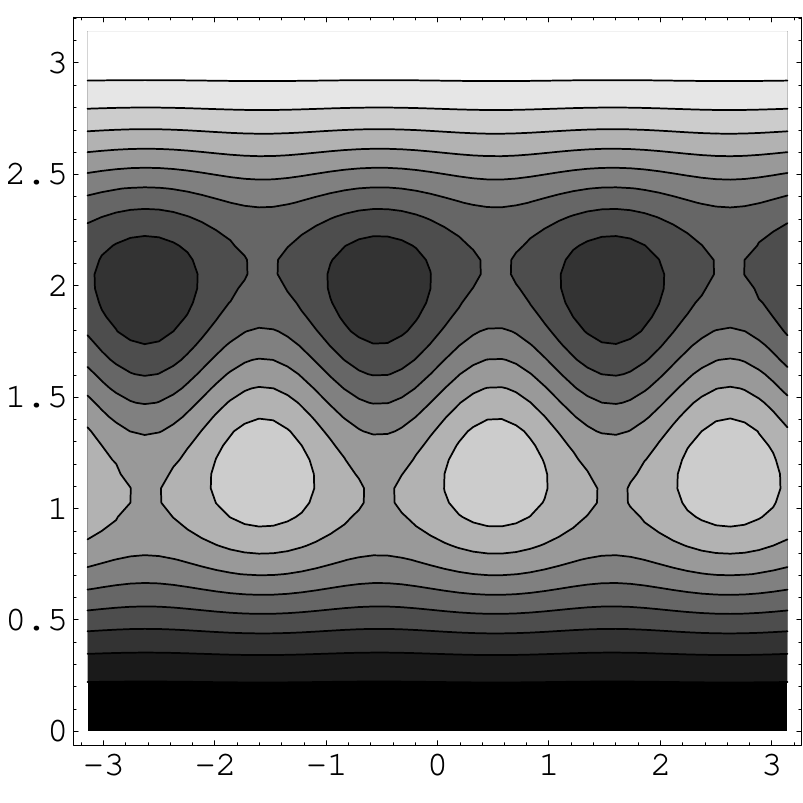} &
		\includegraphics[width=150pt]{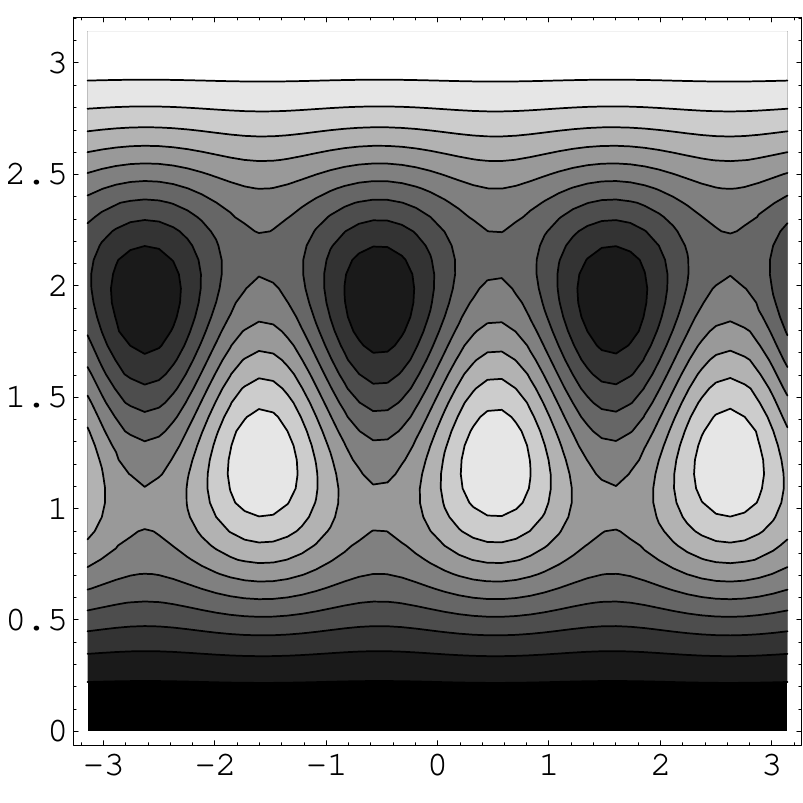} \\
		$K = 1/4$ & $K = 1/2$ \\
		\includegraphics[width=150pt]{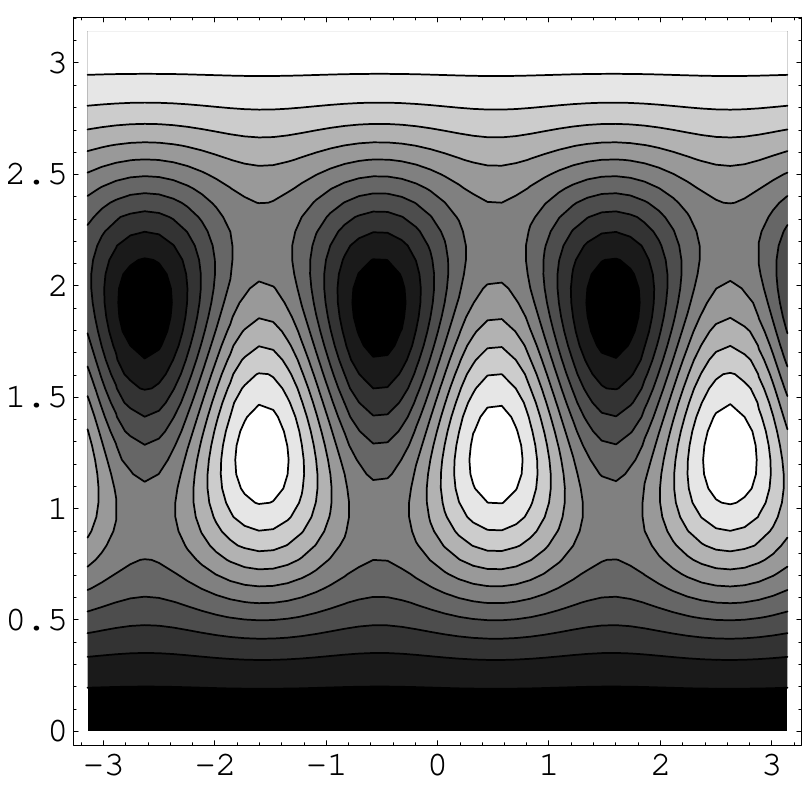} &
		\includegraphics[width=150pt]{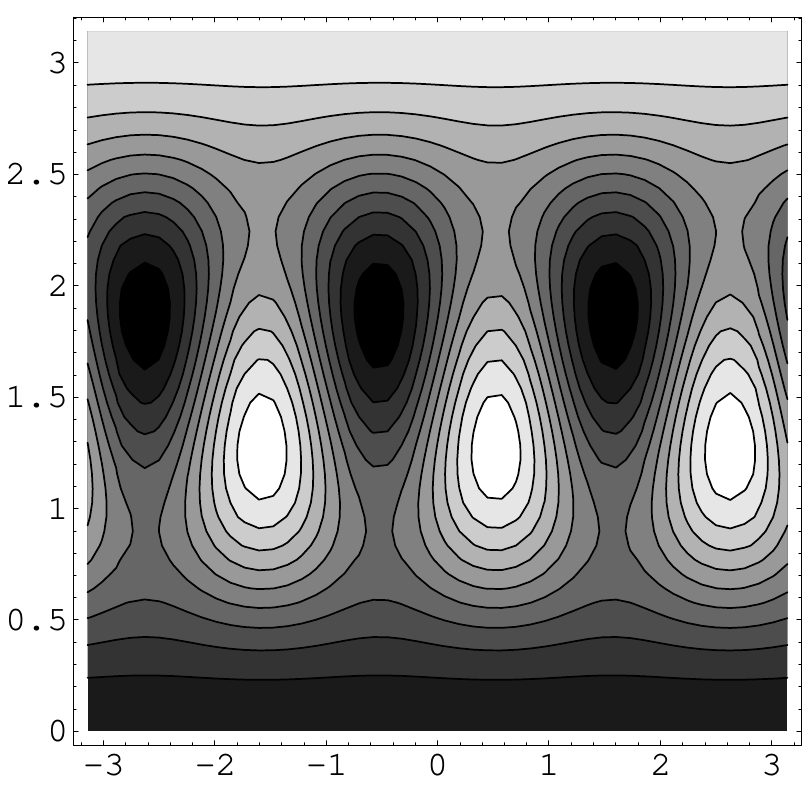} \\
		$K = 3/4$ & $K = 1$  \end{tabular} \\
	\caption{The same as in Fig.~\ref{fig:d3hSP} but with contour plots.
		Note again that for low $K$ the maximum in the North Pole is the highest, while
		for higher $K$ it becomes smaller than the three symmetric maxima in the
		southern hemisphere.}\label{fig:d3hCP}
\end{figure}

On the axis $\mathcal{A}$ of the cylinder (i.e. for $\rho = 0$),
the potential in \eqref{eq:potential_53} reads \beq \Phi_3 =K  y  (y^2 - 3 x^2)  +
 z  \left( z^2 - \frac32 (x^2 + y^2) \right)  , \eeq and the potential expressed in \eqref{eq:potev} in 
angular coordinates is \beq \Psi_3 =\sin^3 \vth_1  -  K 
\cos^3 \vth_1  \sin 3 \vth_2  -  \frac32  \cos^2
\vth_1  \sin \vth_1  . \eeq Of course, we need only consider
points different from the special points $\centre$ and $\tetra$
considered before.
Here $K \ge0$ is a free parameter. Graphical illustrations of both $\Phi_3$ and $\Psi_3$ are shown in Figs.~\ref{fig:d3hSP} and \ref{fig:d3hCP}.

The potential is invariant under the inversion $x \to - x$ and
under rotations by $2 \pi /3$ around the $z$ axis (i.e. shift by
$2 \pi / 3$ in $\vth_2$; thus it would suffice to study the
potential in the sector $0 \le \vth_2 \le 2 \pi/3 $; that is, its
symmetry group is fully described by the matrices
\begin{eqnarray} & &  \left(
\begin{array}{lll}
 1 & 0 & 0 \\
 0 & 1 & 0 \\
 0 & 0 & 1
\end{array}
\right)  , \  \left(
\begin{array}{lll}
 -\frac{1}{2} & -\frac{\sqrt{3}}{2} & 0 \\
 \frac{\sqrt{3}}{2} & -\frac{1}{2} & 0 \\
 0 & 0 & 1
\end{array}
\right)  , \  \left(
\begin{array}{lll}
 -\frac{1}{2} & \frac{\sqrt{3}}{2} & 0 \\
 -\frac{\sqrt{3}}{2} & -\frac{1}{2} & 0 \\
 0 & 0 & 1
\end{array}
\right)  ;\nonumber \\
\\
& & \left(
\begin{array}{lll}
 -1 & 0 & 0 \\
 0 & 1 & 0 \\
 0 & 0 & 1
\end{array}
\right)  , \  \left(
\begin{array}{lll}
 \frac{1}{2} & -\frac{\sqrt{3}}{2} & 0 \\
 -\frac{\sqrt{3}}{2} & -\frac{1}{2} & 0 \\
 0 & 0 & 1
\end{array}
\right)  ,\   \left(
\begin{array}{lll}
 \frac{1}{2} & \frac{\sqrt{3}}{2} & 0 \\
 \frac{\sqrt{3}}{2} & -\frac{1}{2} & 0 \\
 0 & 0 & 1
\end{array}
\right). \nonumber\end{eqnarray}

The potential is of course also covariant under inversion,
\beq \Phi_3 (- \vth_1 , - \vth_2) =-  \Phi_3 (\vth_1 , \vth_2)
 , \eeq which corresponds to $(x,y,z) \to (-x,-y,-z)$; thus it
suffices to consider $\vth_1 >0$ (i.e. the northern hemisphere, as
we already know from general discussion).

The critical point equations are now
\begin{eqnarray}
\cos \vth_1  ( \cos^2 \vth_1 - 4 \sin^2 \vth_1 - K
\sin 2 \vth_1 \sin 3 \vth_2 ) &=& 0, \nonumber \\
\label{eq:d3hcrit}\\
K  \cos^3 \vth_1  \cos 3 \vth_2 &=& 0.\nonumber
\end{eqnarray}
The second of these implies that critical points are either at the
poles, with $\vth_1 = \pm \pi/2$ (there is of course a maximum,
value $\Phi_3 = 1$, at the North Pole; and a minimum, value
$\Phi_3 =-1$, at the South Pole), or on the meridians identified
by \beql{eq:d3hth2crit} \vth_2 \
=  \pm \frac{\pi}{6}  ,    \pm \frac{\pi}{2}  ,    \pm
\frac{5 \pi}{6}  . \eeq
As $\vth_2$ enters the equations only through terms $3 \vth_2$,
it is obvious that cases with $\vth_2$ differing by $2 \pi / 3$
are equivalent (so the cases $\vth_2 = - \pi/2, + \pi/6, + 5
\pi/6$ are equivalent to each other, and so are the cases $\vth_2
= - 5 \pi/6, - \pi/6, \pi/2$); it thus suffices to consider 
$\vth_2 =\pm {\pi}/{6}$  , which yields $\sin 3
\vth_2 = \pm 1$.
With this choice for $\vth_2$, the first of \eqref{eq:d3hcrit}
reads \beq \cos \vth_1  \(3  -  5  \cos 2 \vth_1  \pm \
2  K  \sin 2 \vth_1 \) =0  . \eeq
Assuming $\vth_1 \not= \pm \pi/2$, this further reduces to
\begin{equation}\label{eq:plotKaxisP}
\cos^2 \vth_1  -  4  \sin^2 \vth_1  \mp \
2  K  \sin\vth_1  \cos \vth_1 =0  .
\end{equation}
Writing $\eta = \tan \vth_1$, and taking off an overall factor $\cos^2 \vth_1 \not= 0 $, this reads
\beql{eq:d3heta} 1  -  4  \eta^2  \mp  2  K  \eta =0  .
\eeq
Thus for $\vth_2 = - 5 \pi/6, - \pi/6 , \pi/2$ we get
\beql{eq:d3hetasolm} \eta^-_{\pm} =-  \frac{K}{4}  \( 1  \pm  \sqrt{1 + 4/K^2} \)  , \eeq while for $\vth_2 = -\pi/2 , \pi/6, 5 \pi/6$ we get
\beql{eq:d3hetasolp} \eta^+_{\pm} =\frac{K}{4}  \( 1  \pm  \sqrt{1 + 4/K^2} \)  . \eeq
This means that we have in total twelve nontrivial critical points, i.e. fourteen counting also those at the poles.
We know (see Sect.~\ref{sec:index}) that four of these will be
maxima, other four will be minima, and the remaining six will be
saddles.

Recalling  that $- \pi/2 < \vth_1 < \pi/2$, one then easily shows that critical points can be represented as $\vth_1 = \xi^\pm_\pm$, where
\beql{eq:d3hxi} \xi^+_\pm =\arcsin \( \eta^+_\pm / \sqrt{1 + (\eta^+_\pm)^2} \)  ,\quad    \xi^-_\pm =\arcsin \( \eta^-_\pm / \sqrt{1 + (\eta^-_\pm)^2} \)  . \eeq 

To characterize the nature of critical points, we can either consider the potential on the meridians \eqref{eq:d3hth2crit} and on the parallels identified by  \eqref{eq:d3hxi} (with the aid of \eqref{eq:d3hetasolm} and \eqref{eq:d3hetasolp}); or study the eigenvalues of the Hessian at these critical points. In fact, the eigenvalues of the Hessian on $\vth_2 = \pi/6$ (and equivalent meridians) are
\beql{eq:d3hlap} \Lambda^+ =\left\{ 9  K  \cos^3\vth_1  ,  \frac38  \(2 K \cos \vth_1 + 6 K \cos 3 \vth_1 - \sin \vth_1 + 15 \sin 3 \vth_1 \) \right\}  , \eeq while those on $\vth_2 = - \pi/6$ (and equivalent meridians) are
\beql{eq:d3hlam} \Lambda^- =\left\{ - 9  K  \cos^3 \vth_1  ,  - \frac38  \(2 K \cos \vth_1 + 6 K \cos 3 \vth_1 + \sin \vth_1 - 15 \sin 3 \vth_1 \) \right\}  . \eeq
Recalling that $\cos \vth_1  > 0$ and $K>0$, which fixes the sign of the first eigenvalue in both cases, we see that on $\vth_2 = \pi/6$ (and equivalent meridians) we can only have saddles or minima, and on $\vth_2 = -\pi/6$ (and equivalent meridians) we can only have maxima or saddles.

The above formulas \eqref{eq:d3hlap}, \eqref{eq:d3hlam} also allow us to  compute explicitly the eigenvalues at critical points, and hence the nature of these.
The results are summarized in Table~\ref{table:IV} (which also includes the critical points at the poles, considered as trivial in the previous discussion); we stress that there is no change of stability as $K$ is varied, as follows from \eqref{eq:d3hlap}, \eqref{eq:d3hlam}.

To improve readability of Table~\ref{table:IV}, we have used the shorthand notations
\begin{eqnarray}
q_\pm & := & \[ 10  +  K^2  \pm  K  \sqrt{4 + K^2} \]^{-1}  , \nonumber \\
\tau_\pm & := & \arcsin \( \frac{\kappa_\pm  q_\pm}{2 } \)  ,\label{eq:notTIV}  \\
\zeta_\pm & := & \sqrt{2  q_\pm^3}  \[ K^3  \pm  K^2  \sqrt{4 + K^2}  +  13  K  \mp  5  \sqrt{4 +K^2} \]  . \nonumber \end{eqnarray}
\begin{table}[H]
	\caption{Critical points for the case $D_{3h}$. The relation prescribed by eq.\eqref{eq:laphi} are satisfied, hence $\la$ is not displayed. The shorthand notation \eqref{eq:notTIV} is used.}
	\label{table:IV}
	\centering
\begin{tabular}{||r||c|c||c||l||}
\hline
$n$ & $\vth_1$ & $\vth_2$ & $\Psi$ &  type \\
\hline
 1 & $- \pi / 2$ & $--$ &  -1 & min \\
\hline
 2 & $-\tau_-$ & $-\pi / 2$ & $- \zeta_-$ & min \\
 3 & $- \tau_-$ & $\pi / 6$ &$- \zeta_-$ & min \\
 4 & $- \tau_-$ & $5 \pi / 6$ &  $- \zeta_-$ & min \\
\hline
 5 & $-\tau_+$ & $- \pi / 2$ & $- \zeta_+$ & saddle \\
 6 & $-\tau_+$ & $\pi / 6$ &  $- \zeta_+$ & saddle \\
 7 & $-\tau_+$ & $5 \pi / 6$ & $- \zeta_+$ & saddle \\
\hline
 8 & $\tau_+$ & $-5 \pi / 6$ & $\zeta_+$  & saddle \\
 9 & $\tau_+$ & $- \pi / 6$ &  $\zeta_+$ & saddle \\
 10 & $\tau_+$ & $\pi / 2$ & $\zeta_+$ & saddle \\
\hline
 11 & $\tau_-$ & $- 5 \pi / 6$ & $\zeta_-$ & max \\
 12 & $\tau_-$ & $- \pi / 6$ & $\zeta_-$ & max \\
 13 & $\tau_-$ & $\pi / 2$ & $\zeta_-$ & max \\
\hline
 14 & $\pi / 2$ & $--$ &  1 & max \\
\hline
\end{tabular} 
\end{table}
\begin{rem}\label{rem:*2}
We note that the value of the potential at the non-orienting maxima -- i.e. at the critical points 11, 12, and 13 in Table~\ref{table:IV}, is in explicit terms
\beq\zeta_-  =\frac{\sqrt{2} \left(K^3-\sqrt{K^2+4} K^2+13 K+5
   \sqrt{K^2+4}\right)}{\left(K^2-\sqrt{K^2+4} K+10\right)^{3/2}}  ; \eeq it is a simple matter to check this is always (positive and) increasing with $K$, and that $\zeta_- = 1$ for $K = K_0 := \sqrt{2}/2\doteq0.71$. Thus, these are secondary maxima for $K < K_0$, and become absolute maxima for $K> K_0$. We thus have a ``global bifurcation'' between the phases $D_{3h}^+$ and $D_{3h}^-$ taking place at $K= K_0$. For this value of $K$, the value taken by the potential at the saddle points 5--10 in Table~\ref{table:IV} is $\zeta_+=0$. It should also be noted that $\zeta_+$ is always increasing with $K$. \EOR
\end{rem}
\begin{rem}\label{rem:19}
Let us focus on the maxima. In those other than
the North Pole, we have $\Phi = \Phi_M$ and $\la = \la_M$. 
These are smaller than the corresponding quantities for the
maximum in the North Pole for $K < K_0$, and greater than
those for $K >K_0$. 
In other words, albeit at $K = K_0$ there is no local bifurcation, we have a rearrangement
of the maxima and saddles in terms of their ordering according to
value of the potential. When choosing the orienting
maximum, we could also choose it in such a way that it should always be the largest;
this means that the case with $K > K_0$ necessarily
corresponds to a case present with other values of the parameters,
in which the orienting maximum is the largest. 
Such a choice would break the $D_{3h}$ symmetry around the
orienting axis, and we would be left with a $D_{2h}$ symmetry
around the new orienting axis. \EOR
\end{rem}
\begin{rem}\label{rem:20}
Inverting the relation between $K$ and $\la$,
we can express $K$ in terms of $\la$; similarly, we can invert
the relation between $K$ and $\vth_1$. Eliminating $K$ from these
two equations according to the two variants in \eqref{eq:d3hxi}, they
provide the corresponding (not too simple) relations linking $\la_S$ and
$\la_M$ with $\vth_1$, for saddles and maxima, respectively; more precisely, we have
\beq \la_S =\frac12  \frac{2 - 3 \cos^2 \vth_1}{\sin \vth_1}  ,\quad   \la_M  = \
\frac{8 - 9 \cos^2 \vth_1}{\cos \vth_1  \sqrt{16 - 15 \cos^2
\vth_1 }}  . \eeq  \EOR
\end{rem}

\subsubsection{Symmetry breaking: from $D_{\infty h}$ to $D_{3h}$}
It is interesting to consider the situation in the $D_{3h}$ phase
for values of $K$ near zero; in other words, to consider the
symmetry breaking from $D_{\infty h}$ (case $K = 0$, the center of
the disk $\mathcal{D}$) to $D_{3 h}$.

We will write
$K =\eps $ and work at first order in $\eps$.
With this, we obtain easily
\beq\cos\vth_1 =\frac{10 + \eps}{5  \sqrt{5}}  ; \eeq in
the same way we also get
\beq \la_M =\frac{15 + 24 \eps}{5  \sqrt{5} }  . \eeq
In particular, denoting by a ``0'' the limit for $K \to 0$, we have
\beq \frac{\la_M - \la_M^0}{\cos \vth_1 - \cos \vth_1^0}  =  24  +  O (\eps)  . \eeq

\subsubsection{Symmetry breaking: from $T_d$ to $D_{3h}$}
We can also consider the symmetry breaking from the tetrahedral
phase to $D_{3h}$. In this case we will write
$ K =K_0  +  \eps $
and work again at first order in $\eps$. Now we get
\beq \cos \vth_1  =\frac{2  \sqrt{2}}{3}  +  \frac{4}{81}  \eps  ,\quad   \la_M =3  +  \frac{16  \sqrt{2}}{9}  \eps  . \eeq
Thus in this case, denoting now by a superscript $^T$ the tetrahedral limit, we have
\beq \frac{\la_M - \la^T}{\cos \vth_1 - \cos \vth_1^T}  = \
36  \sqrt{2}  +  O (\eps)  . \eeq

\subsection{Simmetry $D_{2 h}$: the disk $\mathcal{D}$}
\label{sec:disk}
\begin{figure}
	\begin{tabular}{ccc}
		\includegraphics[width=105pt]{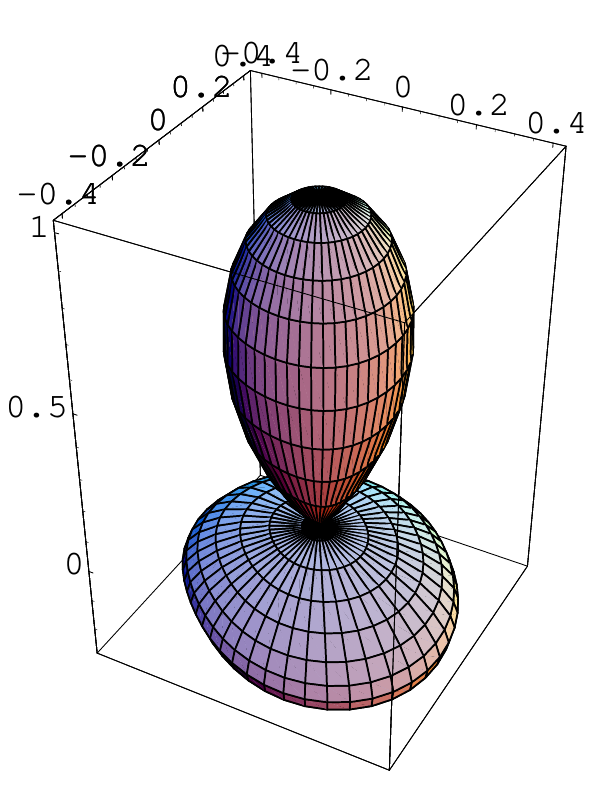} &
		\includegraphics[width=105pt]{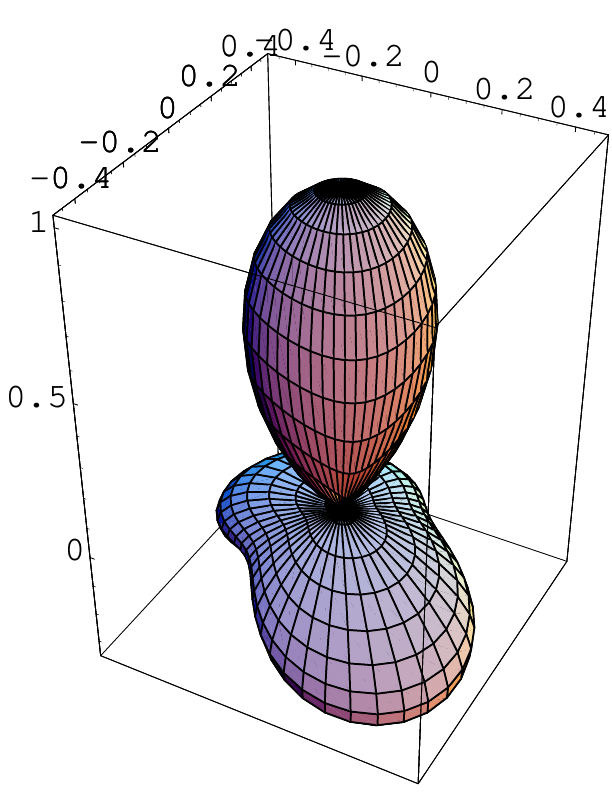} &
		\includegraphics[width=105pt]{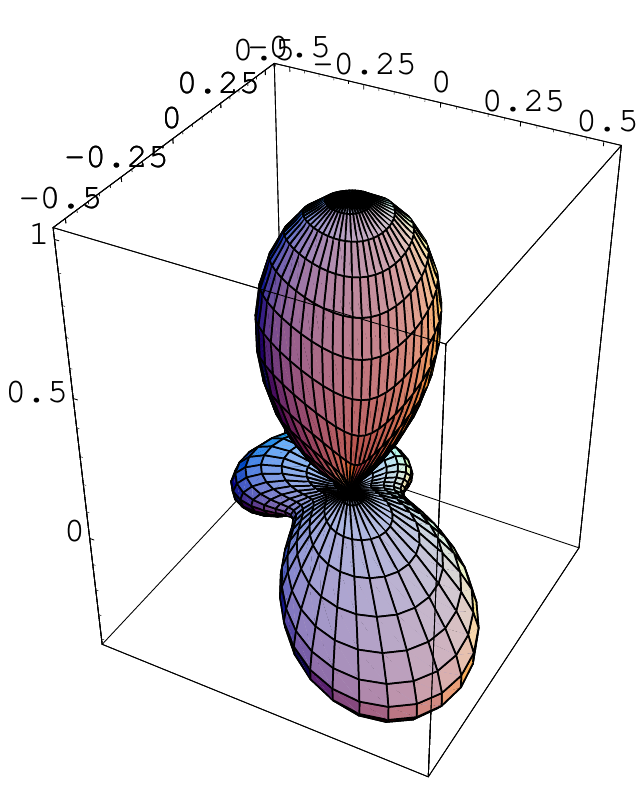} \\
		$\rho = 0.1$ & $\rho = 0.3$ & $\rho = 0.5$ \\
		\includegraphics[width=105pt]{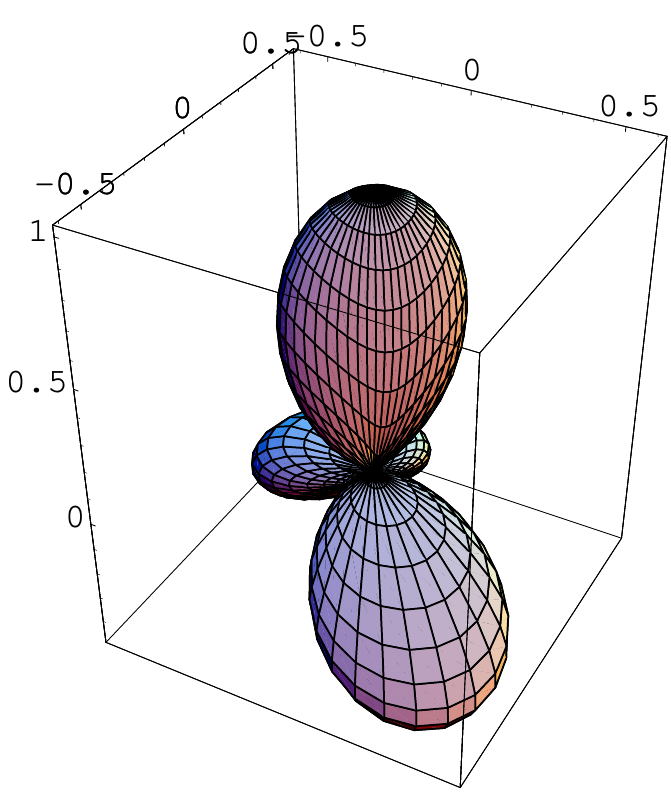} &
		\includegraphics[width=105pt]{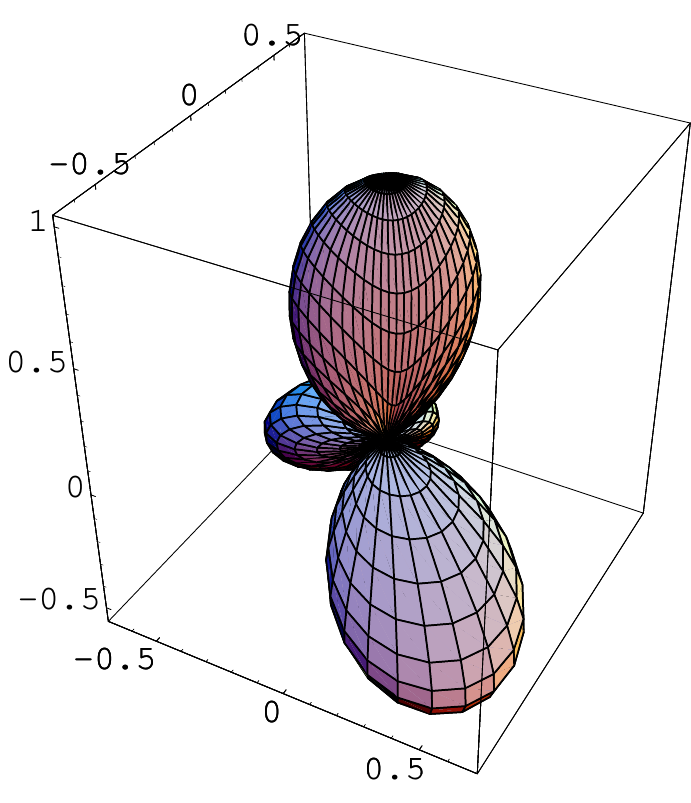} &
		\includegraphics[width=105pt]{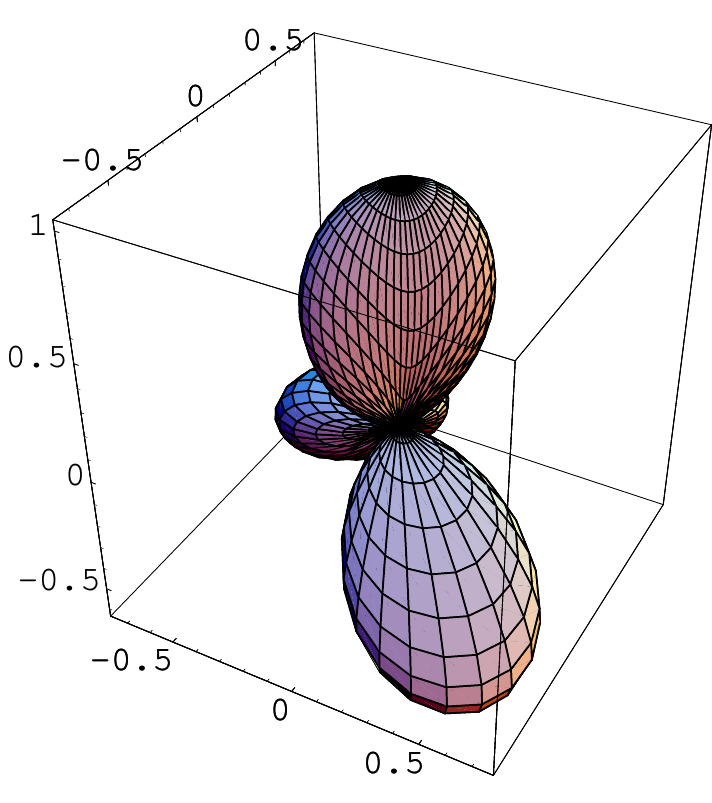} \\
		$\rho = 0.7$ & $\rho = 0.9$ & $\rho = 1.1$ \\
		\includegraphics[width=105pt]{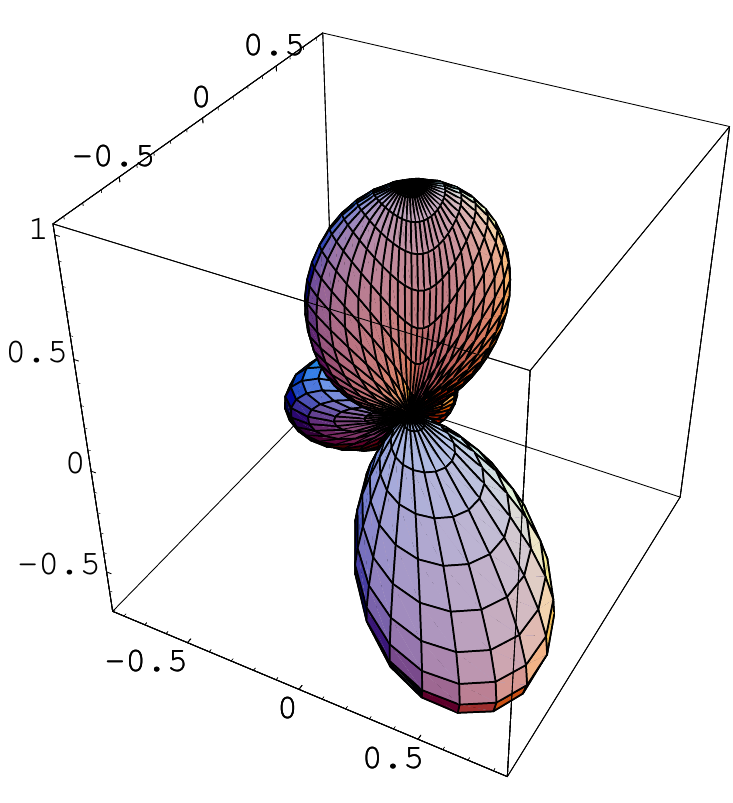} &
		\includegraphics[width=105pt]{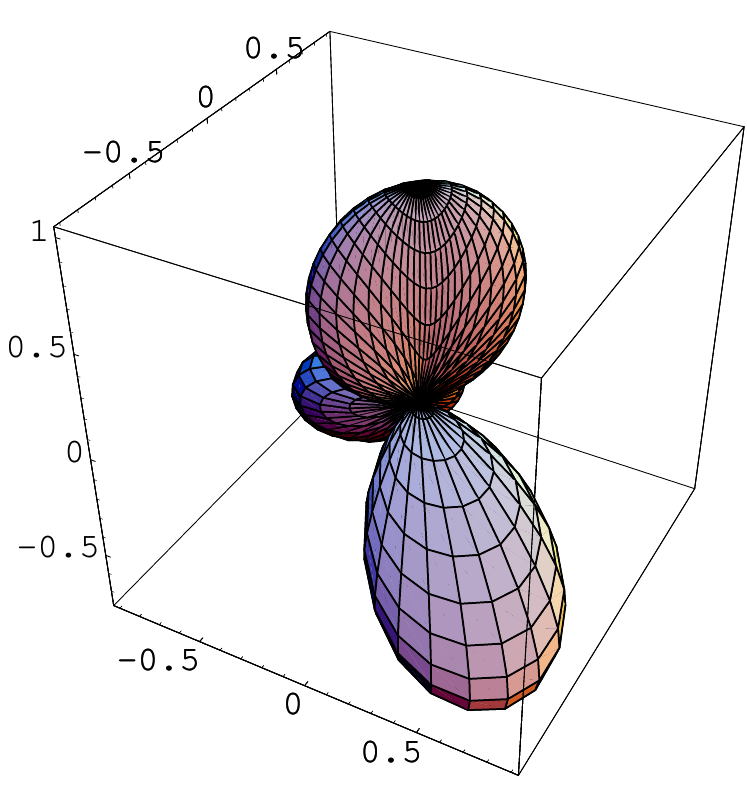} &
		\includegraphics[width=105pt]{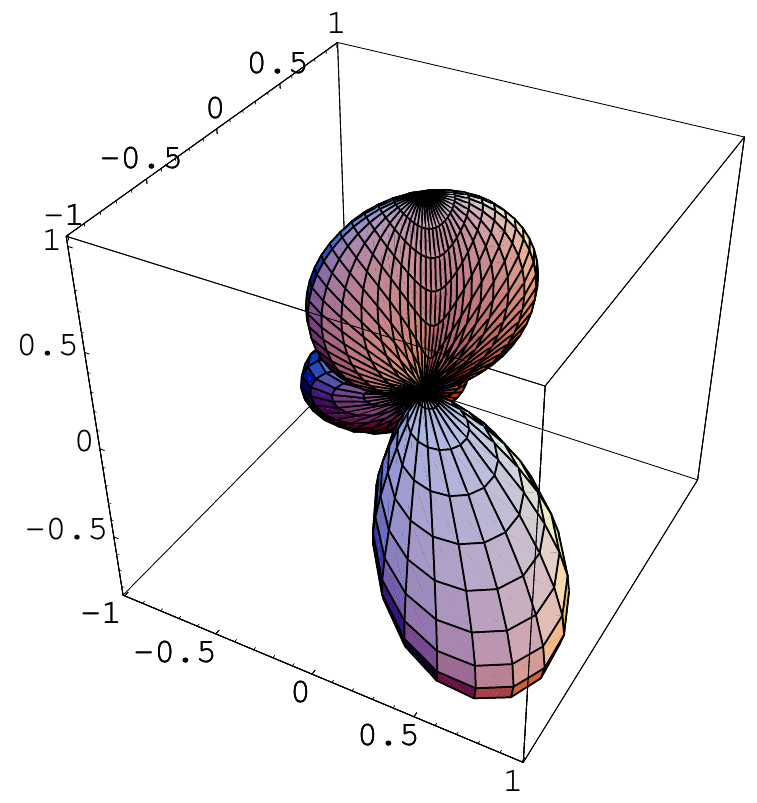} \\
		$\rho = 1.3$ & $\rho = 1.5$ & $\rho = 1.7$ \end{tabular} \\
	\caption{The potential $\Phi_2$ for $\chi = 0$ and different values of the
		parameter $\rho$.}\label{fig:d2hSP}
\end{figure}

\begin{figure}
	\begin{tabular}{ccc}
		\includegraphics[width=105pt]{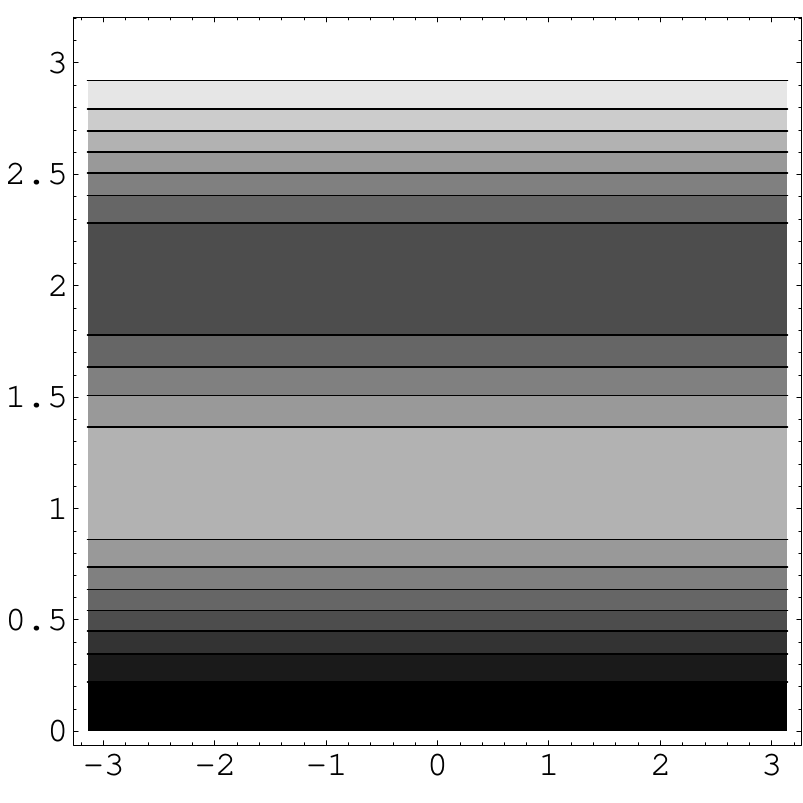} &
		\includegraphics[width=105pt]{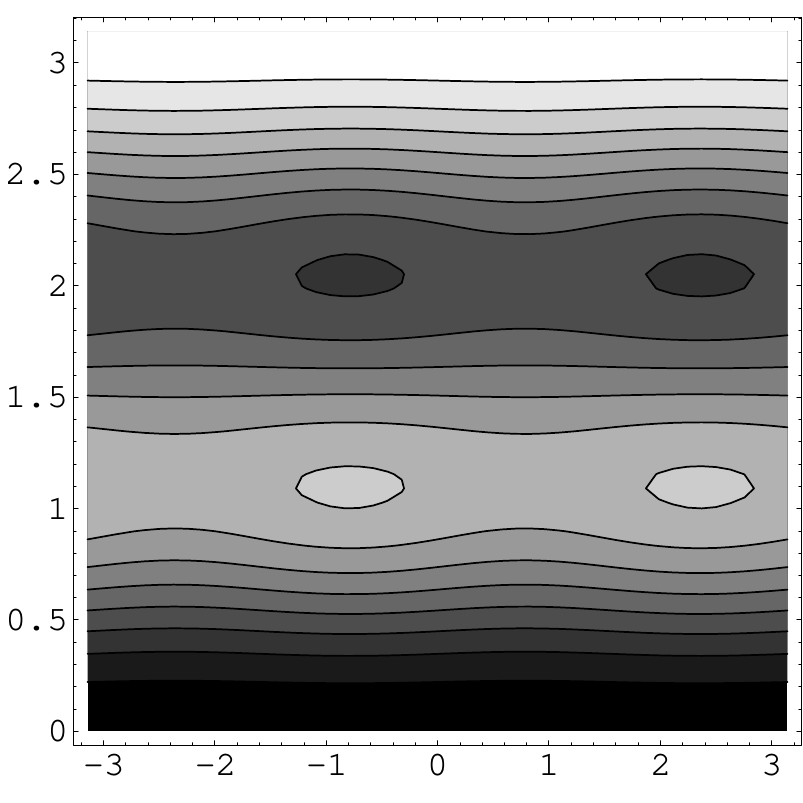} &
		\includegraphics[width=105pt]{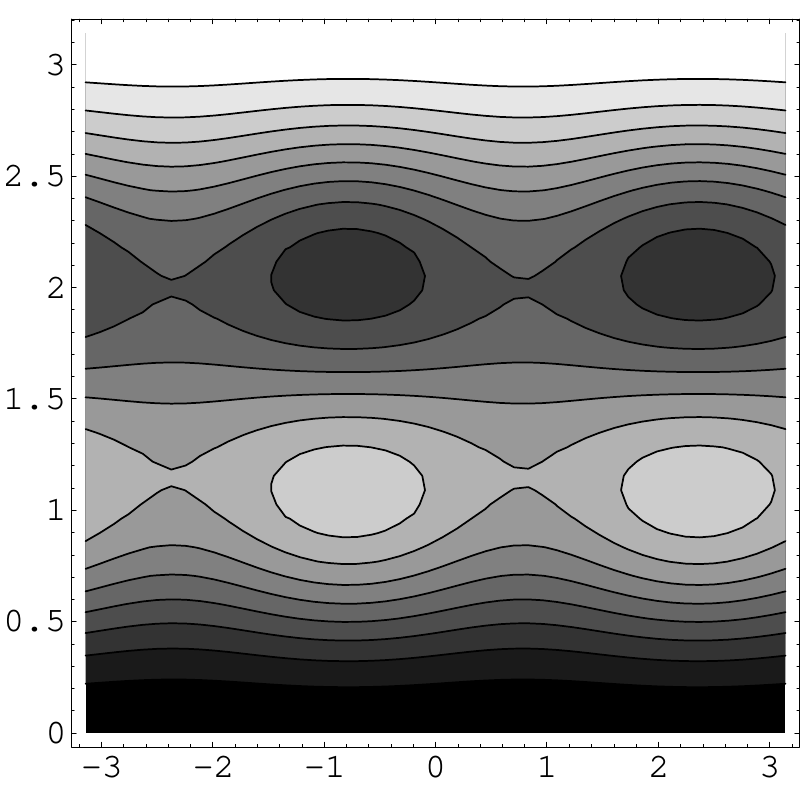} \\
		$\rho = 0$ & $\rho = 0.1$ & $\rho = 0.3$ \\
		\includegraphics[width=105pt]{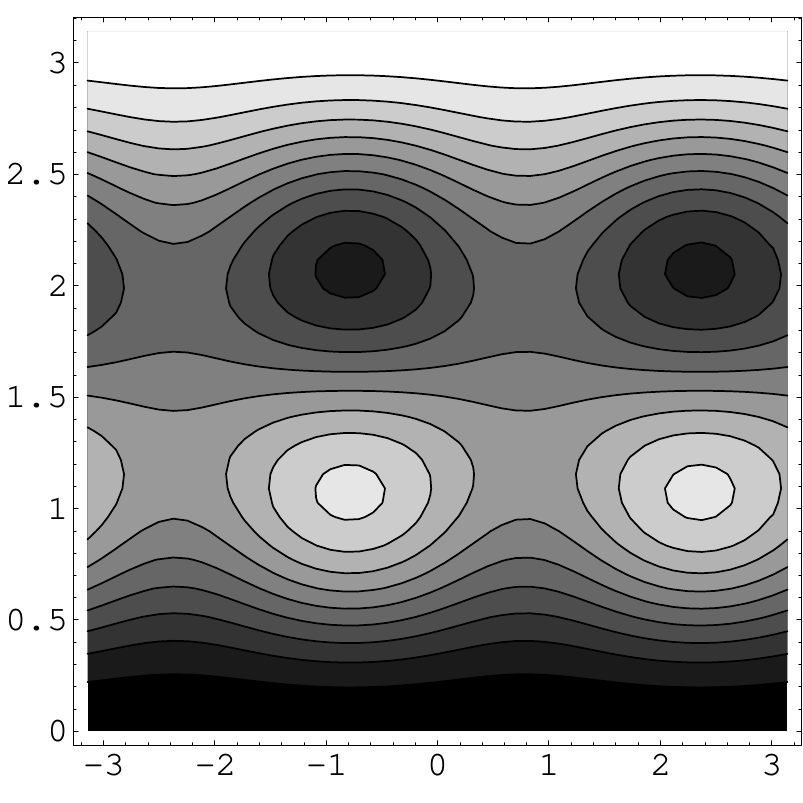} &
		\includegraphics[width=105pt]{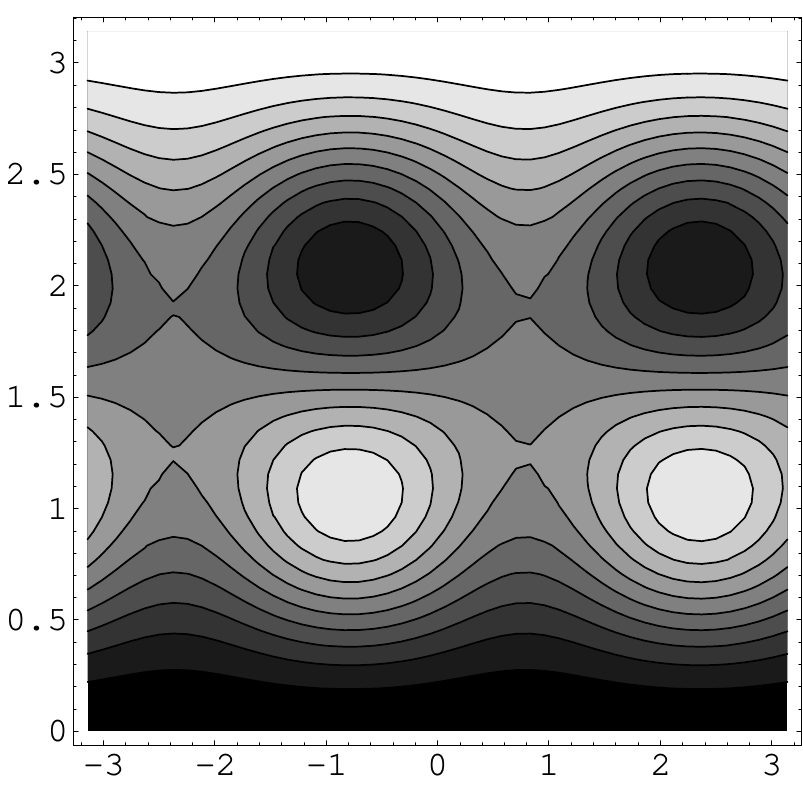} &
		\includegraphics[width=105pt]{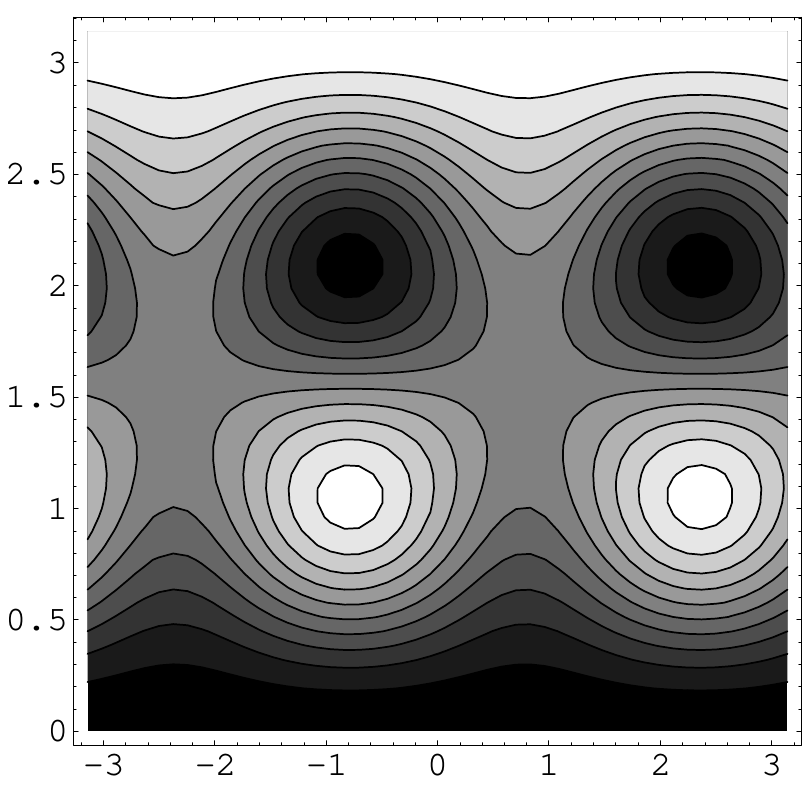} \\
		$\rho = 0.5$ & $\rho = 0.7$ & $\rho = 0.9$ \\
		\includegraphics[width=105pt]{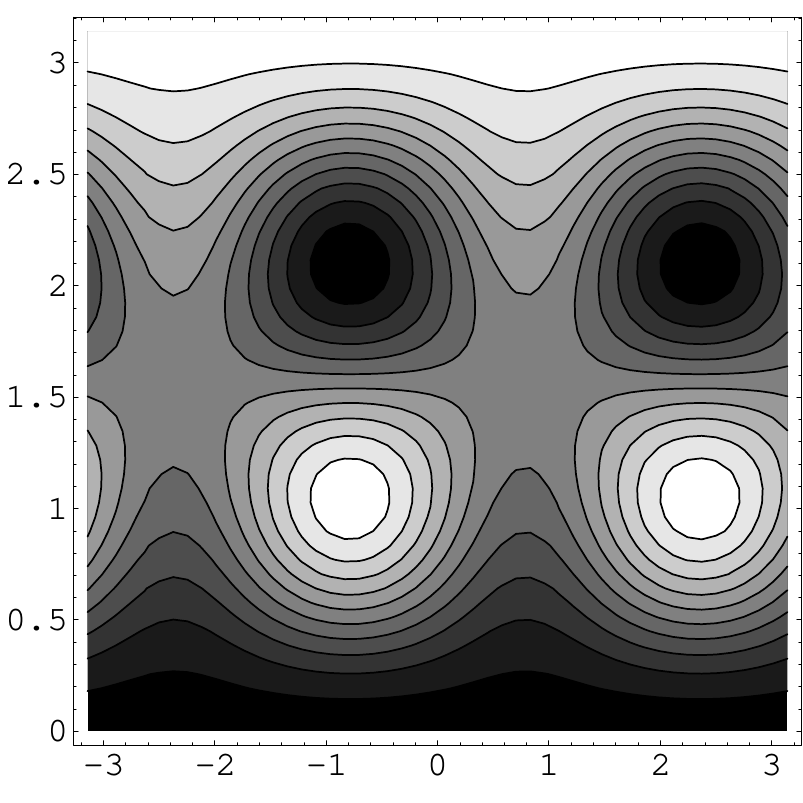} &
		\includegraphics[width=105pt]{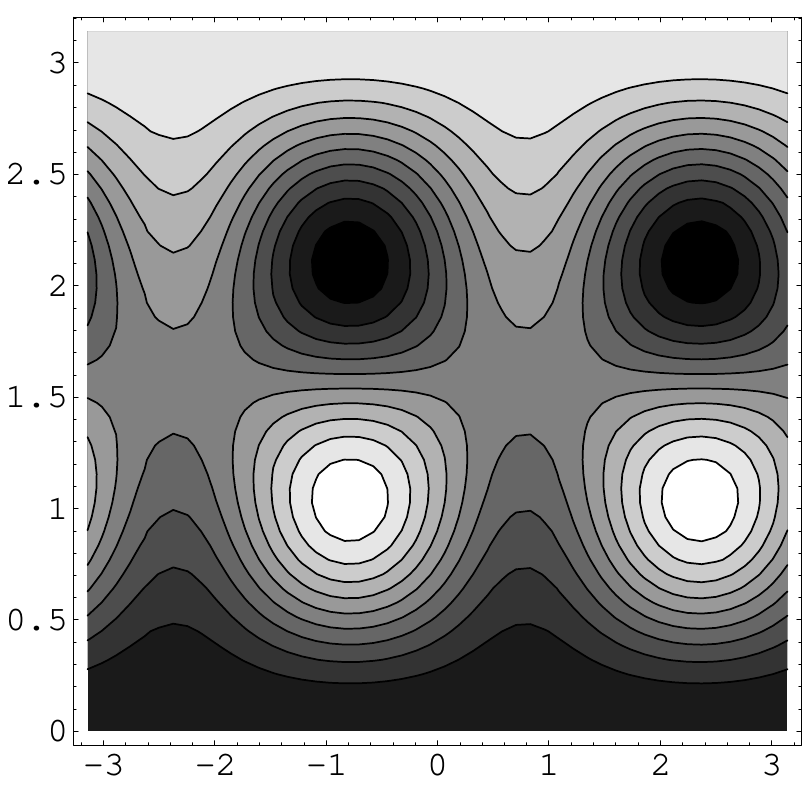} &
		\includegraphics[width=105pt]{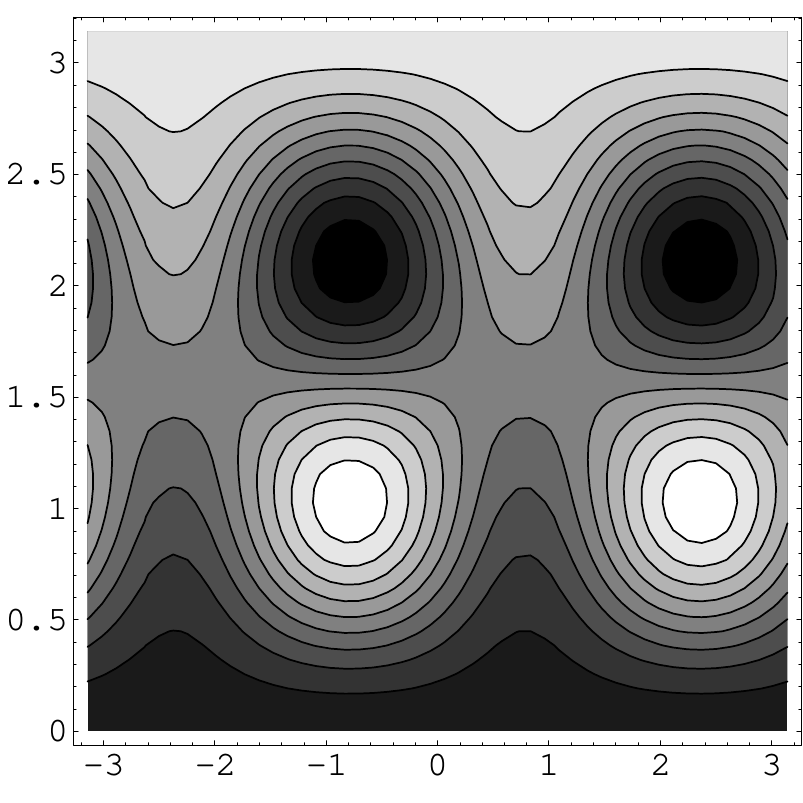} \\
		$\rho = 1.1$ & $\rho = 1.3$ & $\rho = 1.5$ \\
		\includegraphics[width=105pt]{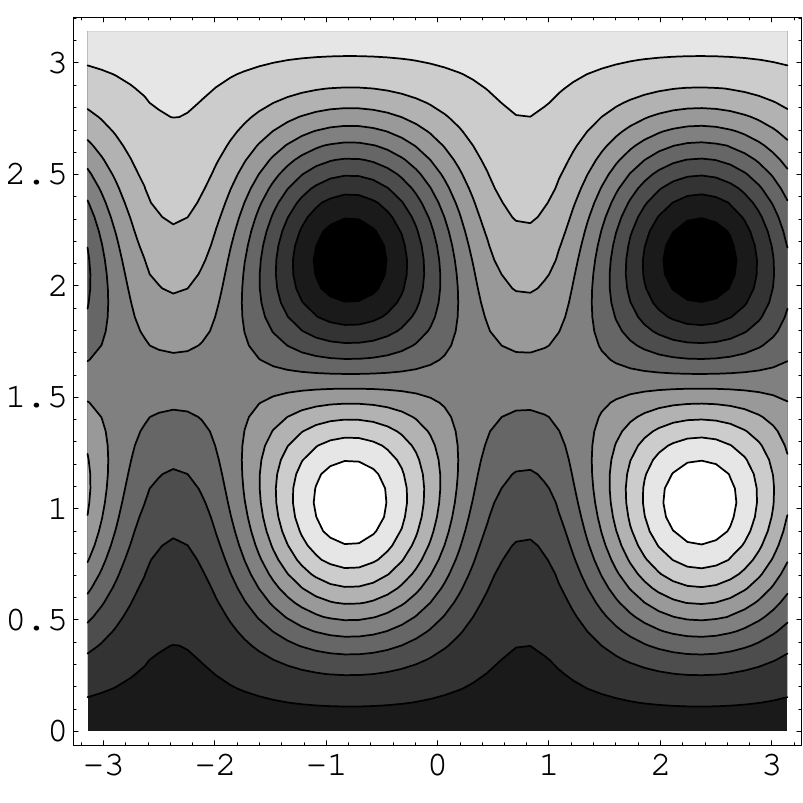} &
		\includegraphics[width=105pt]{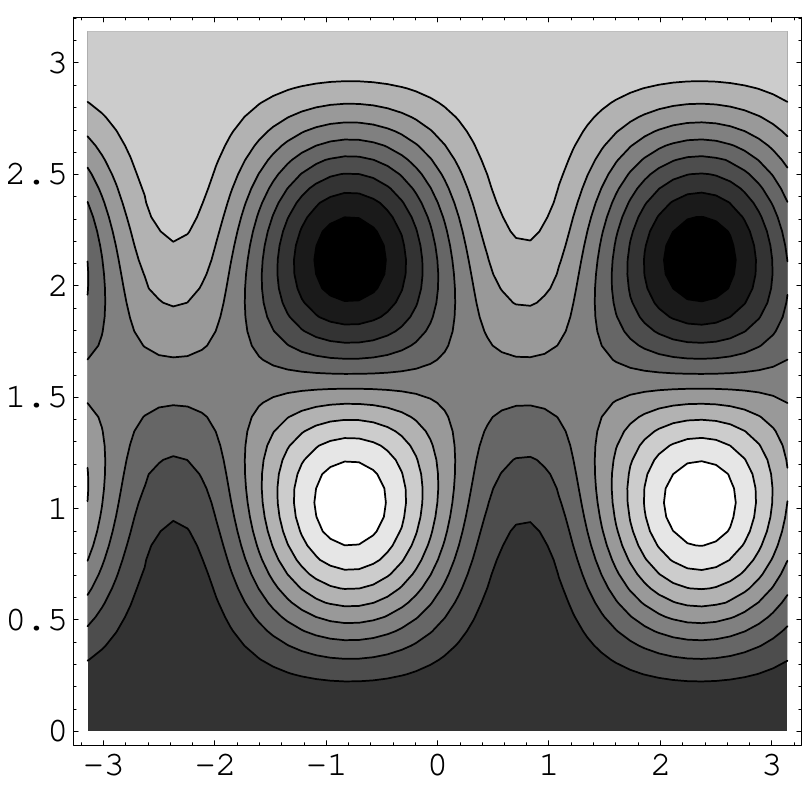} &
		\includegraphics[width=105pt]{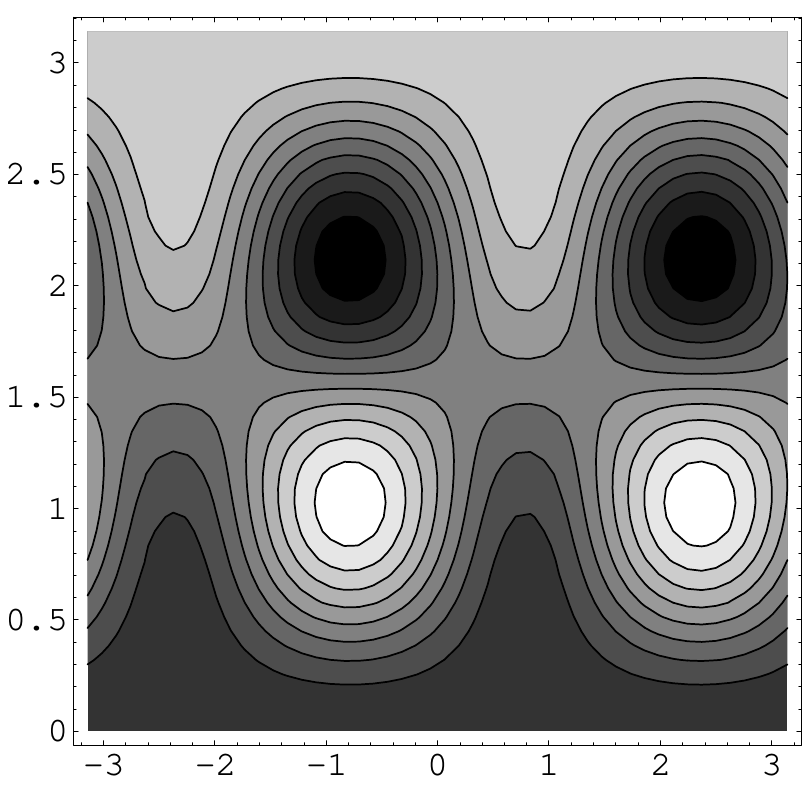} \\
		$\rho = 1.7$ & $\rho = 1.9$ & $\rho = 2$  \end{tabular} \\
	\caption{Contour plot of the potential $\Psi_2$ for $\chi = 0$ and for various values of $\rho > 0$.
		There always are three maxima, three minima and four saddles; for $\rho 1.0$,
		pairs of saddles originally one on top of the other collide and re-emerge side to side.
		See Fig.~\ref{fig:d2hCPBif} for more details.}\label{fig:d2hCP}
\end{figure}

\begin{figure}
	\begin{tabular}{ccc}
		\includegraphics[width=105pt]{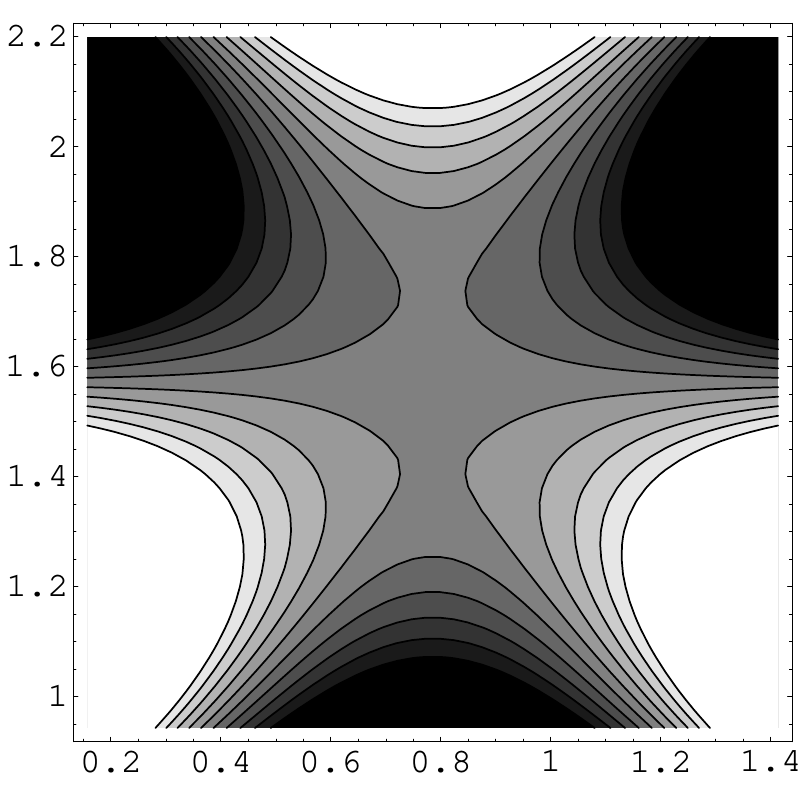} &
		\includegraphics[width=105pt]{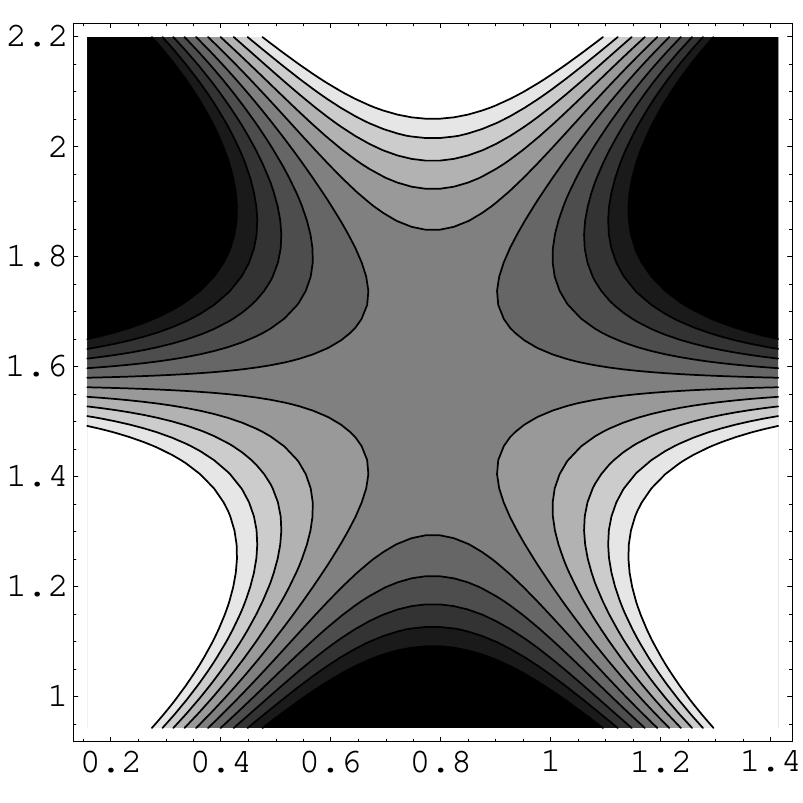} &
		\includegraphics[width=105pt]{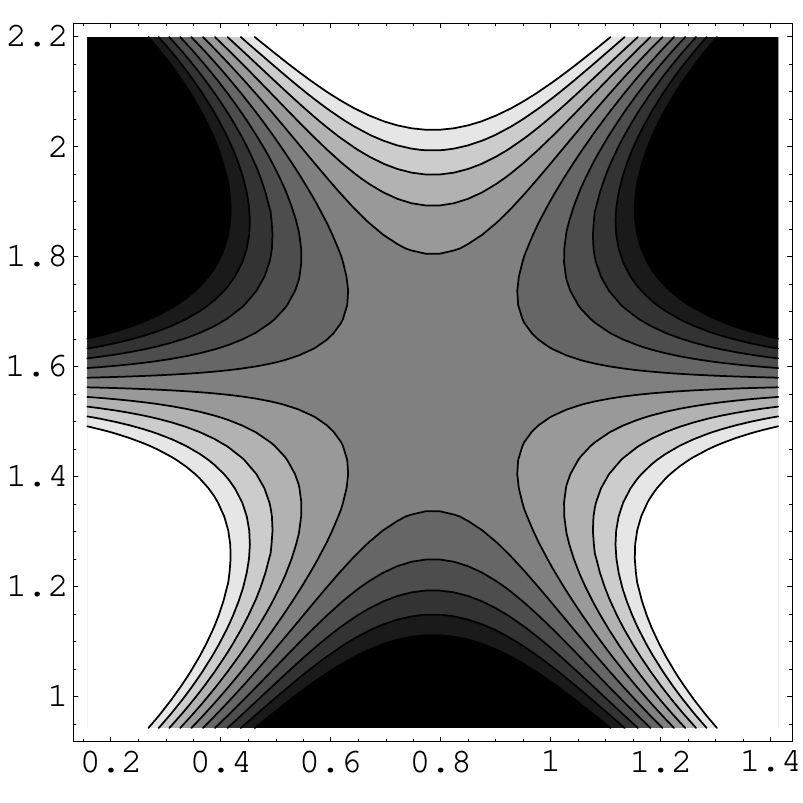} \\
		$\rho = 0.95$ & $\rho = 0.97$   &	$\rho = 0.99$\\
		\includegraphics[width=105pt]{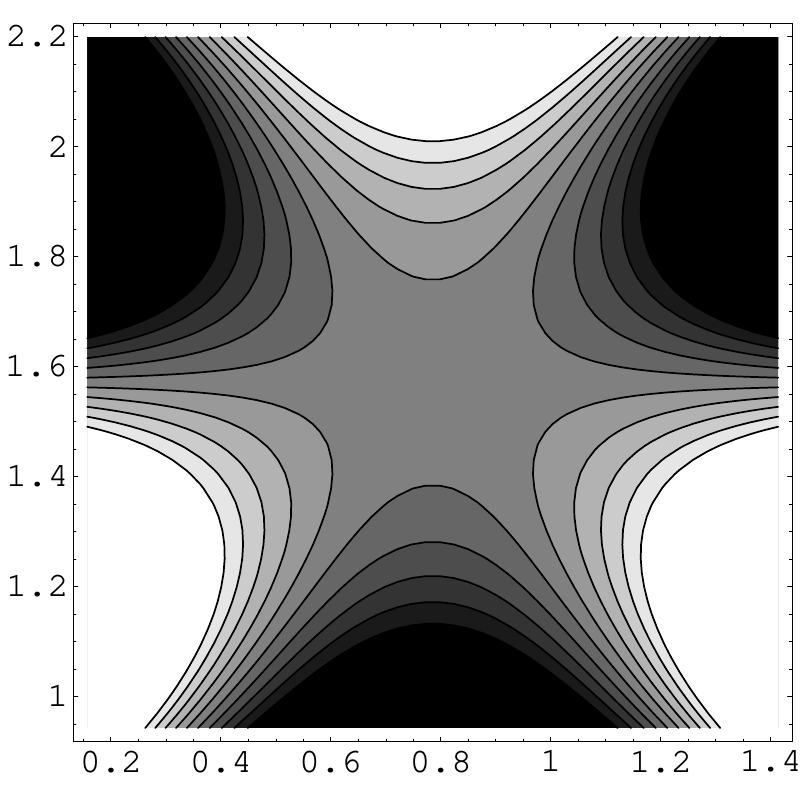} &
		 	\includegraphics[width=105pt]{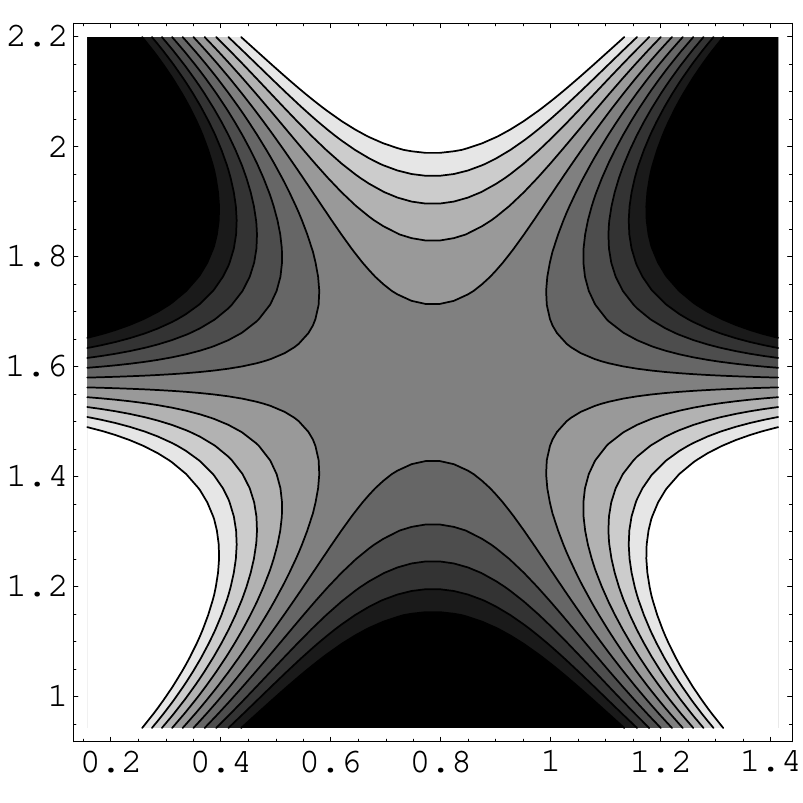}& 
		 \includegraphics[width=105pt]{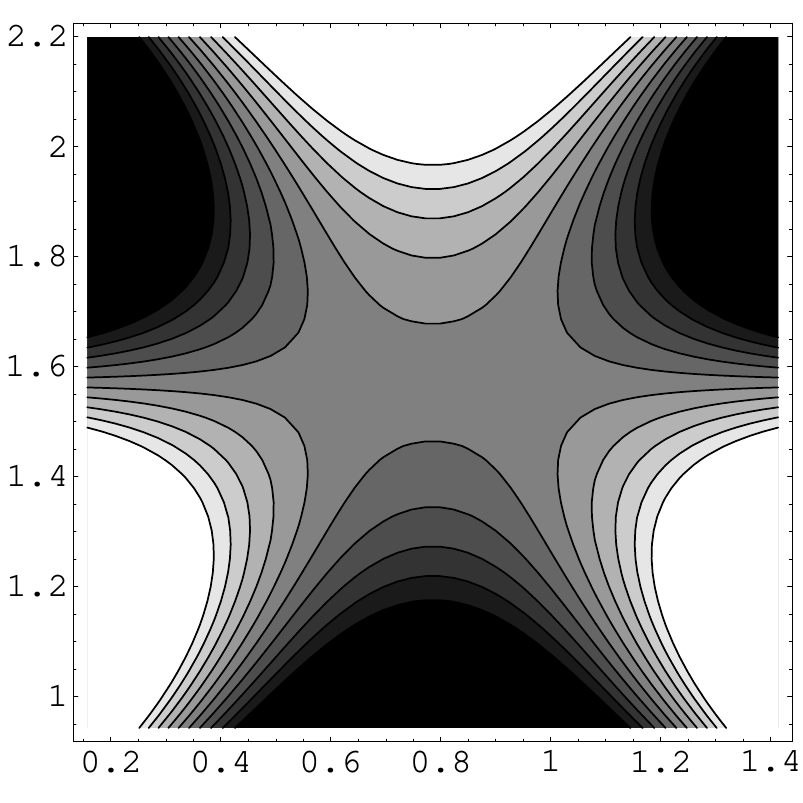}	\\
	 $\rho = 1.01$  & $\rho = 1.03$ & $\rho = 1.05$\\
\end{tabular} \\
	\caption{Contour plot of the potential $\Psi_2$ in the region where the saddle/saddle
		bifurcation takes place (here $\vth_1 \in [-\pi/5,\pi/5]$,
		$\vth_2 \in [\pi/4 - \pi/5 , \pi/4 + \pi/5]$) for $\chi = 0$ and for various values of $\rho$.
		We observe the saddle/saddle bifurcation taking place; in this a
		pairs of saddles originally one on top of the other collide (at $\rho = 1.0$)
		and re-emerge side to side. At $\rho = 1.0$ there is a monkey saddle.}\label{fig:d2hCPBif}
\end{figure}

On the disk $\mathcal{D}$ identified by $K = 0$ (with $\rho \not=
0$, or we would be at the center $\centre$ and in case $D_{\infty h}$
considered above), we have a symmetry $D_{2 h}$. In fact, the
potential in \eqref{eq:potential_53} reduces to \beq \Phi_2 = z^3-3y^2z+3\rho\cos\chi xyz+\frac32(\rho\sin\chi-1)\left(x^2- y^2\right)z  ,\eeq
 and similarly its counterpart in  
angular coordinates is readily obtained from \eqref{eq:potev} as
\beq \label{eq:phi2}
\Psi_2 =\sin\vth_1\[\sin^2\vth_1  +  \frac{3}{2}  \cos^2 \vth_1
 (\rho  \sin (\chi +2 \vth_2)   -  1)\]  .
\eeq
Graphical illustrations for both $\Phi_2$ and $\Psi_2$ in the styles introduced above are provided in Figs.~\ref{fig:d2hSP}, \ref{fig:d2hCP}, and \ref{fig:d2hCPBif}.

\begin{figure}
	\includegraphics[width=150pt]{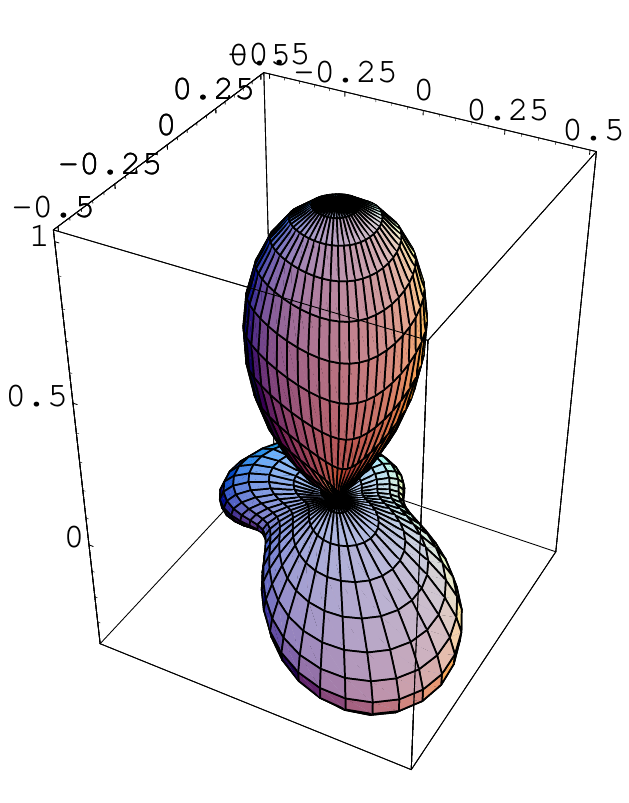} \hfill
	\includegraphics[width=150pt]{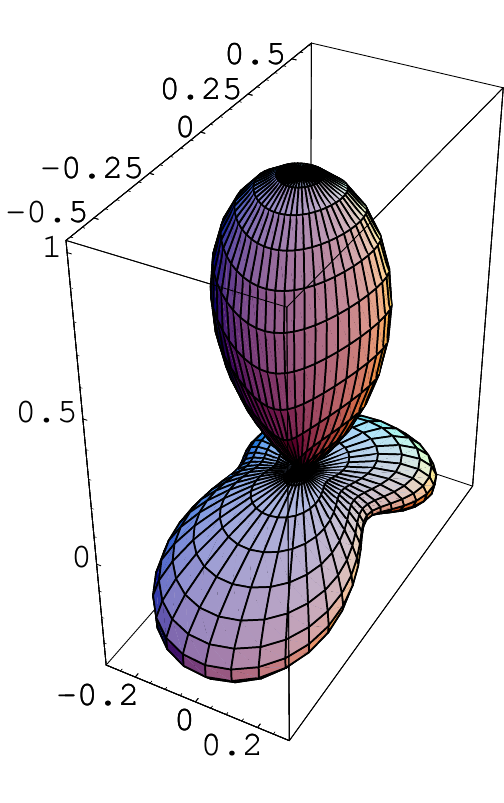} \\
	\caption{Three-dimensional polar plot of the potential $\Phi_2$, see
		\eqref{eq:phi2}, for $\rho = 0.4$. Here we have chosen $\chi = 0$ (left)
		and $\chi = \pi/2$ (right). It is clear that the rotations by $\pi/2$ in $\chi$
		corresponds to a rotation by $\pi/4$ in $\vth_2$.}\label{fig:d2h3d}
\end{figure}

\begin{rem}\label{rem:21}
It is clear from this expression that a shift in
$\chi$ corresponds to a shift (of half the amplitude) in $\vth_2$;
see also Fig.~\ref{fig:d2h3d} for a visual demonstration. \EOR
\end{rem}
\begin{rem}\label{rem:22}
It is also apparent from \eqref{eq:phi2} that
$\Psi_2$ is identically zero on the Equator $\vth_1 =0$ (i.e.
$\Phi_2$ is identically zero for $z=0$); this implies that any
critical point lying on the Equator will necessarily be
non-hyperbolic, hence degenerate. \EOR
\end{rem}
\begin{rem}\label{rem:*3}
The invariance under rotations by $\pi$ in $\vth_2$ is evident. We
also have invariance under rotations by any $\delta$ in $\vth_2$
accompanied by a rotation by $2 (\pi - \delta)$ in $\chi$. In
particular, we have invariance under a rotation by $\pi/2$ in
$\vth_2$ accompanied by a rotation by $\pi$ in $\chi$; thus, we can
just consider $-\pi/2\le\chi\le\pi/2$.

We could also restrict the domain in which we study the potential
by noting that $\Psi_2 \to - \Psi_2$ under $\vth_1 \to - \vth_1$;
this also follows by the usual skew-symmetry of the potential
under the antipodal map combined with invariance under $\vth_2 \to
\vth_2 + \pi$. \EOR
\end{rem}
\subsubsection{Symmetry of the potential}
The potential is invariant under a subgroup $O(2) \ss T_d$; this
is explicitly given by the matrices
 \beq\label{eq:symmD2h1} M_0 \
=  \left(
\begin{array}{lll}
 1 & 0 & 0 \\
 0 & 1 & 0 \\
 0 & 0 & 1
\end{array}
\right)  ,\quad   M_\pi =\left(
\begin{array}{lll}
 -1 & 0 & 0 \\
 0 & -1 & 0 \\
 0 & 0 & 1
\end{array}
\right)
  ; \eeq
\beq\label{eq:symmD2h2} R_1 =\left(
\begin{array}{ccc} - \sin\chi & - \cos\chi & 0 \\
- \cos\chi  & \sin\chi & 0 \\ 0 & 0 & 1
\end{array} \right)   ,  \
 R_2 =\left(
\begin{array}{ccc} \sin\chi & \cos\chi & 0 \\ 
\cos\chi & - \sin\chi & 0 \\ 0 & 0 & 1
\end{array} \right)  . \eeq
Both $M_0$ and $M_\pi$ have
determinant $+1$, while both $R_1$ and $R_2$ have determinant $-1$.

The matrices $R_i$ describe a reflection in a vertical plane. More
precisely, recalling that the matrix describing a reflection in the
vertical plane $y = m x$ is given by
\begin{equation}
\label{eq:R_m_a_b}
R_m =\left(
\begin{array}{lll}
 a & \ b & 0 \\
 b & - a & 0 \\
 0 & \ 0 & 1
\end{array}
\right)  ,\quad   a = \frac{1-m^2}{m^2+1} ,\quad  b = \frac{2 m}{m^2+1}
, 
\end{equation}
and that conversely a matrix $R_m$ as in \eqref{eq:R_m_a_b} with $a^2+b^2=1$ represents a reflection through the plane $y=mx$ with
\beq
m=\frac{1-a}{b},
\eeq
the planes of reflection  for the matrices $R_i$ (orthogonal to one another) have  equations
$y = m_i x$, where
\beq
m_1=-\frac{1+\sin\chi}{\cos\chi},\quad m_2=\frac{1-\sin\chi}{\cos\chi},
\eeq
which clearly satisfy $m_1m_2=-1$.

We have special cases for $\chi =\pm\pi/2$ or
$\chi = 0, \pi$ (note that when both these conditions are met, we are
back to the case $\rho = 0$ considered above; in fact, we have
characterized $D_{2 h}$ by also requiring $\rho \not= 0$).
\begin{rem}\label{rem:23}
To  appreciate better our previous Remark~\ref{rem:21}, consider the case $\chi = 0$. Now  $R_i$ read as
\beq R_1  = \( \begin{array}{ccc} 0 & - 1 & 0 \\ - 1 & 0 & 0 \\ 0
& 0 & 1 \end{array} \)  ,\quad   R_2 = \(
\begin{array}{ccc} 0 & 1 & 0 \\ 1 & 0 & 0 \\ 0 & 0 & 1 \end{array}
\)  . \eeq
These represent reflections in the vertical planes $y=x$
and $y = - x$, as illustrated in the left panel of Fig.~\ref{fig:d2h3d}.
\EOR
\end{rem}

\subsubsection{Critical points}
The critical points are identified as solutions to the equations
\begin{eqnarray}
\frac{3}{4} \cos \vth_{1}
   [-5 \cos 2 \vth_{1}+\rho  (3
   \cos 2 \vth_{1}-1) \sin (\chi
   +2 \vth_{2})+3] &=& 0  , \nonumber \\
   \label{eq:ecpd2h} \\
3 \rho  \cos^2\vth_{1} \cos (\chi +2
   \vth_{2}) \sin
   \vth_{1} &=& 0  . \nonumber\end{eqnarray}
   \begin{rem}\label{rem:new_2}
   It is immediately apparent that any change in $\chi$ can be
   		compensated by a change in $\vth_2$, and conversely; in other
   		words, the relevant angle is $\Theta = (\vth_2 + \chi/2)$. Thus it
   		suffices to study the problem with a given value of $\chi$, e.g.,
   		$\chi = \pi/2$ (hence $\a_0 = \rho$, $\ga = 0$) or $\chi = 0$
   		(hence $\a_0 = 0$, $\ga = \rho$); the general case (i.e. the case
   		of general $\chi$) will be obtained via a suitable rotation in
   		$\vth_2$.
   	\end{rem}
Solving equations \eqref{eq:ecpd2h} in general is a
matter of standard algebra and trigonometry (and some patience).
The results are reported in Table~\ref{table:V} below; it should be  stressed that some of the solutions exist only for certain ranges of $\rho$.

Let us first focus on the second of \eqref{eq:ecpd2h}; discarding as usual the ''trivial'' (in this context) solutions for $\vth_1 = \pm \pi/2$, which corresponds to the poles, we need either $\cos (\chi + 2 \vth_2) = \pm \pi/2$; or $\vth_1 = 0$. But inserting $\vth_1 = 0$ in the first of \eqref{eq:ecpd2h},
we obtain \beq -  \frac32  \[ 1  -  \rho  \sin (\chi + 2
\vth_2 ) \]=0  . \eeq This admits solutions \emph{only} for $\rho
\ge 1$; in view of our general restriction on $\rho$, see \eqref{eq:rho}, this means
$\rho \in [1,2]$.

On the other hand, if $\cos (\chi + 2 \vth_2) = 0$, i.e. $\chi + 2
\vth_2 = \pm \pi/2$, and hence $\sin (\chi + 2 \vth_2) = \pm 1$,
the first of \eqref{eq:ecpd2h} reads (after taking away the
inessential factor $(3/4) \cos \vth_1$) \beq  3  -  5  \cos
2 \vth_1  \mp  \rho  (1  -  3  \cos2 \vth_1)  =0
 ; \eeq this means
\beq \cos 2 \vth_1  =\frac{3 \mp \rho}{5 \mp 3 \rho}  . \eeq
While the solution with the plus sign
does exist for all values of $\rho$, the solution with the minus sign
exists only for $\rho=2$ and $\rho \le 1$.

In other words, we have some family of solutions existing through
the whole range $\rho \in [0,2]$, while for other families we will
have to consider separately the subranges $\rho \in [0,1]$ and
$\rho \in [1,2]$. It has to be expected (and it will indeed
result) that a multiple bifurcation takes place at $\rho =
1$. It should be noted that at the bifurcation point the saddles are
degenerate, so carrying a different index.

\begin{figure}
  \begin{tabular}{ccc}
  \includegraphics[width=105pt]{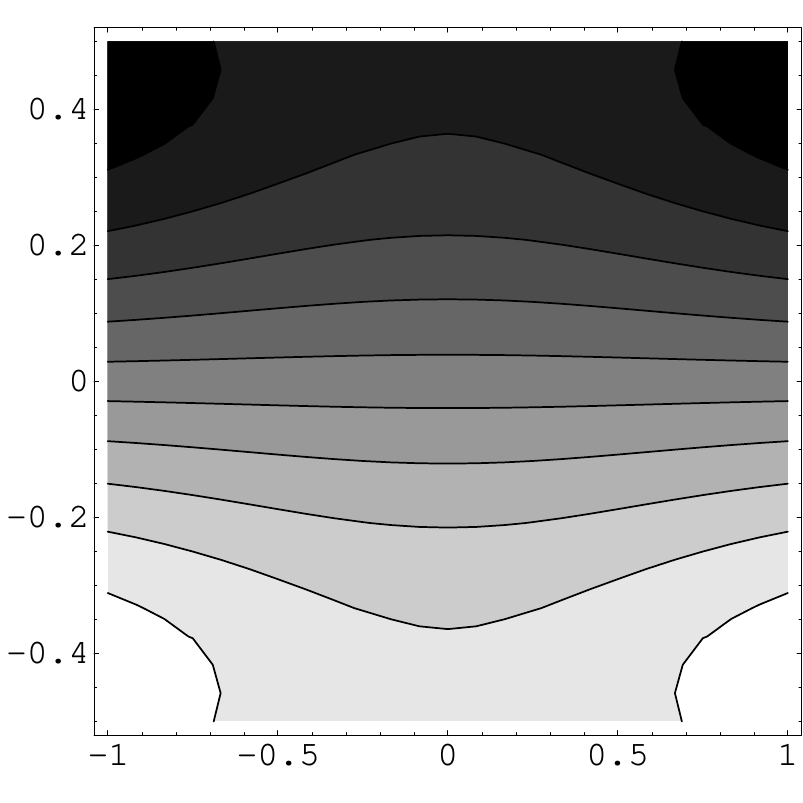} &
  \includegraphics[width=105pt]{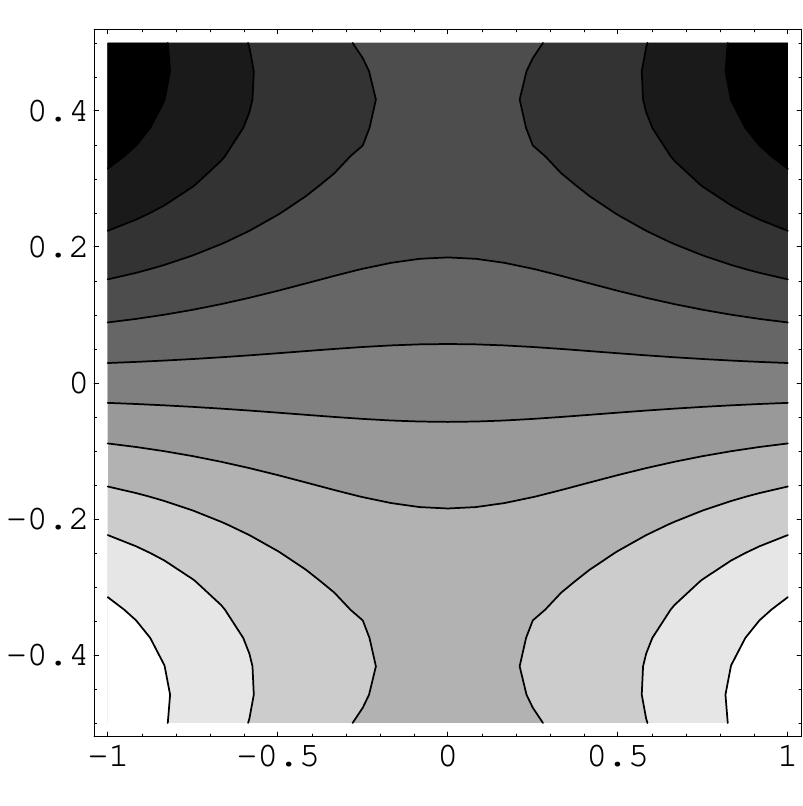} &
  \includegraphics[width=105pt]{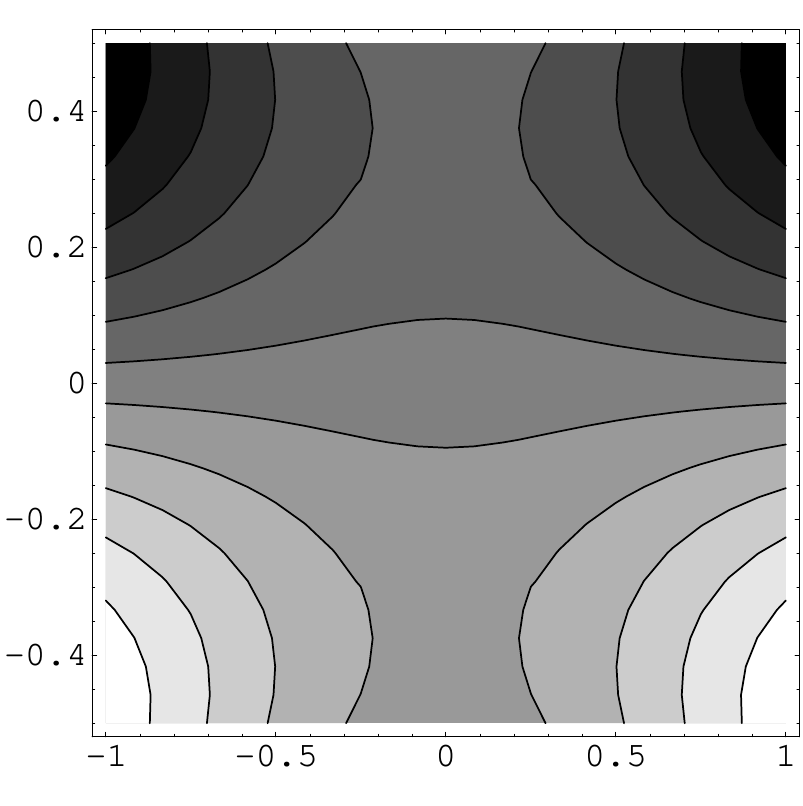} \\
  $\rho = 0.2$ & $\rho = 0.4$ & $\rho = 0.6$ \\
  \includegraphics[width=105pt]{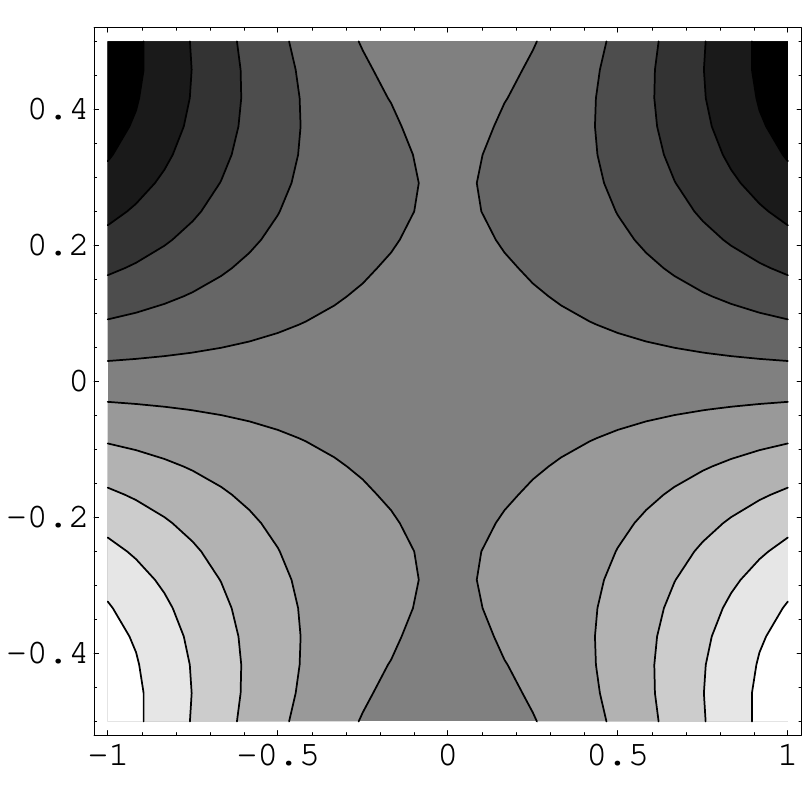} &
  \includegraphics[width=105pt]{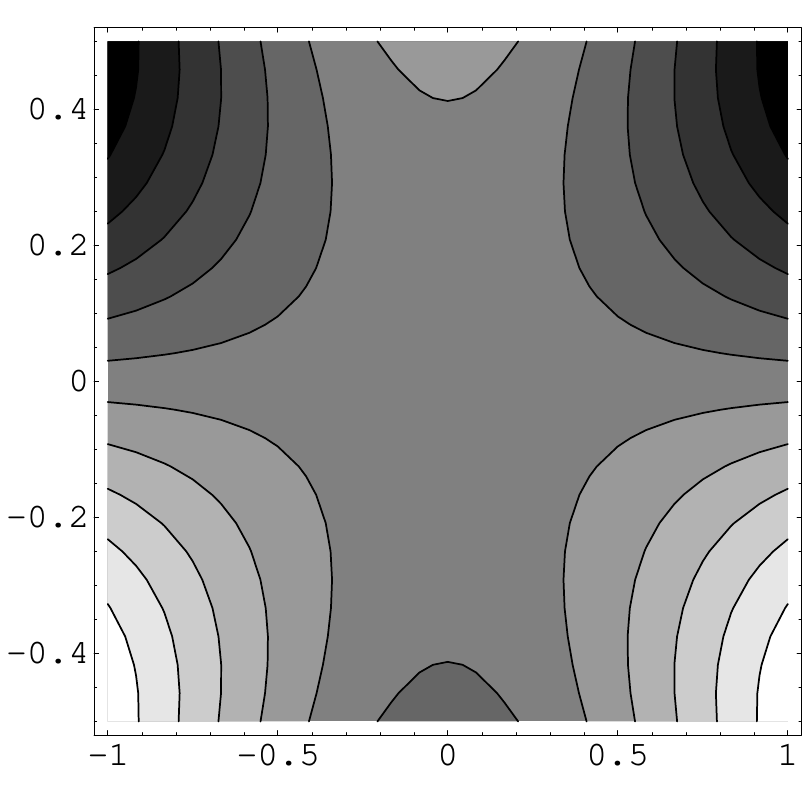} &
  \includegraphics[width=105pt]{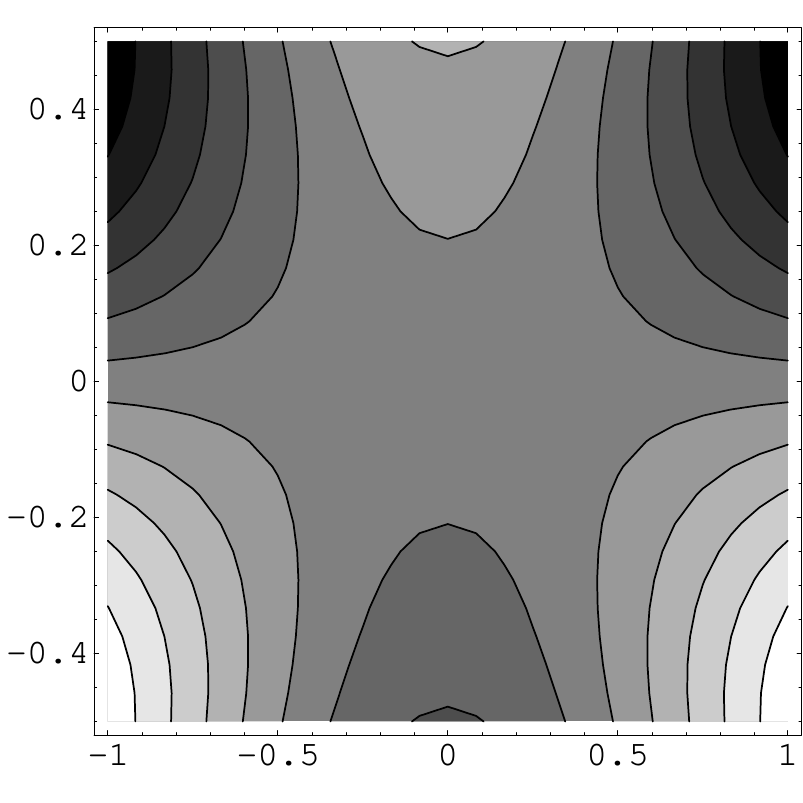} \\
  $\rho = 0.8$ & $\rho = 1.0$ & $\rho = 1.2$ \\
  \includegraphics[width=105pt]{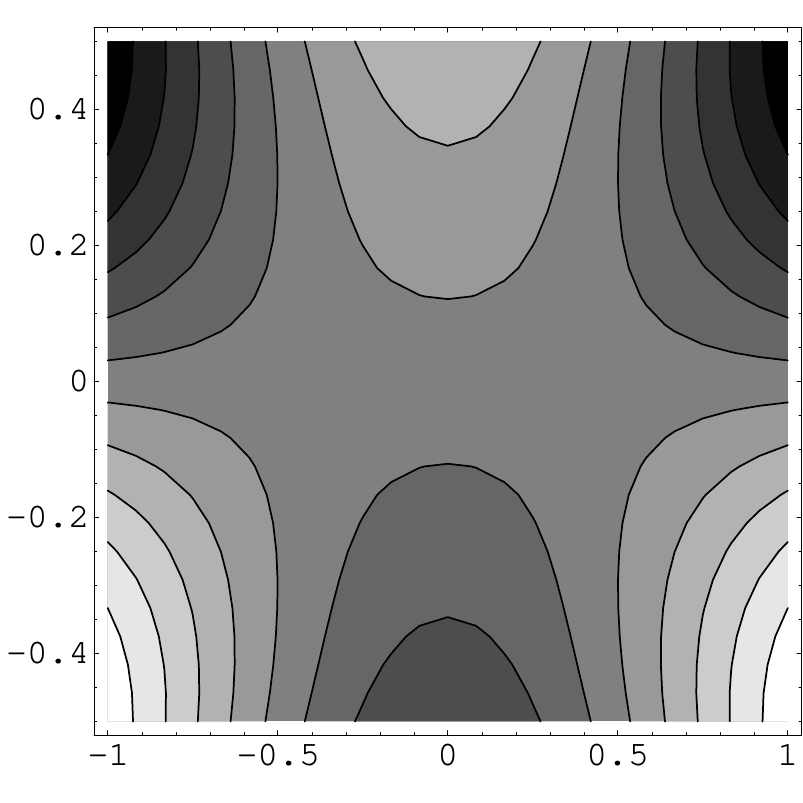} &
  \includegraphics[width=105pt]{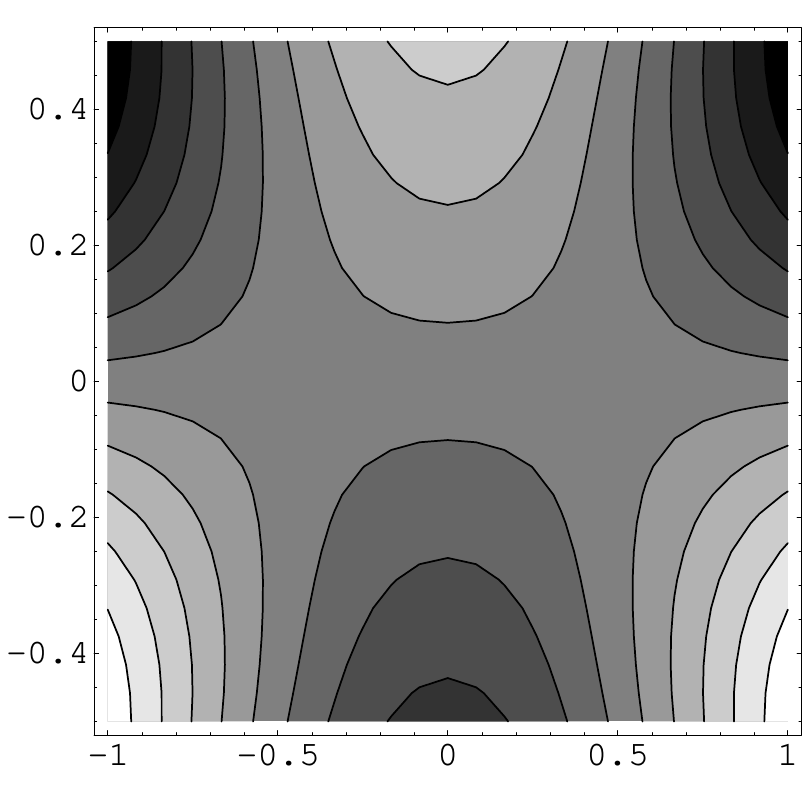} &
  \includegraphics[width=105pt]{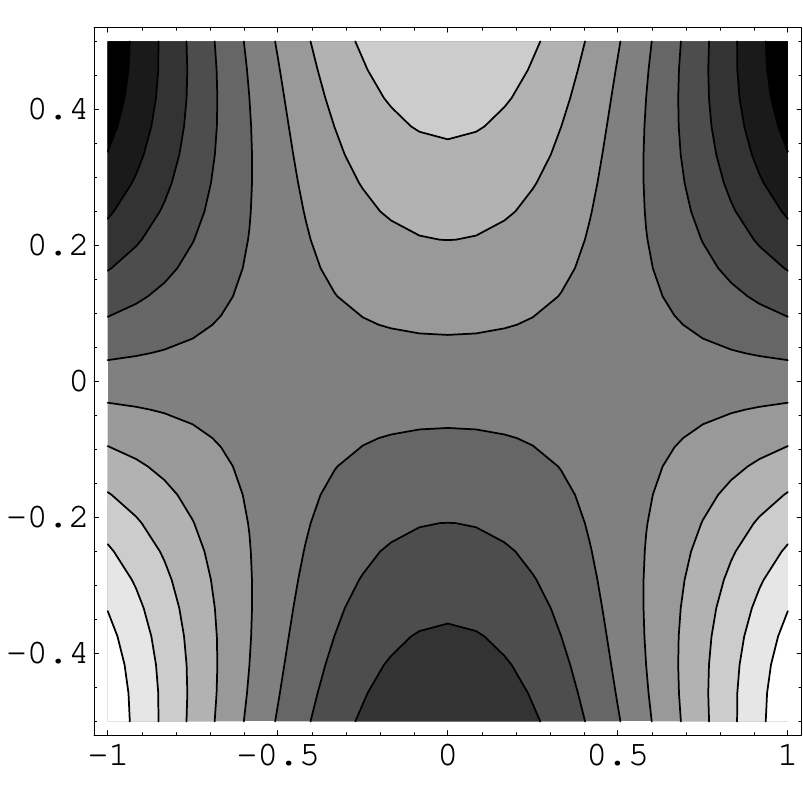} \\
  $\rho = 1.4$ & $\rho = 1.6$ & $\rho = 1.8$ \end{tabular} \\
  \caption{The saddle-saddle bifurcation in the $D_{2h}$ case taking place at $\rho = 1$.
  We have chosen $\chi= \pi/2$ and plotted a neighborhood of the point
  $(\vth_1 , \vth_2) = (0,0)$ -- more precisely the region $|\vth_1| \le 1/2$, $|\vth_2| \le 1$
  -- as $\rho$ is varied. The two saddles located (symmetrically w.r.t. the Equator)
  on the meridian $\vth_2 = 0$ collide at $\rho = 1$, giving raise to a new pair of saddles
  located on the Equator, symmetrically w.r.t. the meridian $\vth_2 = 0$.}\label{fig:d2h_bif1}
\end{figure}

The situation is rather clear if we think of a fixed value of
$\chi$, say $\chi = \pi/2$, and focus only on families of
solutions not existing for the whole range of $\rho$ (it turns out that
all of these are saddles). For $\rho \in [0,1)$ there are
solutions on the meridians identified by $\chi + 2 \vth_2 =(k
+ 1/2) \pi$ and approaching the Equator as $\rho \to 1$, but none
on the equator; on the other hand, for $\rho
\in (1,2]$ there are no solutions on those meridians, but we have instead families of solutions on the Equator, drifting away from the meridians identified by $\chi + 2 \vth_2 =(k + 1/2) \pi$ as $\rho$ grows away from $\rho = 1$. This is clearly illustrated in Fig.\ref{fig:d2h_bif1}.

We now return to the solutions to the critical point equations
\eqref{eq:ecpd2h}. The nature of these critical points is easily
ascertained by considering the eigenvalues of the Hessian at them; again the result of
this analysis is reported in Table~\ref{table:V}.
Here again we resort to some shorthand notation to make the table more
readable; in that we have defined
\begin{eqnarray}
\rho_\pm & := &  \frac{\rho \pm 3}{3 \rho \pm 5}  ,  \quad
r_\pm  :=  \arccos \( \rho_\pm \)  ,\quad   r_0  :=  \arccos \( 1 / \rho \)  , \nonumber \\
\label{eq:d2hshortnot}\\
\mu_\pm & := & \frac{\chi}{2}  \pm  \frac{\pi}{4}  ,  \quad
\nu_\pm  :=  \frac{\chi}{2}  \pm  \frac{3  \pi}{4}  ,  \qquad\
\omega_\pm  :=  \frac{\chi  \pm  r_0}{2}  .
\nonumber
\end{eqnarray}
\begin{table}[H]
	\caption{Critical points for the case $D_{2h}$. The shorthand notation \eqref{eq:d2hshortnot} is used here.}
	\label{table:V}
	\centering
\begin{tabular}{||r||c|c||c||c||c||}
\hline
$n$ & $\vth_1$ & $\vth_2$ & $\Psi$ & range & type \\
\hline
1 & $-{\pi }/{2}$ & --- &  $-1$ &
   always & min  \\
\hline
 2 & $r_+$ & $-\mu_+$ & $-(\rho +1)
   \sin r_+$ & always & min  \\
 3 & $r_+$ & $-\nu_-$ & $-(\rho +1)
   \sin r_+$ & always & min  \\
\hline
 4 & $r_-$ & $-\mu_-$ &  $(\rho -1)
   \sin r_-$ & $\rho \le 1$ & saddle \\
 5 & $-r_-$ & $-\mu_-$ &  $-(\rho
   -1) \sin r_-$ & $\rho \le 1$ & saddle \\
 6 & $-r_-$ & $-\nu_+$ & $-(\rho
   -1) \sin r_-$ & $\rho \le 1$ & saddle \\
 7 & $r_-$ & $-\nu_+$ & $(\rho -1)
   \sin r_-$ & $\rho \le 1$ & saddle \\
\hline
 8 & 0 & $-\omega_-$ & 0 & $\rho \ge 1$ & saddle \\
 9 & 0 & ${\pi }/{2}-\omega_+$ & 0 & $\rho \ge 1$ & saddle \\
 10 & 0 & $\pi -\omega_-$ &  0 & $\rho \ge 1$ & saddle \\
 11 & 0 & $-\omega_+-{\pi }/{2}$ &0 & $\rho \ge 1$ & saddle \\
\hline
 12 & $-r_+$ & $-\mu_+$ & $(\rho +1)
  
   \sin r_+$ & always & max  \\
 13 & $-r_+$ & $-\nu_-$ & $ (\rho +1)
   \sin r_+$ & always & max  \\
\hline
 14 & ${\pi }/{2}$ & --- & 1 & always
   & max \\
\hline
\end{tabular}
\end{table}
\begin{rem}\label{rem:24}
Looking at Table~\ref{table:V}, we note that maxima and minima belong to families running through the whole range of admitted values for $\rho$, while the saddles undergo bifurcations. For $\rho \le 1$ the four saddles are at
symmetric points on two opposite meridians (for $\chi = \pi/2$ these are identified by $y=0$) and drift
towards the Equator as $\rho$ approaches the critical value $\rho
= 1$, while for $\rho > 1$ the four saddles are at symmetric
points on the equator and drift away from the previously mentioned meridians as
$\rho$ increases. This means that there is a (saddle/saddle) \emph{local} bifurcation. \EOR
\end{rem}
\begin{rem}\label{rem:25}
Note also that a \emph{global} change takes place
at the same value $\rho = 1$. That is, for $\rho < 1$ the
orienting local maximum in the North Pole is also the absolute
maximum, the other two being (degenerate and) lower than this; for
$\rho > 1$, on the other hand, the other two maxima are
(degenerate and) higher than the orienting one. Similarly to what we have done for the $D_{3h}$ phase, we will distinguish these as $D_{2h}^+$ and $D_{2h}^-$ phases. \EOR
\end{rem}
\begin{rem}\label{rem:26}
We could have defined the orientation requiring
that the North Pole is not only a local maximum, but actually the
absolute maximum. In this case the parameter range would be
further restricted from the cylinder $\cyl$ to a subset
$\cyl_0$; and  Remark~\ref{rem:25} shows that the
intersection of $\cyl_0$ with the disk $\mathcal{D}$ (of
radius $\rho = 2$) would just be the disk of radius $\rho = 1$.
This would however introduce rather complex mappings involving
both the physical and the parameter space \cite{GV2016,CQV}, and
we prefer not to discuss it here; the reader is referred to
\cite{CQV} for a detailed discussion. \EOR
\end{rem}
\begin{rem}\label{rem:27}
If we look at the potential for $\rho = 1$, say
for the ``reference case'' $\chi = \pi/2$, it turns out this is invariant under a
subgroup of the group $O(2)$ acting in the $(y,z)$ plane; this is
generated by the matrices
\begin{eqnarray}
M_1 &=& \left(
\begin{array}{ll}
 1 & 0 \\
 0 & 1
\end{array}
\right)  ,\   M_2 =\left(
\begin{array}{ll}
 -\frac{1}{2} & -\frac{\sqrt{3}}{2} \\
 \frac{\sqrt{3}}{2} & -\frac{1}{2}
\end{array}
\right)  , \  M_3 =\left(
\begin{array}{ll}
 -\frac{1}{2} & \frac{\sqrt{3}}{2} \\
 -\frac{\sqrt{3}}{2} & -\frac{1}{2}
\end{array}
\right)  ; \nonumber\\
\\
M_4 &=& \left(
\begin{array}{ll}
 - 1 & 0 \\
 0 & 1
\end{array}
\right)  , \  M_5 =\left(
\begin{array}{ll}
 \frac{1}{2} & \frac{\sqrt{3}}{2} \\
 \frac{\sqrt{3}}{2} & - \frac{1}{2}
\end{array}
\right)  , \  M_6 =\left(
\begin{array}{ll}
 \frac{1}{2} & -\frac{\sqrt{3}}{2} \\
 -\frac{\sqrt{3}}{2} & -\frac{1}{2}
\end{array}
\right)  . \nonumber\end{eqnarray} These satisfy
\begin{eqnarray}
&M_2^3 = M_3^3 = M_4^2=M_1 = I  ;\nonumber\\  &M_2^2 = M_3  ,\quad  M_5 = M_4 M_2 
,\quad  M_6 = M_4 M_3  ; \\& M_5^2 = M_6^2 = I  . \nonumber
\end{eqnarray} 
The $SO(2)$
matrices $\{M_1,M_2,M_3 \}$ span the group $S_3$ of rotations
through multiples of $2 \pi / 3$, while $\{ M_4,M_5,M_6 \}$
(having determinant $-1$) are reflections in $y$ -- that is,
in three-dimensional terms, through the plane $(x,z)$ -- and
through planes obtained from this by rotations of $2 \pi / 3$ about the $x$ axis.
Thus the six matrices provide a representation
of the group $D_3$. The potential is also obviously invariant under
reflections in $x$, i.e. through the $(y,z)$ plane, hence \emph{at
the bifurcation point} we have a $D_{3h}$ symmetry (as on the axis
$\mathcal{A}$). \EOR
\end{rem}
\begin{rem}\label{rem:28}
Note also that the group $S_3$ acts mapping
maxima to maxima and minima to minima; the reflections map maxima into minima
and minima into maxima. Saddle points are obviously invariant, as
the transformations we are considering do not act on the $x$
coordinate. \EOR
\end{rem}
\begin{rem}\label{rem:29}
Our discussion in the last
two Remarks has been conducted in the ``reference case'' $\chi =
\pi/2$. For different values of $\chi$, we have the same situation but with an overall
rotation of the whole picture (see
Remark~\ref{rem:21}). \EOR
\end{rem}


\subsubsection{Bifurcations}
We are again interested in  relations between the
eigenvalues and the direction of eigenvectors, in particular near
the bifurcation points. In this case,
$ \Phi_+ =(1 + \rho)  \sin \xi_+$,
where $\Phi_+$ is the height of the ``secondary'' maxima (which
for $\rho > 1$ are actually higher than the orienting one) and
$(\xi_+ + \pi/2)$ is the angle between these maxima and the
orienting one (``secondary maxima'' can be recognized by
the fact they are always degenerate). We express $\rho$ in
terms of  $\xi_+$ through the equation \beq
\label{eq:d2hrhoxip} \rho =-  \frac{3 - 5 \cos2 \xi_+}{1
- 3 \cos2 \xi_+}  . \eeq

At the bifurcation point $\rho = 1$, we have $\xi_+ = \pi/6$ and
 $\Phi_+ = 1$. Thus, using \eqref{eq:d2hrhoxip} and with
some trivial algebra, we see that
\beq \Phi_+ = 4  \frac{\sin^3 \xi+}{3 \cos 2 \xi_+ - 1}  . \eeq By series
expansion at the bifurcation point, i.e. for $\xi_+ = \pi/6
+\eps$, we get \beq \Phi_+ (\eps) - \Phi_+ (0) =9  \sqrt{3}
 \eps  +  O (\eps^2 )  . \eeq 

Let us also consider the bifurcation from the $D_{\infty h}$ to
the $D_{2 h}$ phase, taking place at $\rho = 0$. Using again
\eqref{eq:d2hrhoxip}, and writing $\xi_+ = \xi_+^{(0)} + \eps$
where \beq \xi_+^{(0)} =\frac12  \arccos \( \frac35 \) \doteq0.46\eeq is
the value taken by $\xi_+$ for $\rho = 0$, we get \beq \Phi_+
(\eps) - \Phi_+ (0) =( 15  \cos \xi_+^{(0)}   \sin^2
\xi_+^{(0)}  +  30  \sin^3 \xi_+^{(0)} )   \eps  +  O
(\eps^2)  \doteq  5.37\, \eps  +  O ( \eps^2 )  . \eeq

\subsection{Reflection symmetry: special planes in the Bulk $\bulk$}
\label{sec:refl}
As suggested by the classification of $T_d$ subgroups, see
Appendix \ref{app:tetra}, we expect that there are specific values of
the parameters such that the potential is invariant under a
reflection in a vertical plane, i.e. under a $Z_2 = D_h$ group.

This is indeed the case for $\chi = \pm \pi/2$, $\chi = \pm
\pi/6$, $\chi = \pm 5 \pi /6$. Let us just consider the first case
(the others are just obtained from this by a $2 \pi / 3$
rotation, see below).

For $\chi = \pm \pi/2$, the potential  \eqref{eq:potential_53}
reduces to
\beql{eq:PhiRP0}
\Phi_{or} = z^3  +  K  y  (y^2 - 3 x^2) -
\frac32 (x^2 + y^2) z  \pm \frac32\rho  (x^2 - y^2 )  z ;
\eeq this is manifestly invariant under the reflection in the
$(y,z)$ plane, i.e. under $x \to - x$. Equivalently, we have
invariance under the subgroup of $T_d$ generated by $M_{13}$ (see Appendix \ref{app:tetra} for the matrices $M_i$).

Similar considerations apply for 
the subgroups generated by
$M_{14}$, with invariance subject to the condition $\chi = -
\pi/6$ or $\chi = 5 \pi / 6$; in this case the reflection is
through the plane $x = - \sqrt{3} y$.
In comolete analogy with this is
the subgroup generated by
$M_{15}$, where now one has to require $\chi = - 5 \pi / 6$ or
$\chi = \pi/6$; the reflection plane is then $x = \sqrt{3} y$.

We will refer to these planes (collectively) as $\mathcal{P}$;
when we want to be more specific (see Table~\ref{table:VII} at the end of this section) we will
call them, respectively, $\mathcal{P}_0$, $\mathcal{P}_-$,
$\mathcal{P}_+$. These reflection symmetries had evaded our previous studies \cite{GV2016,CQV} and  have proved quite significant in the present one.

\begin{figure}
	\begin{tabular}{ccc}
		\includegraphics[width=105pt]{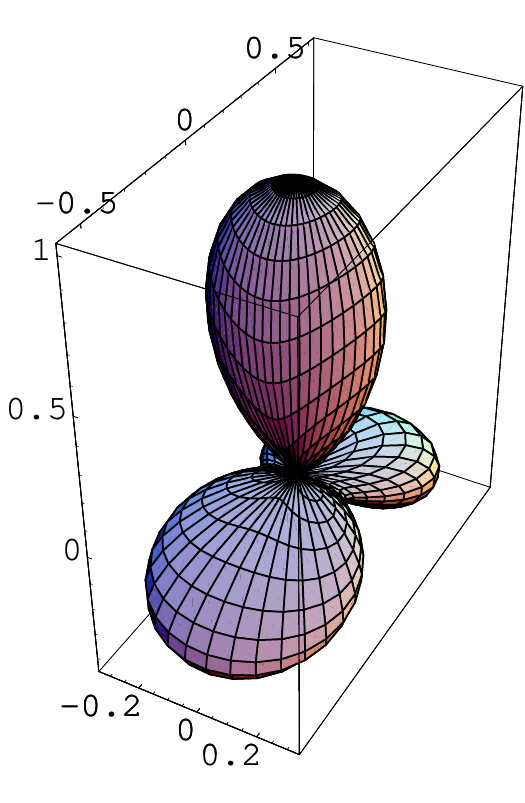} &
		\includegraphics[width=105pt]{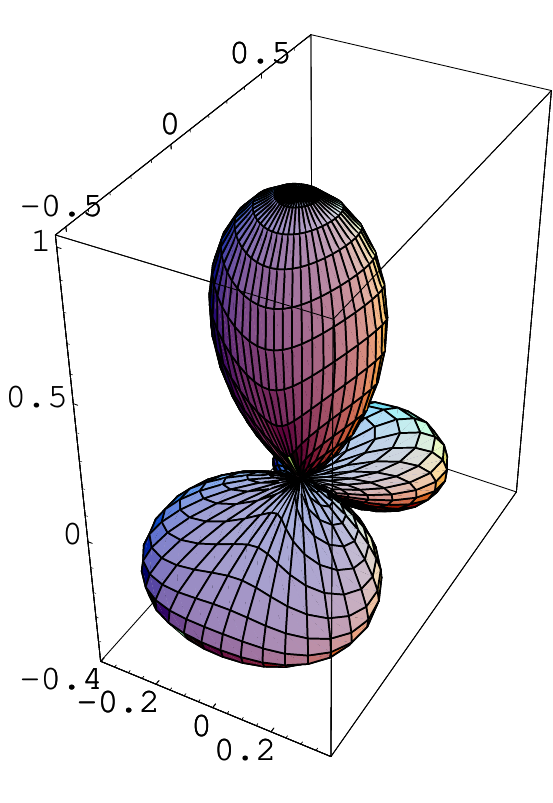} &
		\includegraphics[width=105pt]{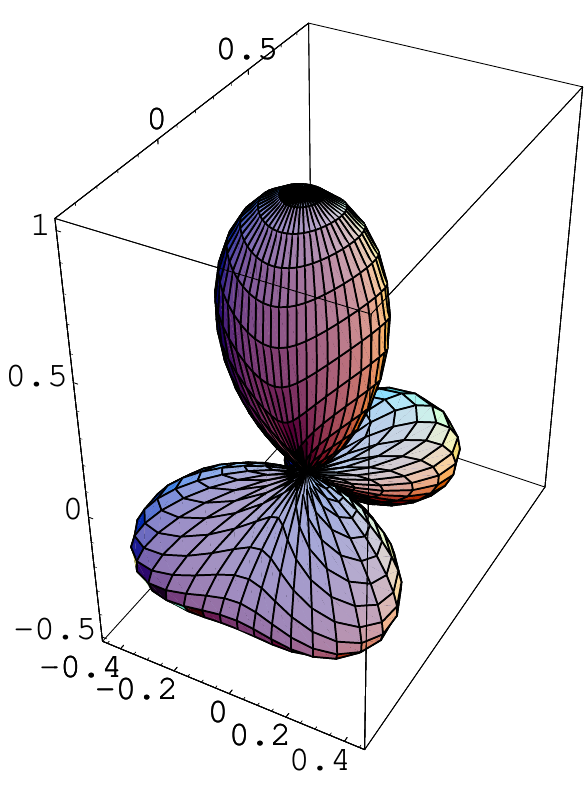} \\
		$K = 0.1$ & $K = 0.2$ & $K = 0.3$ \\
		\includegraphics[width=105pt]{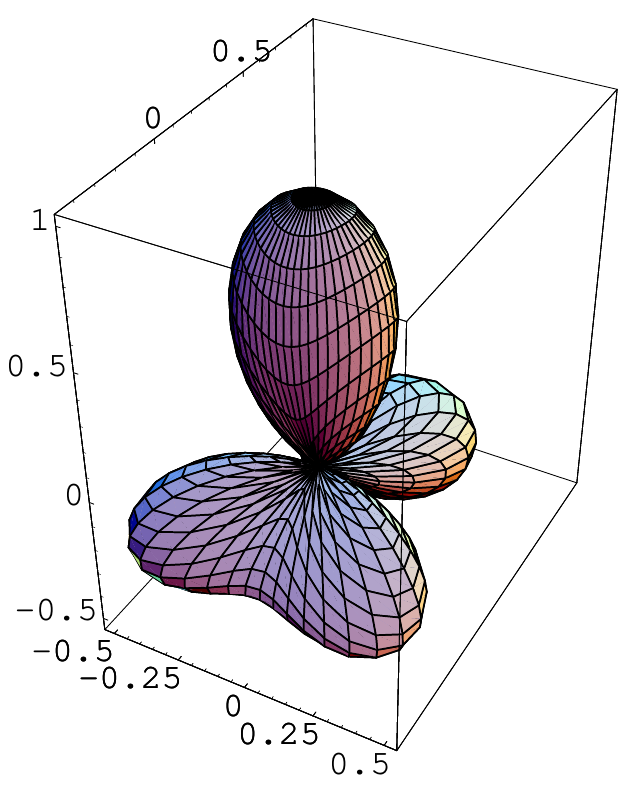} &
		\includegraphics[width=105pt]{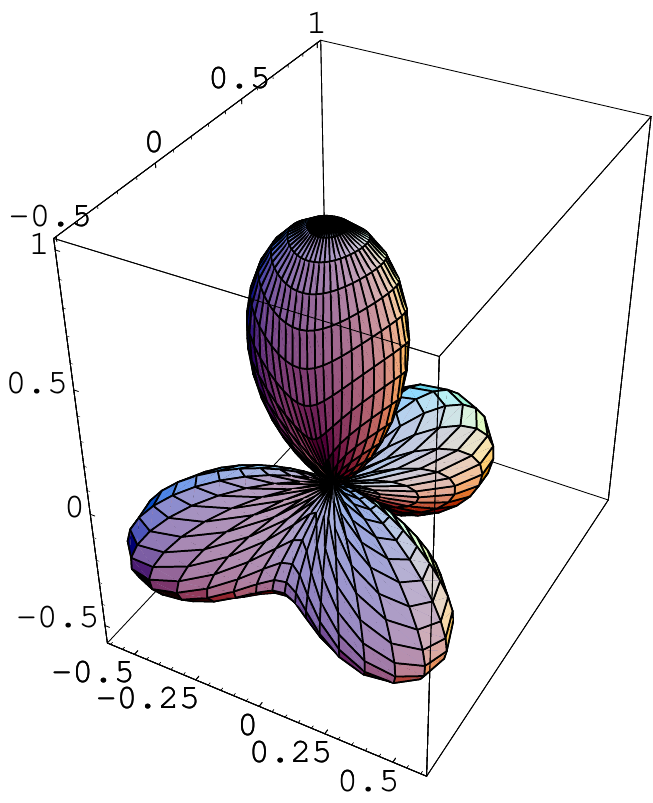} &
		\includegraphics[width=105pt]{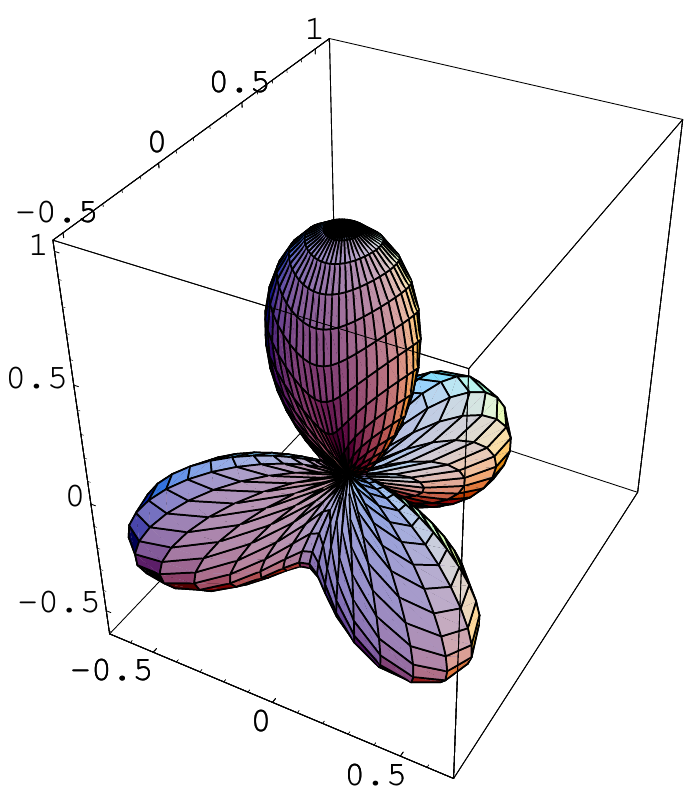} \\
		$K = 0.4$ & $K = 0.5$ & $K = 0.6$ \end{tabular} \\
	\caption{The potential for $\chi = \pi/2$ and $\rho = 1/2$, for different values of $K>0$.
		We observe the transition from a phase with three maxima at low $K$ to a phase with four
		maxima at higher $K$. The octupolar potential is always reflection invariant through a
		vertical plane spanned by the axes $y$ and $z$.}\label{fig:rpSP}
\end{figure}

\begin{figure}
	\begin{tabular}{ccc}
		\includegraphics[width=105pt]{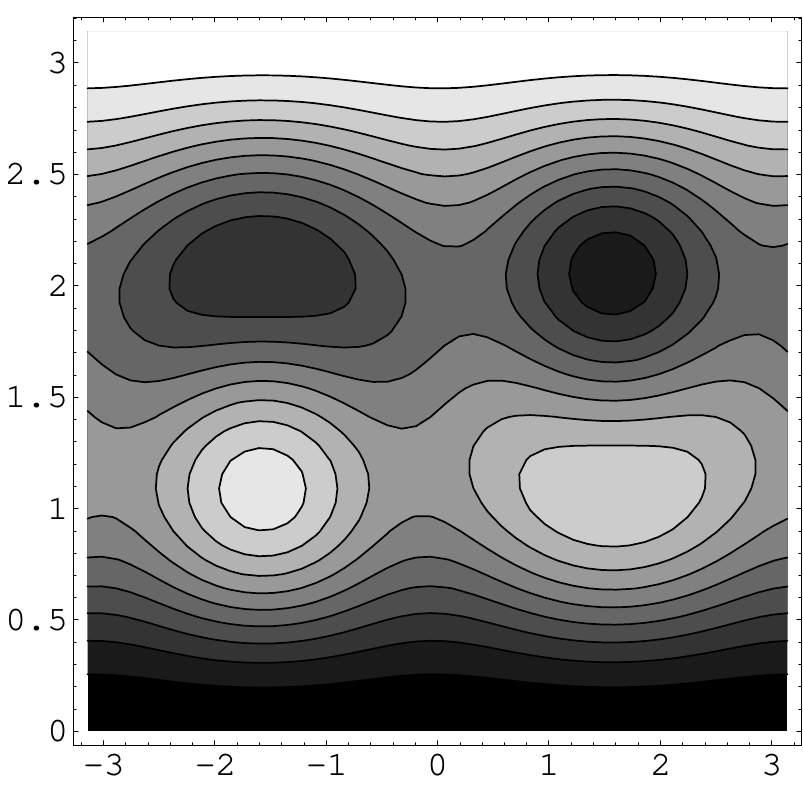} &
		\includegraphics[width=105pt]{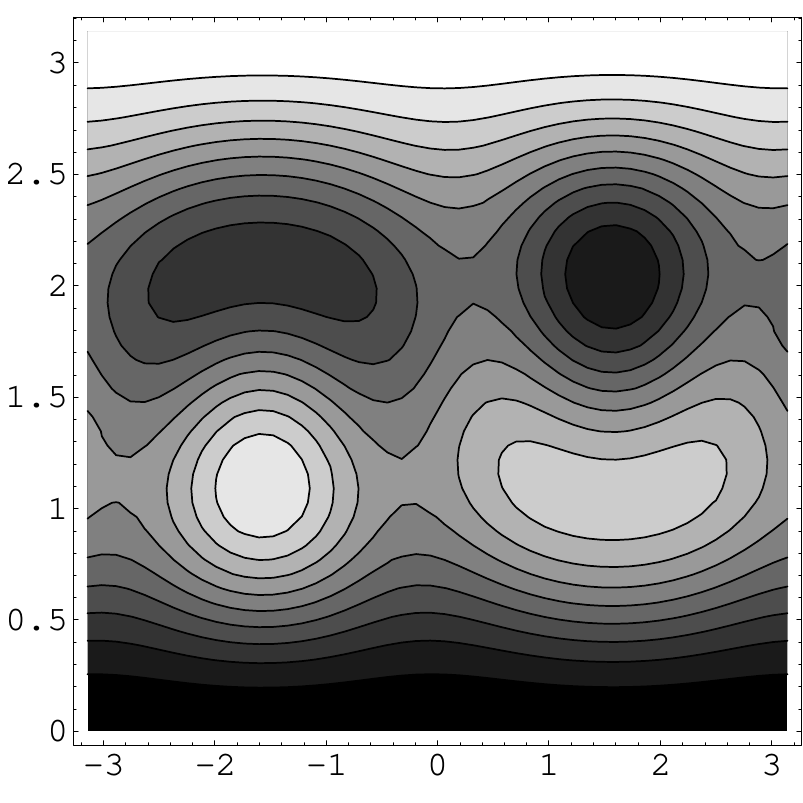} &
		\includegraphics[width=105pt]{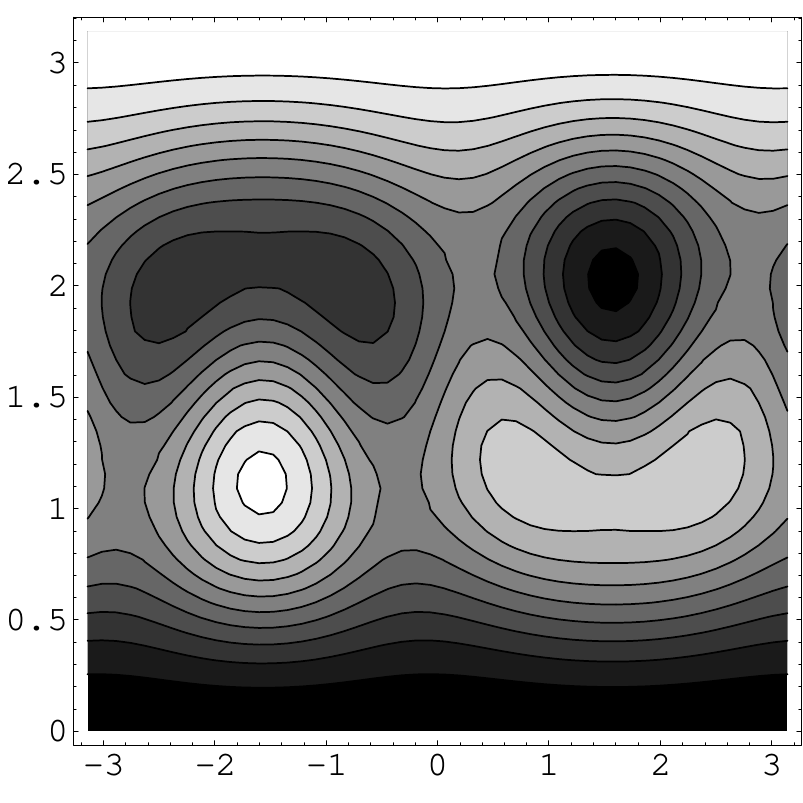} \\
		$K = 0.1$ & $K = 0.2$ & $K = 0.3$ \\
		\includegraphics[width=105pt]{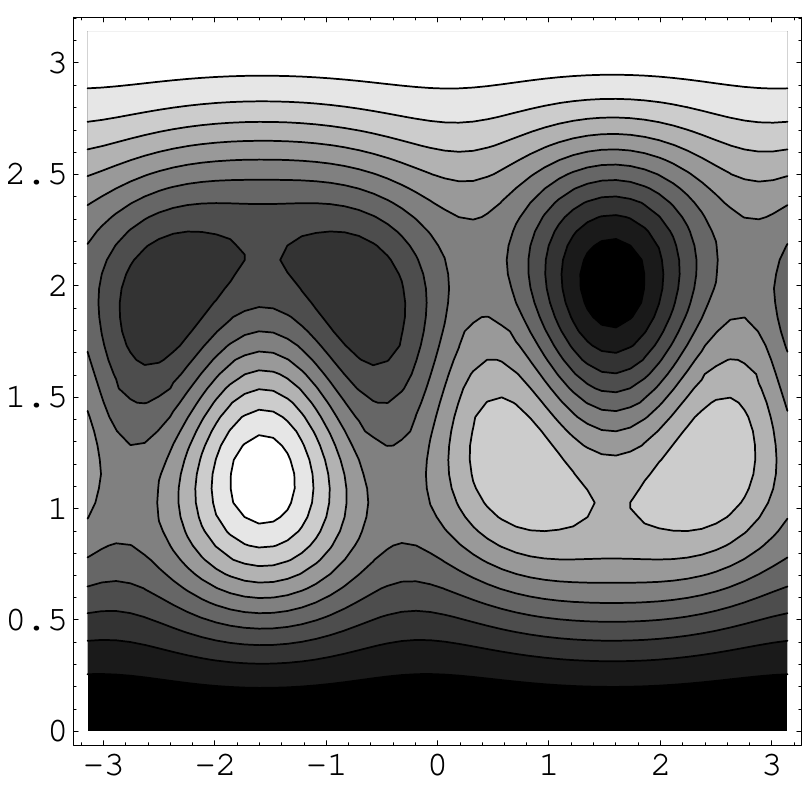} &
		\includegraphics[width=105pt]{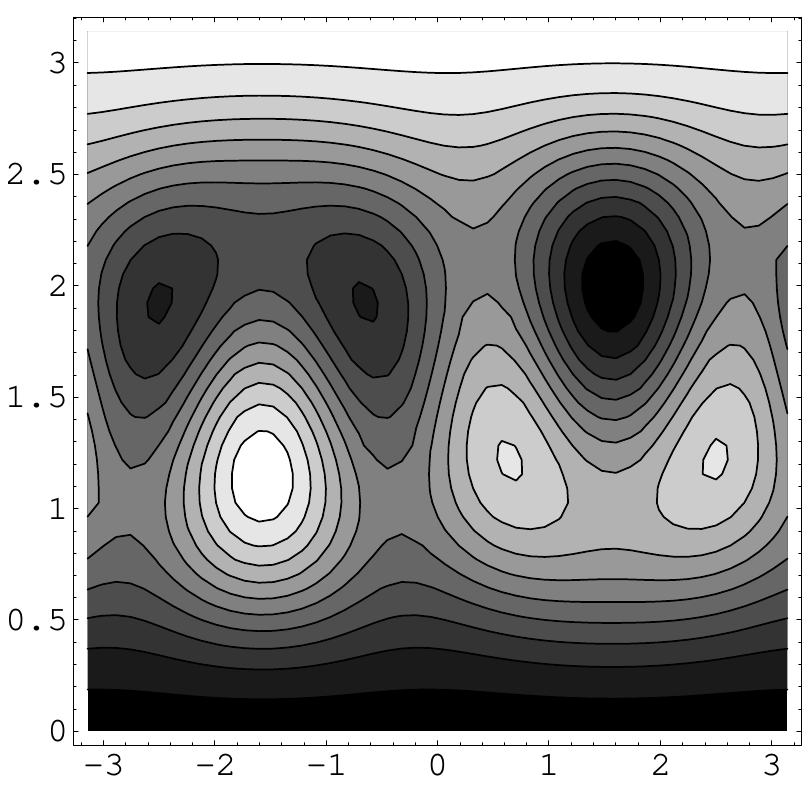} &
		\includegraphics[width=105pt]{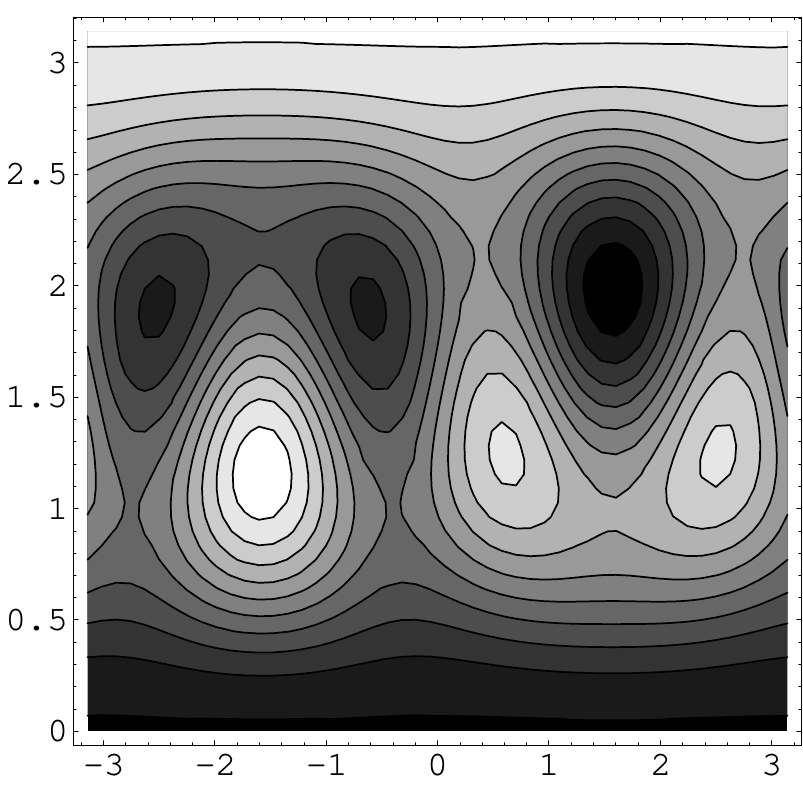} \\
		$K = 0.4$ & $K = 0.5$ & $K = 0.6$ \end{tabular} \\
	\caption{Same as Fig.~\ref{fig:rpSP} but with contour plots.}\label{fig:rpCP}
\end{figure}

For the special $D_h$ phases considered here, as for all others, the transition   from the octupolar potential having four maxima to that having only three takes place for parameters chosen on the \emph{separatrix}
identified in our previous work \cite{GV2016,CQV} and also recalled
in the following Sect.~\ref{sec:bulk}. This is
illustrated in Figs.~\ref{fig:rpSP} and \ref{fig:rpCP}. The separatrix is a surface in parameter space that marks the border between  the subregions $\bulk_3$ and $\bulk_4$ of $\bulk$, where $\pot_{or}$ has three or four maxima, respectively.

To discuss in more detail the critical points in these phases, we find it more convenient to expressing the octupolar potential in angular coordinates ans in \eqref{eq:potev}.
\begin{rem}\label{rem:chimenchi} Since, by $\gamma_3$ in \eqref{eq:parsym1}, the potential in \eqref{eq:potev} is invariant under the simultaneous shifts
	$$ \chi \, \to \, \chi \, + \, \frac{2 m}{3} \pi , \ \ \vth_2 \, \to \, \vth \, + \, \frac{2 m}{3} \pi , $$ by studying the case $\chi = \pi/2$ we also obtain information about the cases $\chi = - 5\pi/6 $ and $\chi = -\pi/6$; and by studying the case $\chi = -\pi/2$ we also obtain information about the cases $\chi = \pi/6$ and $\chi = 5\pi/6$.
\end{rem}
We will only consider the cases $\chi = \pm \pi/2$.

\subsubsection{The case $\chi = \pi/2$}

By setting $\chi = \pi/2$, the potential \eqref{eq:potev} reduces to
\beql{eq:PsiRP0} \Psi_{or} = \sin^3\vth_1 - K \cos^3\vth_1 \sin 3 \vth_2  - \frac32 \sin\vth_1 \cos^2\vth_1( 1 -  \rho  \cos2 \vth_2) . \eeq
The conditions for a critical point are then
\beq
\frac{\pa \Psi_{or}}{\pa \vth_1} = 0  , \quad
\frac{\pa \Psi_{or}}{\pa \vth_2} = 0  ; \eeq
both equations have a power of $\cos \vth_1 $ as an overall factor, which of course vanishes only at the poles; quotienting this factor (and constants) out, we remain with
\begin{eqnarray}
K \sin 2 \vth_1  \sin 3 \vth_2  - \cos^2 \vth_1  ( 1  -  \rho \cos 2 \vth_2) 
 + 2 \sin^2 \vth_1  ( 2  - \rho \cos 2 \vth_2) &=& 0 ,\nonumber\\ \label{eq:gr1} \\
K \cos \vth_{1} \cos 3 \vth_{2} + \rho \sin \vth_{1}  \sin 2
\vth_{2}  &=& 0  .\nonumber \end{eqnarray}

The second equation \eqref{eq:gr1} is  solved for $\vth_2 = \pm \, \pi/2 $ and moreover (assuming $\vth_2 \not= \pm \pi/2$) by
\beql{eq:solvth1gen} \tan \vth_1 = - \frac{K}{\rho} \
\frac{\cos 3 \vth_2}{\sin 2 \vth_2}  , \eeq which determines uniquely $\vth_1 \in (- \pi/2 , \pi/2)$ once $\vth_2$ is given.
In the following it will be useful to express this, and more specifically $X= \sin \vth_1$, in terms of $Y=\sin \vth_2$. With some trigonometry, it turns out that
\beql{eq:S1plus} X = - K \frac{1 - 4  Y^2}{2 \rho Y \sqrt{1 + \frac{K^2 (1 - 4 Y^2)^2}{4 \rho^2 Y^2}}} 
 . \eeq
Since the right hand of \eqref{eq:S1plus} is an odd function of $Y$, this equation delivers so it gives the same result for $X$, and hence for $\vth_1 = \arcsin X \in [-\pi/2,\pi/2]$, for the two determinations of $\vth_2 = \arcsin Y \in [-\pi,\pi]$.

\paragraph{The solutions $\vth_2 = \pm \pi/2$.}
Let us consider first the solutions $\vth_2= \pm \pi/2$. Inserting this into the first equation \eqref{eq:gr1}, we obtain
\beq 3 + \rho (5 + 3 \rho) \cos 2 \vth_1 \mp 2 K \sin  2 \vth_1 = 0 . \eeq
This is better written in terms of $X$ as 
\beql{eq:solth2pi2_new} 10  X^2 - 2 + \rho  (6 X^2 - 2) = \pm  4  K  X \sqrt{1 -X^2} . \eeq

The solutions to \eqref{eq:solth2pi2_new} for the case $\vth_2= \pi/2$ are
\beql{eq:xsolgr1p_new} X_1 = \sqrt{\frac{\a + \b}{\ga} }  , \quad
X_2 = - \sqrt{\frac{\a - \b}{\ga} } ; \eeq
in the case $\vth_2 = - \pi/2$ we have instead
\beql{eq:xsolgr1m_new} X_3 = -X_2 , \quad
X_4 = - X_1 . \eeq
In the formulas \eqref{eq:xsolgr1p_new} and \eqref{eq:xsolgr1m_new} we have set
\begin{eqnarray}
	\a &=& 5 +2 K^2 + 8 \rho +3 \rho^2 , \nonumber\\
	\b &=& 2 \, K \sqrt{4 +K^2 +6 \rho + 2 \rho^2} , \\
	\ga &=& 4 K^2 \, + \, (5 + 3 \rho)^2 .\nonumber \end{eqnarray}
It is obvious that the argument of the square root is always positive (hence $\b$ is always real), and the same applies for $\a$ and $\ga$; moreover $\a \ge \b$ in our range $0 \le \rho \le 2$. Thus the four solutions are all real. It is easy to check, even numerically, that the solutions \eqref{eq:xsolgr1p_new} and \eqref{eq:xsolgr1m_new} also satisfy (for $0 \le \rho \le 2$) the condition $|X| \le 1$, necessary to be in accord with the definition of $X$ as $\sin\vth_1$.

By looking at the Hessian of the potential computed in these critical points, we can ascertain their nature.

For $\vth_2 = \pi/2$, it turns out that $X_2$ is a maximum for all values of $K$ and $\rho$ in the considered range, while $X_1$ undergoes a bifurcation along a certain curve $K = f (\rho)$, and is a minimum (for $K < f (\rho)$) or a saddle (for $K > f(\rho)$) depending on the values of the parameters.

Similarly, for $\vth_2 = - \pi/2 $, it turns out that $X_3$ is a minimum for all allowed values of $K$ and $\rho$, while $X_4$ undergoes a bifurcation along the same curve $K = f (\rho)$, and is a  maximum (for $K < f(\rho )$) or a saddle  (for $K > f(\rho )$) depending on the values of the parameters.

The function $f(\rho)$ is given explicitly  by
\beql{eq:frhoRP} f(\rho) = \sqrt{\frac{2 \, \rho^2 (1+\rho)}{3 (6 +\rho )}}  \eeq
and plotted in Fig.~\ref{fig:g_graph} below.

\paragraph{The other solutions ($\vth_2 \not= \pm \pi/2$).}
The other solutions, i.e. those with $\vth_2 \not= \pm \pi/2$, are characterized by \eqref{eq:solvth1gen} as solution to the second equation \eqref{eq:gr1}. Plugging this into the first equation \eqref{eq:gr1}, recalling again that $\vth_1 \in [- \pi/2,\pi/2]$, applying some standard trigonometry and writing for ease of notation $Y = \sin \vth_2$, we obtain
\beql{eq:rpy} K^2 (\rho - 2) + \[ 4 \, K^2 \, (4 - \rho) + 2 \, \rho^2 \, (1 - \rho) \]  Y^2 + 4 (\rho^3  -  8 )  Y^4 = 0  . \eeq

The solutions to this equation are
\beql{eq:rpysol} Y= \pm \sqrt{ \frac{A  \pm  B}{C} }  ,\eeq
where we have written
\begin{eqnarray}
	A &=& 2  K^2  (4 - \rho)  +  \rho^2  ( 1 -  \rho)  ,\nonumber \\
	B&=& \rho \sqrt{ 4 K^4 \rho^2 + 4 K^2 \rho^2 (4 - 3 \rho) + \rho^4 (1 - \rho)^2 } := \sqrt{b} ; \\
	C&=& 4 (8  K^2  -  \rho^3)  .\nonumber \end{eqnarray}

For these to be real we need that $b \ge 0$; and moreover, if this condition is satisfied, that $(A \pm B)/C \ge 0$. One easily checks, e.g. numerically, that (for $0 \le \rho \le 2$) indeed $b \ge 0$, hence $B$ is real (and non-negative, as we take the positive determination of the root).

Actually, we have
\begin{eqnarray} b &=& A^2 + 4  K^2  (2 - \rho)  (\rho^3 - 8 K^2) ;  \nonumber\label{eq:betaL}\\
\\ B &=& |A| \sqrt{1 + \frac{4  K^2 (2 - \rho) (\rho^3 - 8 K^2)}{A^2}} .\nonumber  \end{eqnarray}
Thus we can have $A \pm B = 0$ only for $K=0$, for $\rho=2$, and on the curve $K^2 = (\rho/2)^3$. Note that $A \ge 0$ for \beq K^2 \ge \frac{\rho^2 (\rho - 1)}{2 (4 - \rho)} ; \eeq this implies in particular that $A$ is always positive for $\rho < 1$.

In order to study if the solutions
\begin{equation}
\label{eq:a_b}
 Y_a = \pm \sqrt{\frac{A-B}{C}} , \quad Y_b = \pm \sqrt{\frac{A+B}{C} }
 \end{equation}
are real, we consider the signs of $A \pm B$ and of $C$.
It turns out that $A+B \ge 0$ for all values of $K$ and for all $0 \le \rho \le 2$. As for $A-B$, it follows from \eqref{eq:betaL} that it has the same sign as $C = 8 K^2 -\rho^3$. Finally, it is obvious that $C >0$ for $K^2 > (\rho/2)^3$.

This means that, by requiring $Y$ to be real, for $K^2 < (\rho / 2)^3$ we only have the solutions $Y_a $, while for $K^2 > (\rho/2)^3$ we have both $Y_a$ and $Y_b$.

This is not enough: in fact, $Y = \sin\vth_2$ requires also $|Y| \le 1$. This condition is always satisfied by $Y_a$, while for $Y_b$ it requires 
$ K \ge f(\rho)$. It follows from \eqref{eq:frhoRP} that,  in the relevant range for $\rho$, $f$ satisfies $f(\rho)\ge (\rho/2)^{3/2}$.

Finally we note that each solution for $Y = \sin\vth_2$ corresponds to two solutions for $\vth_2 \in [- \pi, \pi]$; and, as mentioned above, once $\vth_2$ is given, $\vth_1$ is uniquely determined.

Thus we conclude that:
\begin{enumerate}
	\item For $K > f(\rho)$ we have four solutions for $Y$, two of type $(a)$ and two of type $(b)$ as in \eqref{eq:a_b}, and hence eight critical points beside the two at the poles and the four with $\vth_2 = \pm \pi/2$, for a total of \emph{fourteen} critical points;
	\item For $K < f(\rho)$ we have two solutions for $Y$, of type $(a)$, and hence four critical points beside those at the Poles and for $\vth_2 = \pm \pi/2$, for a total of \emph{ten} critical points;
	\item On the curve $K = f(\rho)$ there is a bifurcation, in which the two solutions of type $(b)$ disappear as $K$ is reduced. On the curve $K=f(\rho)$ the solutions $Y_b$ have $\vth_2 = \pm \pi/2$, hence merge with those studied before.
\end{enumerate}
\begin{rem}\label{rem:*4}
	Clearly, the simple expression $K = f(\rho)$ for the bifurcation curve was possible only because we have fixed the value of $\chi$. This curve is the section of the separatrix with the plane $\chi=\pi/2$. The separatrix in the full three-dimensional $(K,\rho,\chi)$ parameter space has been studied in \cite{GV2016} and \cite{CQV}, but it has an awkward analytic expression, which duly reduces to \eqref{eq:frhoRP} for $\chi=\pi/2$ (\emph{modulo} the different scaling of $\rho$, as shown by (25) of \cite{CQV}). \EOR
\end{rem}

\subsubsection{The case $\chi = - \pi/2$.}
The case $\chi = - \pi/2$ is analyzed in the same way, though it entails a somewhat dissimilar outcome, on which se shall particularly concentrate.

\begin{figure}[h]
	\centering
\includegraphics[width=.75\linewidth]{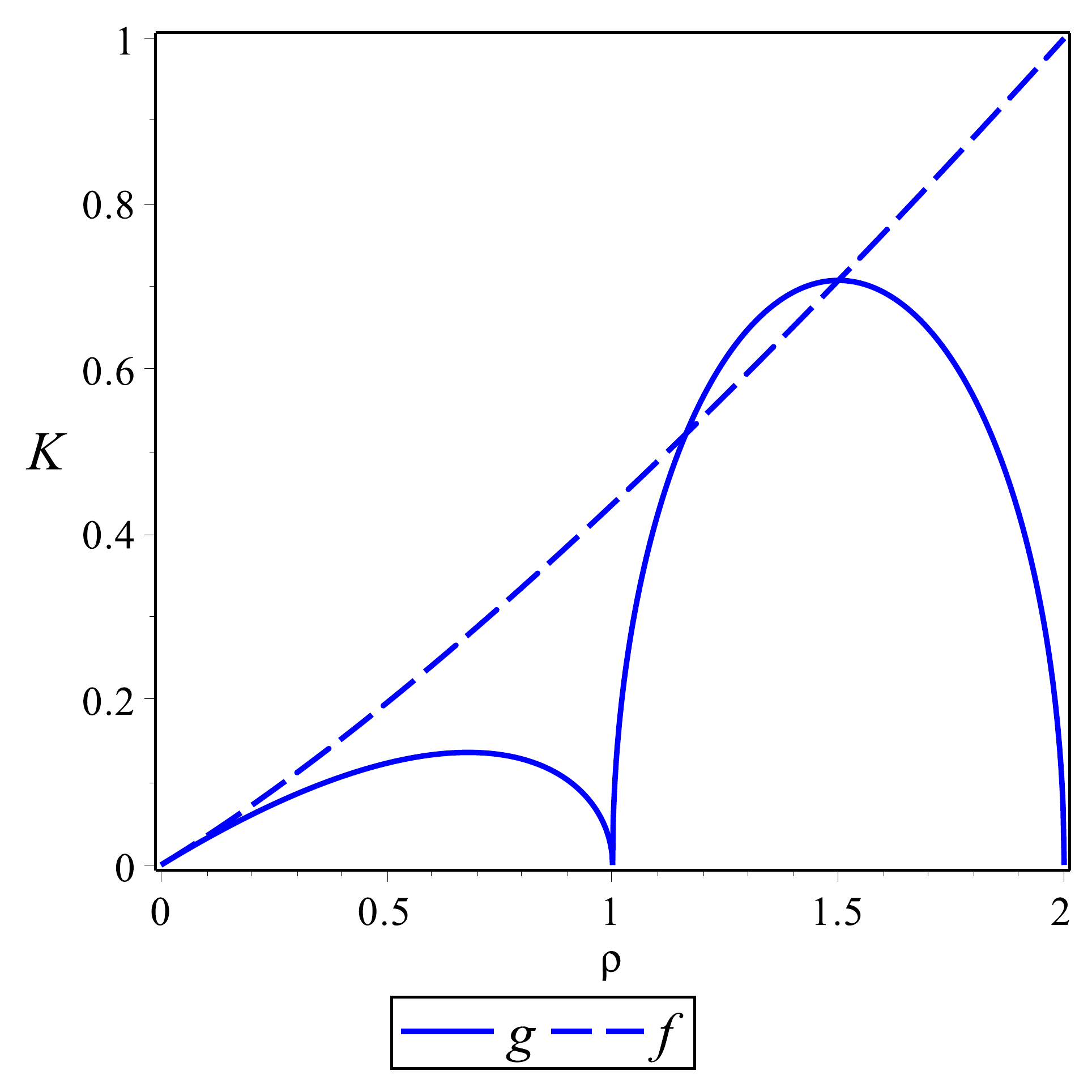}
\caption{The graph of the function $g$ defined by \eqref{eq:g_definition} is superimposed to the graph of the function $f$ defined by \eqref{eq:frhoRP}.}
\label{fig:g_graph}
\end{figure}

Paralleling $f$ in \eqref{eq:frhoRP}, there is a continuous function $g$ defined for $0\le\rho\le2$ by
\begin{equation}\label{eq:g_definition}
g(\rho)=
\cases{
\sqrt{\frac{2\rho^2(1-\rho)}{3(6-\rho)}} & for $0\le\rho\le1$\cr\cr
\sqrt{2(2-\rho)(\rho-1)} & for $1\le\rho\le2$\cr},
\end{equation}
whose graph is reproduced in Fig.~\ref{fig:g_graph} for the reader's ease.
\begin{enumerate}
	\item For $K > g(\rho)$ we have eight generic critical points beside the two at the poles and  four on the special meridians with $\vth_2 = \pm \pi/2$, for a total of \emph{fourteen} critical points. Four are maxima, four minima, and the remaining six are saddles.  
	\item For $K < g(\rho)$, we have a total   of \emph{ten} critical points, of which three are maxima, three minima, and the remaining four are saddles.
	\item For $K=g(\rho)$, two different scenarios present themselves, according to whether $0<\rho<1$ or $1<\rho<2$. In the former case, the critical points are ten, whereas in the latter case they are \emph{twelve}. In both cases, the total number of maxima is three, as many as the minima; only the number of saddles differs: there are four for $0<\rho<1$ and six for $1<\rho<2$. In the former case, two saddles are degenerate, but all four have index $\iota=-1$. In the latter case, two out of the six saddles are degenerate and have index $\iota=0$ (see Remark~\ref{rem:new_1}), while the remaining four are not degenerate and have the usual index $\iota=-1$.
	\item A special note is deserved by the limiting values $\rho=1$ and $\rho=2$. For the former value, the total number of critical points is \emph{eight}, whereas it is ten for the latter. For $\rho=1$, three maxima and three minima are accompanied by two degenerate saddles, each with index $\iota=-2$. For $\rho=2$, the same number of maxima and minima is accompanied by four degenerate saddles, each  with index $\iota=-1$, for a total of ten critical points. The total number of critical points for both these limiting cases were predicted by our taxonomic analysis in Sect.~\ref{sec:index}: the case $\rho=1$ falls under row $(c_2)$ in Table~\ref{table:Ib}, while the case $\rho=2$ falls under row $(c)$ in Table~\ref{table:Ia}.
\end{enumerate}
We now describe in more detail how the different components of this varied landscape of critical points are combined together.
The critical points for $K<g(\rho)$ are related to those for $K>g(\rho)$ in two different ways, corresponding to the two branches of $g$ in \eqref{eq:g_definition}, according to whether $0<\rho<1$ or $1<\rho<2$.

For $0<\rho<1$, the four critical points on the special meridians $\vth_2=\pm\pi/2$ survive as $K$ decreases below $g(\rho)$; two saddles, one for each meridian, stay saddles, whereas a maximum on one meridian ($\vth_2=\pi/2$) becomes a saddle, as does a minimum on the other meridian ($\vth_2=-\pi/2$). On the other hand, always for $0<\rho<1$, two generic critical points, which are both saddles, approach each  meridian $\vth_2=\pm\pi/2$ as $K$ approaches $g(\rho)$ from above; they coalesce for $K=g(\rho)$ on the extremal that on the targeted meridian will transform into a saddle, and disappear as $K<g(\rho)$. For $K=g(\rho)$, on each meridian $\vth_2=\pm\pi/2$ the coalesced critical point is a degenerate saddle with index $\iota=-1$.

For $1<\rho<2$, the evolution of critical points is somewhat different, though the final outcome is identical. As $K$ decreases below $g(\rho)$, the four critical points on meridians $\vth_2=\pm\pi/2$ cease to exist, but they do not mingle with the generic critical points, which instead survive. They rather annihilate in pairs   on each meridian for $K=g(\rho)$. The superposition of a maximum with a saddle (for $\vth_2=\pi/2$) and that of a minimum with a saddle (for $\vth_2=-\pi/2$) give rise to a critical point with index $\iota=0$, so that for $K=g(\rho)$ and $1<\rho<2$ the total number of critical points is \emph{twelve}: three maxima, three minima, and six saddles. 

To determine the nature of the latter critical points, we expanded $\Psi_{or}$ in their vicinity. For $\vth_2=-\pi/2$ and $K=g(\rho)$, we found a degenerate saddle located at
\begin{equation}\label{eq:saddle_location}
\vth_1=\vth_S=-\arcsin\sqrt{\frac{\rho-1}{3-\rho}}
\end{equation}
and we computed
\begin{eqnarray}\label{eq:third_order_expansion}
\Psi_{or}(\vth_{1},\vth_{2})&=&-\sqrt{(\rho-1)(3-\rho)}+12(2-\rho)\sqrt{\frac{\rho-1}{3-\rho}}\left(\vth_2+\frac\pi2\right)^2\nonumber\\& +&\frac12\sqrt{2(2-\rho)(3-\rho)}\left(\vth_1-\vth_S\right)^3
\nonumber\\&+&3\sqrt{2}(5\rho-6)\sqrt{\frac{2-\rho}{3-\rho}}\left(\vth_2+\frac\pi2\right)^2\left(\vth_1-\vth_S\right)\nonumber\\
&+&O(4).
\end{eqnarray}
A similar formula applies to the degenerate saddle on the meridian $\vth_2=\pi/2$. It is then easy to conclude that both critical points are degenerate saddles with index $\iota=0$ (see, for example \cite{bolis:degenerate}). They migrate towards the poles as $\rho$ approaches $2$ along the line $K=g(\rho)$ and towards the Equator as $\rho$ approaches $1$. Correspondingly, the North Pole becomes a degenerate maximum (while the South Pole becomes a degenerate minimum) and the Equator hosts two symmetric ``monkey saddles''.

\subsection{Trivial symmetry $\{ e \}$: the Bulk $\mathcal{B}$}
\label{sec:bulk}

We have so far discussed the critical strata that correspond to
nontrivial symmetries in the cylinder $\cyl$  representing the parameter
space of our theory.

\begin{figure}
	\begin{tabular}{ccc}
		\includegraphics[width=105pt]{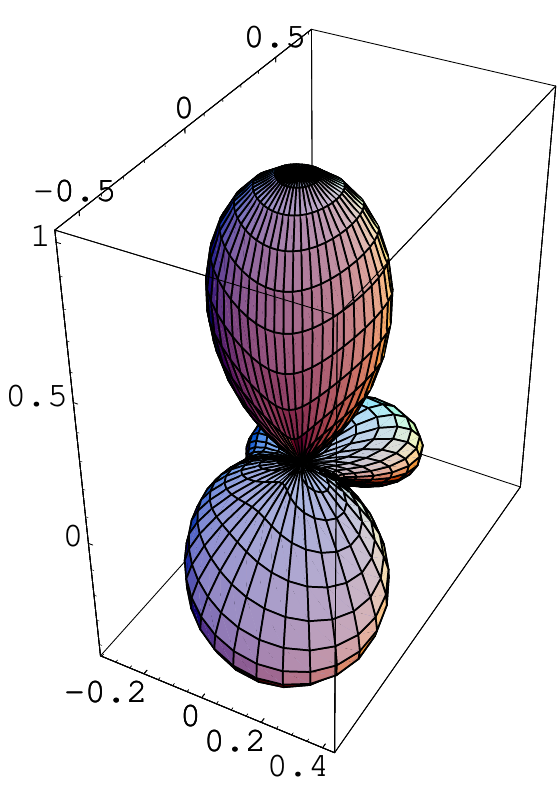} &
		\includegraphics[width=105pt]{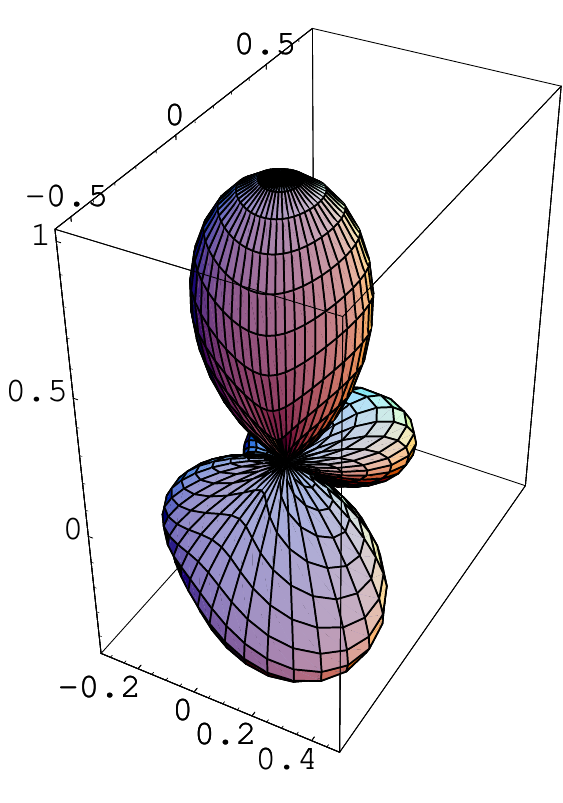} &
		\includegraphics[width=105pt]{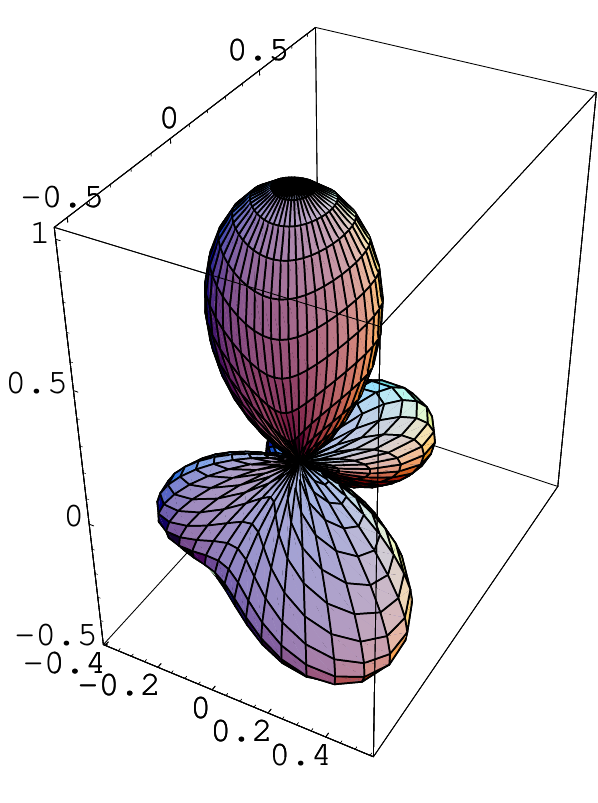} \\
		$K = 0.1$ & $K = 0.2$ & $K = 0.3$ \\
		\includegraphics[width=105pt]{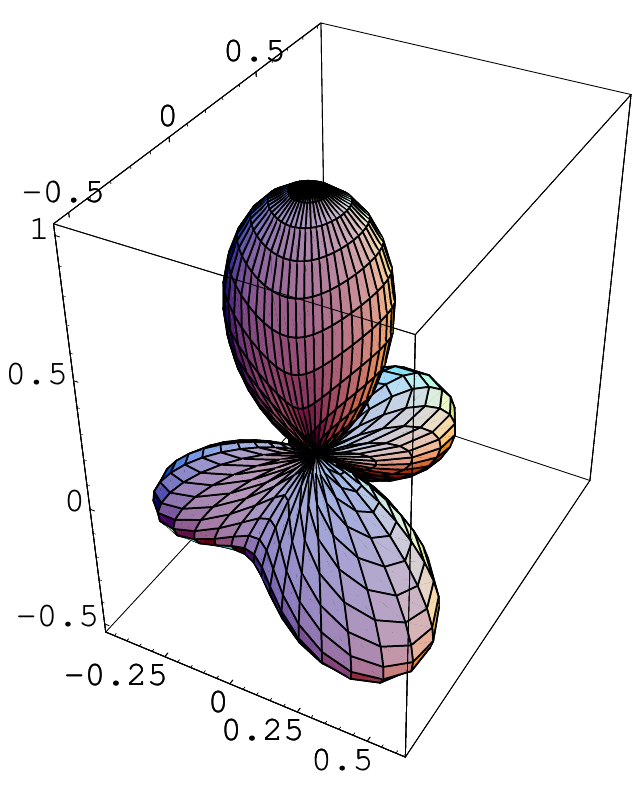} &
		\includegraphics[width=105pt]{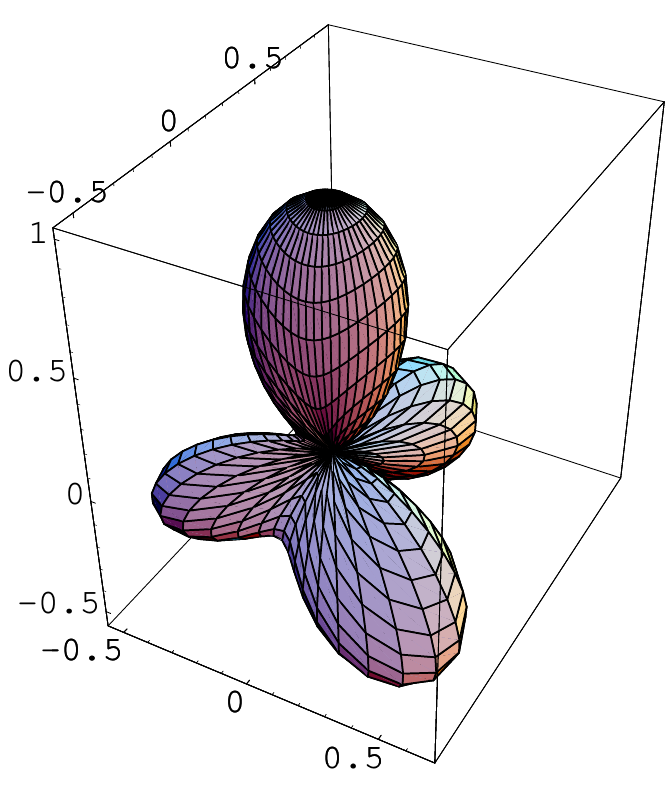} &
		\includegraphics[width=105pt]{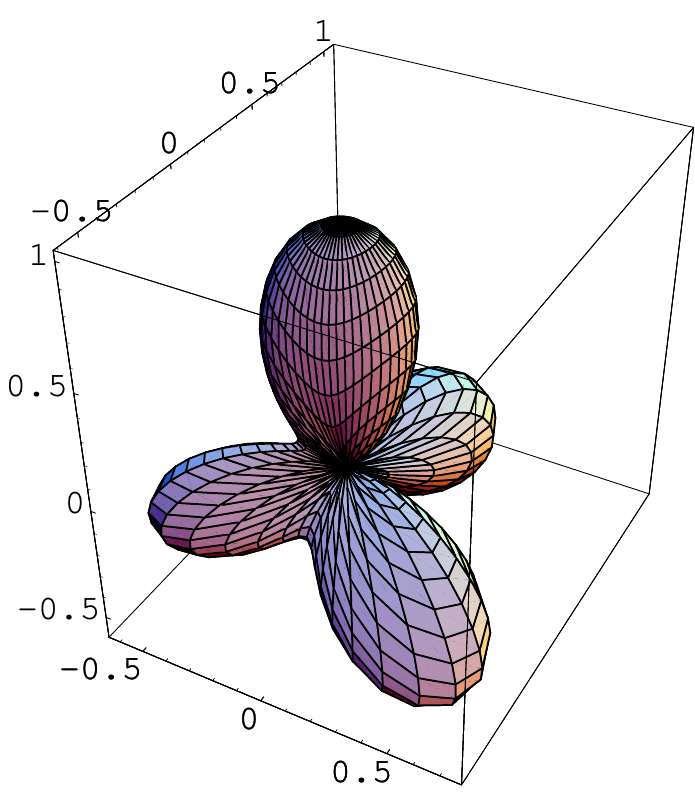} \\
		$K = 0.4$ & $K = 0.5$ & $K = 0.6$ \end{tabular} \\
	\caption{The potential for $\chi = \pi/3$ and $\rho = 1/2$, for different values of $K>0$.
		We observe the transition from a phase with three maxima at low $K$ to a phase with four
		maxima at higher $K$. Generically, the potential is never reflection invariant through a
		vertical plane.}\label{fig:bulkSP}
\end{figure}

\begin{figure}
	\begin{tabular}{ccc}
		\includegraphics[width=105pt]{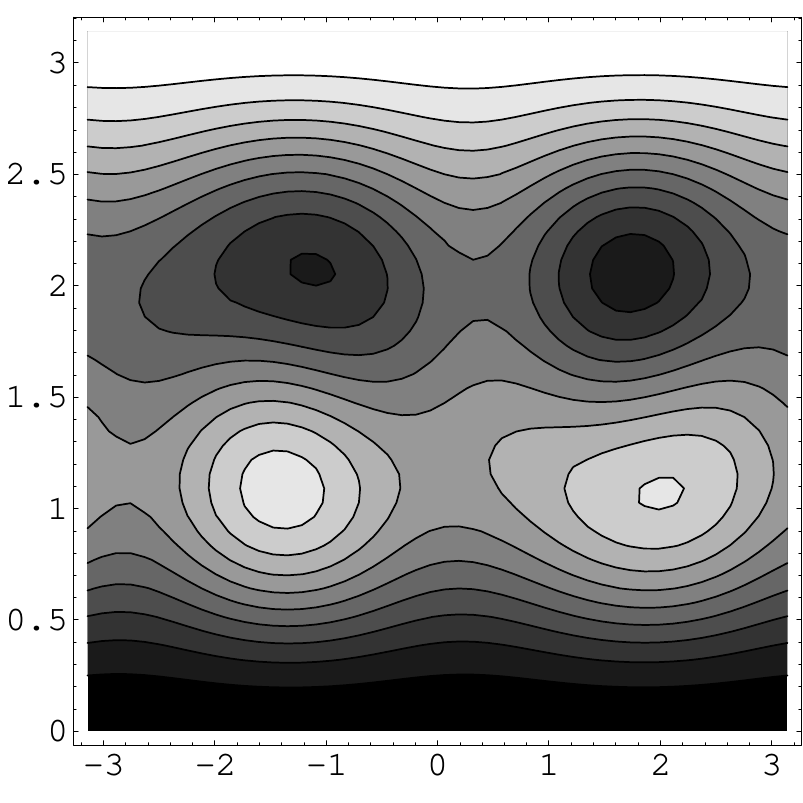} &
		\includegraphics[width=105pt]{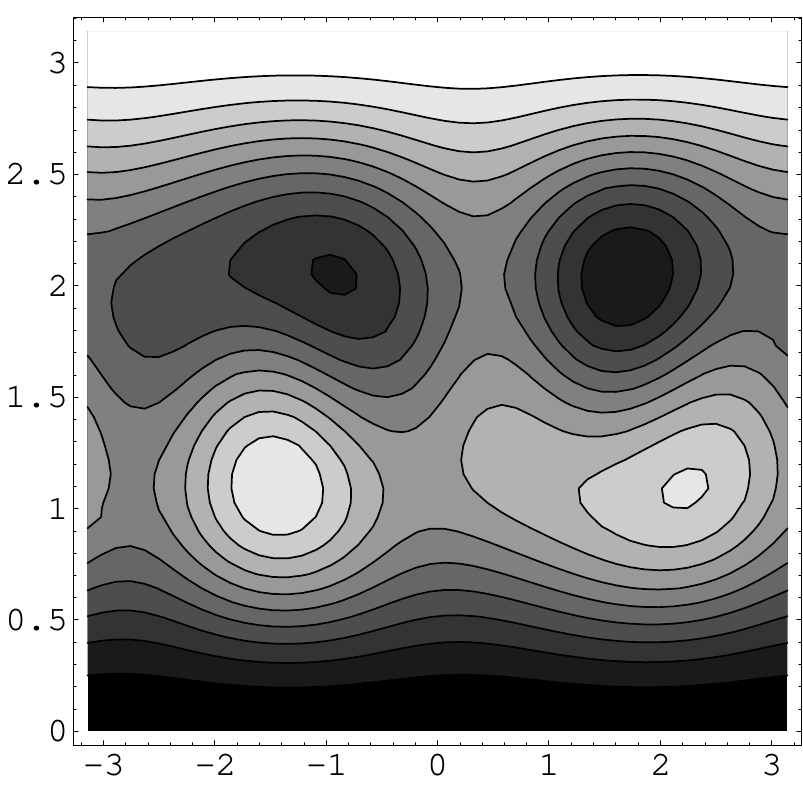} &
		\includegraphics[width=105pt]{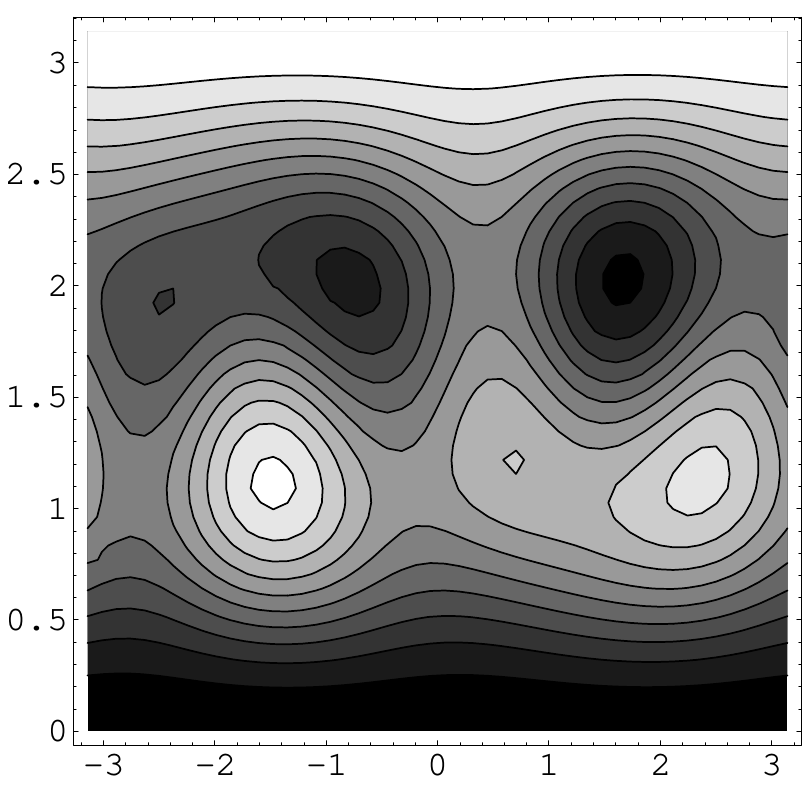} \\
		$K = 0.1$ & $K = 0.2$ & $K = 0.3$ \\
		\includegraphics[width=105pt]{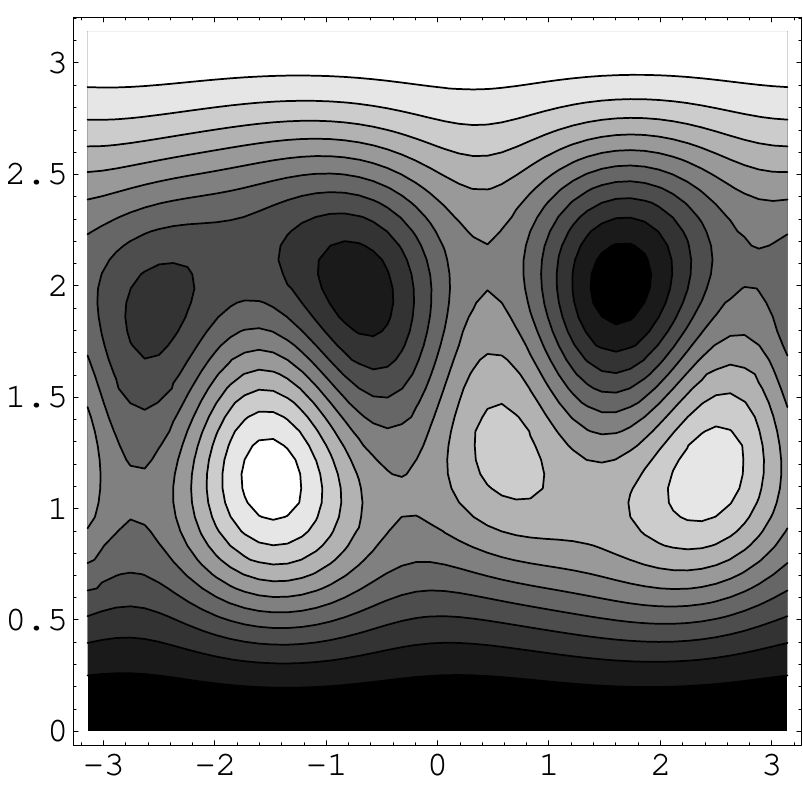} &
		\includegraphics[width=105pt]{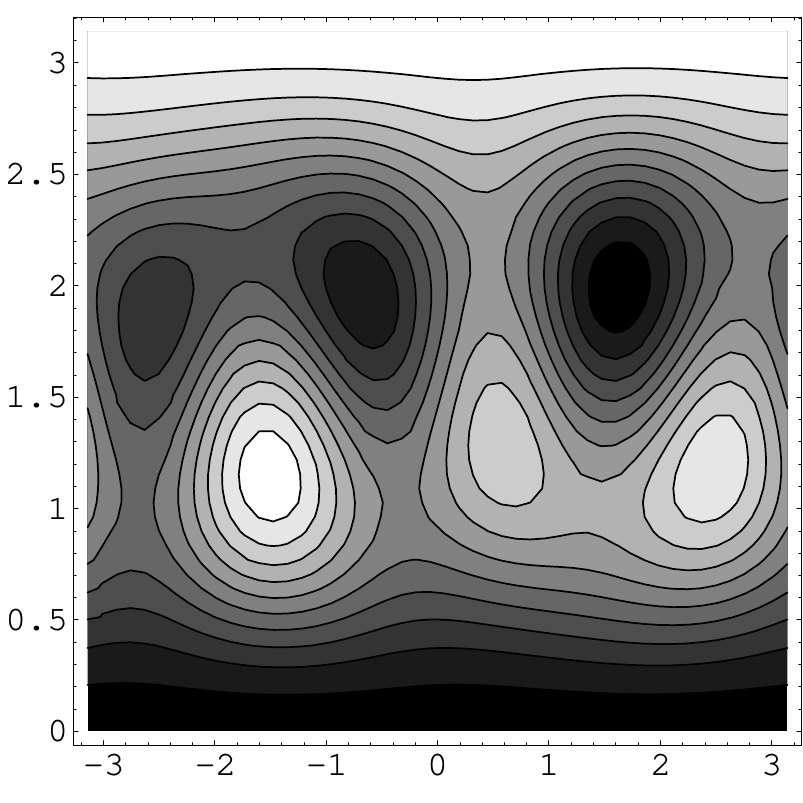} &
		\includegraphics[width=105pt]{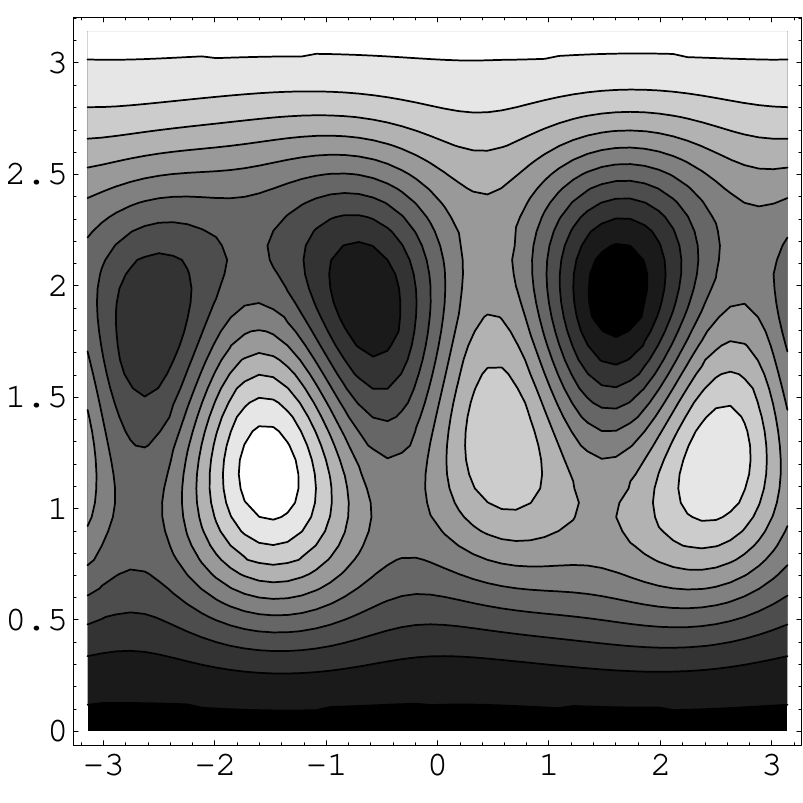} \\
		$K = 0.4$ & $K = 0.5$ & $K = 0.6$ \end{tabular} \\
	\caption{Same as Fig.~\ref{fig:bulkSP} but with contour plots.}\label{fig:bulkCP}
\end{figure}

The situation in the bulk $\mathcal{B}$ of the cylinder, i.e. for
the generic case, has been discussed in detail in our previous
work \cite{GV2016}; there we have shown that -- quite surprisingly
-- there are \emph{two} generic octupolar phases, characterized by
a different number (10 and 14) of critical points, and separated
in parameter space by a \emph{separatrix}. Moreover, one could
also distinguish the cases where the maximum in the North Pole is
the absolute one, and that where it is a local maximum but not the
absolute one; these two regions are separated by a \emph{dome},
having its vertex in one of  the tetrahedral points $\mathcal{T}$ and
meeting the disk $\mathcal{D}$ on the circle of radius $\rho = 1$.
This dome has been further investigated, providing more detailed
information, in \cite{CQV}. The generic incarnations of the octupolar potential below the dome in parameter space are illustrated in Figs.~\ref{fig:bulkSP} and \ref{fig:bulkCP}.

One easily distinguishes two different regions where the potential can be seen
as topologically equivalent to the highly symmetric incarnations  met on the disk $\mathcal{D}$ and at the
tetrahedral points $\mathcal{T}$. In fact, the study of the generic 
configurations in the bulk, and of the separatrix between the two
octupolar phases, was based on continuation techniques starting
from these singular strata. Much of this study has already been presented in our
previous work \cite{GV2016,CQV}.

Here, the more detailed insight gained  through the analysis of the highly symmetric cases $\chi=\pm\pi/2$ allows us to refine our understanding of the separatrix by establishing the existence of an  extension where the total number of critical points for the octupolar potential is either 8 or 12, instead of the 10 that had already been found in \cite{GV2016,CQV}.

\begin{figure}
	\centering
	\includegraphics[width=.9\linewidth]{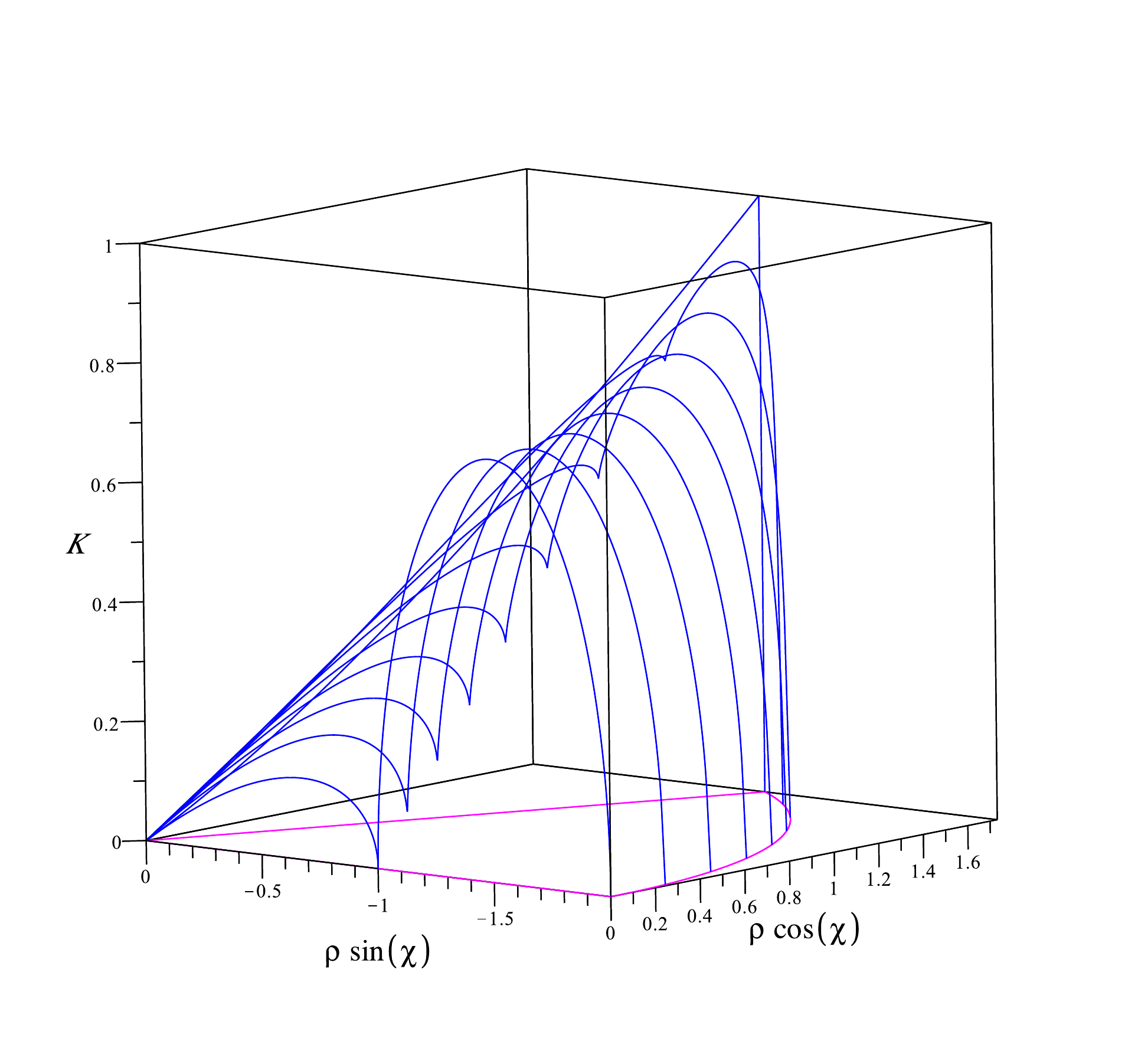} 
	\includegraphics[width=.9\linewidth]{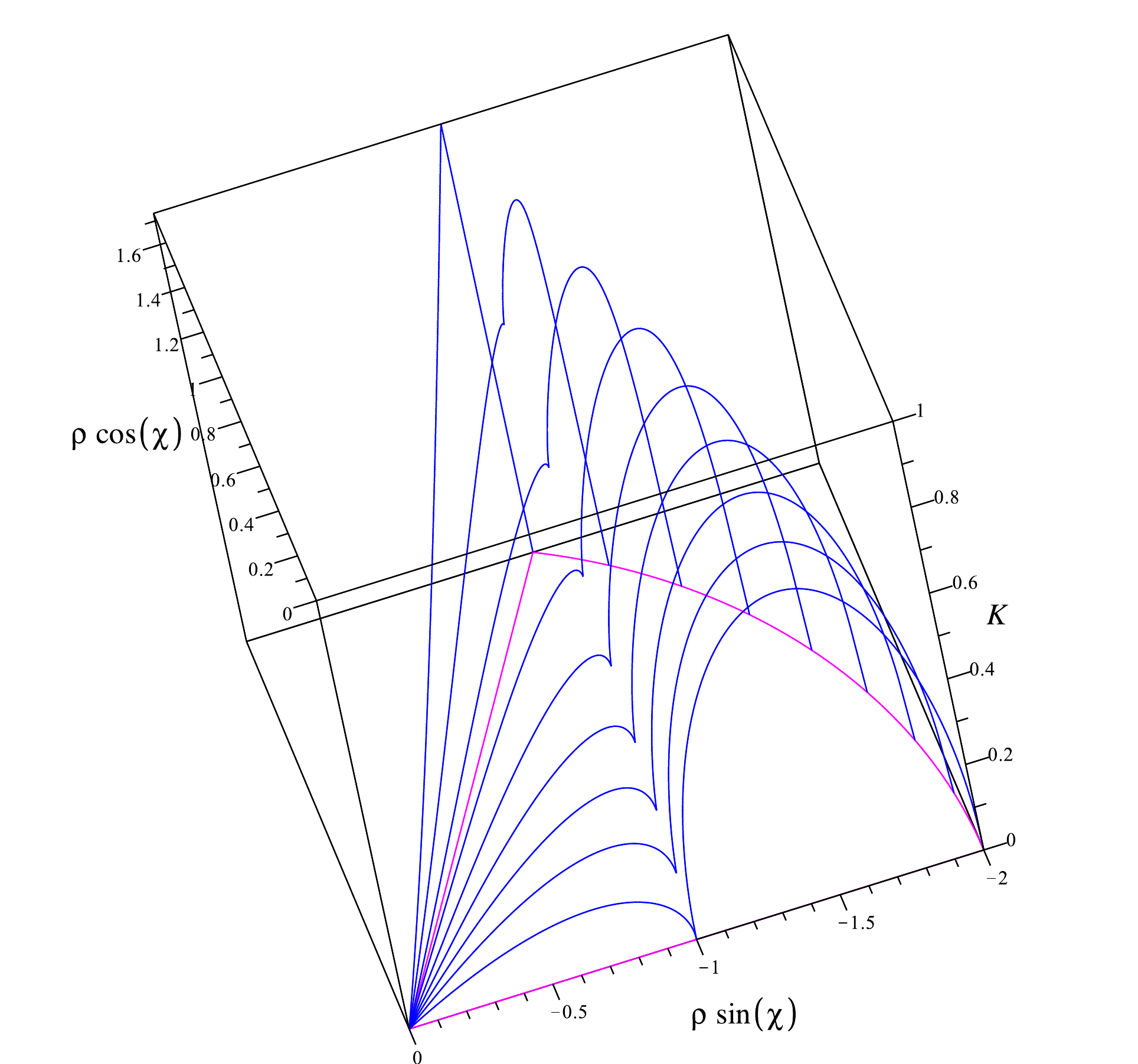} 
	\caption{Two views of the whole separatrix, which is outlined by curves at equally spaced values of $\chi$ in the interval $[-{\pi}/{2},-{\pi}/{6}]$. A line of cusps, which bears 8 critical points for the octupolar potential, separates the inner component of the separatrix with 10 critical points from the outer component with 12 critical points.}
	\label{fig:separatix_views}
\end{figure}

Fig.~\ref{fig:separatix_views} shows the outcome a standard numerical continuation technique applied to the sector $-\pi/2\le\chi\le-\pi/6$ in parameter space $(K,\rho,\chi)$, which in view of the symmetries described above is the only one bearing essential information. The educated eye will discern the graphs of functions $g$ anf $f$, plotted against $\rho$ upon the sections $\chi=-\pi/2$ and $\chi=-\pi/6$, respectively. The remaining curves outline the whole extended separatrix.

A line of cusps connects the point $K=0$, $\rho=1$, $\chi=-\pi/2$  with the point $K=1$, $\rho=2$, $\chi=-\pi/6$: it bears 8 critical points for the octupolar potential. The line that connects the point $K=0$, $\rho=2$, $\chi=-\pi/2$  with the point $K=1$, $\rho=2$, $\chi=-\pi/6$ consists of an arc of circle of radius $2$ in the plane $K=0$ and a straight  segment orthogonal to that plane: it bears instead 10 critical points. The bump delimited by these degenerate lines bears 12 critical points. Always, above the sepratrix the critical points are 14, whereas they are 10 below it. Our previous studies were confined to the cylinder $0\le\rho\le1$, and so they were blind to the outer component of the sepratrix with 12 critical points for the octupolar potential.

\subsection{Special transformations and potential invariance}
\label{app:ori}
Our choice of orientation and scaling allowed to simplify the
potential, passing from seven to three parameters. This also
reduced the allowed transformations. In fact, except at the
special tetrahedral point $\rho = 0, K = K_0$ (see above)
we can only consider transformations which preserve the unit
sphere \emph{and} leave invariant the $z$ axis; we refer to these
as \emph{oriented orthogonal transformations}. They are just maps
in $O(2)$ acting on the $(x,y)$ variables, and hence correspond to
matrices of the form \beq M =\pmatrix{ \cos \ga  & -  s
	\sin \ga \cr \sin \ga & s  \cos \ga  \cr}  , \eeq
where $s$ is just a sign, $s = \pm 1$ and corresponds to the
determinant of $M$.

When we require invariance of $\Phi$ under such transformations,
we obtain -- by standard algebra -- five classes of nontrivial
solutions; each of these can exist only for certain values of the
parameters, i.e. for certain regions $R$ in the parameter space. The results are summarized in Table~\ref{table:VII} below. This table confirms that our classification of different phases is
complete. In fact, the first class gives $G = O(2)$ in
$\mathcal{C}$; the second one provides $G = D_{2 h}$ in
$\mathcal{D}$; the third ones provides $G = D_{3h}$ in
$\mathcal{A}$, while the fourth points out at the fact that a symmetry
$C_3$ is also present in $\mathcal{A}$; finally the fifth class
shows that on the special planes $\mathcal{P}$ there is a
reflection symmetry.

\begin{table}[H]
	\caption{Different oriented transformations as
		symmetries of the oriented potential in different regions $R$ of
		the parameter space. The symbol $-$ means that any value of the
		corresponding parameter is allowed.}
	\label{table:VII}
	\centering
	\begin{tabular}{||l|l|l||c||c||c||}
		\hline
		$K$ & $\rho$ & $\chi$ & $R$ & $s$ & $\ga$ \\
		\hline
		0 & 0 & -- & $\mathcal{C}$ & $\pm 1$ & -- \\
		\hline
		0 & -- & -- & $\mathcal{D}$ & $-1$ & $\pm  \arccos ( \pm \sin \chi)$ \\
		0 & -- & -- & $\mathcal{D}$ & $+ 1$ & $\pm  \pi$ \\
		\hline
		-- & 0 & -- & $\mathcal{A}$ & $- 1$ & $- 5\pi/3$ \\
		-- & 0 & -- & $\mathcal{A}$ & $- 1$ & $- \pi/2$ \\
		-- & 0 & -- & $\mathcal{A}$ & $- 1$ & $- \pi/3$ \\
		-- & 0 & -- & $\mathcal{A}$ & $- 1$ & $+ \pi/3$ \\
		-- & 0 & -- & $\mathcal{A}$ & $- 1$ & $+ \pi/2$ \\
		-- & 0 & -- & $\mathcal{A}$ & $- 1$ & $+ 5\pi/3$ \\
		\hline
		-- & 0 & -- & $\mathcal{A}$ & $+ 1$ & $-  \pi$ \\
		-- & 0 & -- & $\mathcal{A}$ & $+ 1$ & $- 4\pi/3$ \\
		-- & 0 & -- & $\mathcal{A}$ & $+ 1$ & $- 2\pi/3$ \\
		-- & 0 & -- & $\mathcal{A}$ & $+ 1$ & $+ 2\pi/3$ \\
		-- & 0 & -- & $\mathcal{A}$ & $+ 1$ & $+ 4\pi/3$ \\
		-- & 0 & -- & $\mathcal{A}$ & $+ 1$ & $+  \pi$ \\
		\hline
		-- & -- & $- 5\pi/6$ & $\mathcal{P}_+$ & $\pm 1$ & $\pi/3$ \\
		-- & -- & $- \pi/2$ & $\mathcal{P}_0$ & $\pm 1$ & $\pi$ \\
		-- & -- & $- \pi/6$ & $\mathcal{P}_-$ & $\pm 1$ & $- \pi/3$ \\
		-- & -- & $+ \pi/6$ & $\mathcal{P}_+$ & $\pm 1$ & $\pi/3$ \\
		-- & -- & $+ \pi/2$ & $\mathcal{P}_0$ & $\pm 1$ & $\pi$ \\
		-- & -- & $+ 5\pi/6$ & $\mathcal{P}_-$ & $\pm 1$ & $- \pi/3$ \\
		\hline
	\end{tabular}
\end{table}

\section{Summary of critical points}
\label{sec:discu}
In the following Table~\ref{table:VI}, we summarize our findings concerning critical points of the octupolar 
potential $\Phi$  in different regions of the admissible semi-definite cylinder $\cyl_+$ in parameter space. We preliminary leave out the critical points with index $\iota=0$, which we  found on some special bifurcation loci. A more refined description of the landscape of critical points is outlined at the end of the section.
 
\begin{table}[H]
	\caption{Summary of the totality of critical points  in parameter space with index $\iota\neq0$, irrespective of all their possible additional features and characterization.}
	\label{table:VI}
	\centering
\begin{tabular}{||c|c|l||}
  \hline
  $G$ & set & crit.\ pts.  \\
  \hline
  $\{ e \}$ & $\mathcal{B}$ & 10 or  14 \\
  $D_{\infty h}$ & $\mathcal{C}$ & $2 + \infty$ \\
  $D_{2 h}$ & $\mathcal{D}$ & 10  \\
  $D_{3h}$ & $\mathcal{A}$ & 14 \\
  $T_d$ & $\mathcal{T}$ & 14  \\
  \hline
\end{tabular}
\end{table}

The relations between geometric (and symmetry) strata of the
cylinder are summarized in this diagram $$ \matrix{
 & & & & \mathcal{B} & & \cr
 & & & \swarrow & & \searrow & \cr
 & & \mathcal{A} & & & & \mathcal{D} \cr
 & \swarrow & & \searrow & & \swarrow & \cr
 \mathcal{T} & & & & \mathcal{C} & & \cr}  $$
It should be noted  that $\mathcal{C}$ and $\mathcal{T}$ can be reached from the
bulk $\mathcal{B}$ also directly, i.e. without passing through
$\mathcal{A}$ or $\mathcal{D}$.

Similarly, the relation between isotropy subgroups are summarized
in the next diagram $$ \matrix{ D_{\infty h} & & T_d
& & D_{\infty h} \cr
 \downarrow & \swarrow & & \searrow & \downarrow \cr
 D_{3h} & & & & D_{2h} \cr
 & \searrow & & \swarrow & \cr
 & & \{ e \} & & \cr} $$

Here we do not distinguish in $\mathcal{B}$
between the region $\bulk_4$ with four maxima of $\Psi_{or}$ from the region  $\bulk_3$ with three maxima, as  these correspond to different phases
which do not differ in terms of symmetry. However, if we wish to classify the regions in $\cyl_+$ according to the cardinality of the real spectrum of the octupolar tensor $\A$ corresponding to $\Phi$ (which is the number of critical points of $\Phi$ on the unit sphere $S^2$), we need to be more refined. Then even the distinction between $\bulk_3$ and $\bulk_4$ is too gross. 

We have shown that $\bulk_3$ and $\bulk_4$ are separated by a separatrix $\mathcal{S}$, a surface in parameter space which consists of two folds, $\mathcal{S}_1$ and $\mathcal{S}_2$, where $\Phi$ has 10 and 12 critical points, respectively. These subsurfaces are divided by a line of cusps $\mathcal{L}_1$ where $\Phi$ has 8 critical points. Moreover, $\mathcal{S}_2$ is bordered by another line, $\mathcal{L}_2$, where $\Phi$ has 10 critical points.










\section{Conclusions}\label{sec:conclu}
An octupolar tensor $\A$ is any third order, completely symmetric and completely traceless tensor. In 2D, $\A$ has the symmetries of an equilateral triangle and it can be effectively represented by a vector in the plane. In 3D, the symmetries enjoyed by $\A$ outline a much richer landscape. This paper has classified all of them by introducing the octupolar potential $\pot$ associated with $\A$, that is, the cubic form of $\A$ restricted to the unit sphere $S^2$. The maxima (and antipodal minima) of $\pot$ and their locations on the unit sphere describe the whole variety of octupolar tensors and allow for a visual representation of their symmetries.

We showed that a semi-indefinite cylinder $\cyl_+$ in a three-dimensional parameter space suffices to represent all possible incarnations of $\A$ in a three-dimensional physical space. Such a reduction (of the originally seven-dimensional parameter space) is obtained by fixing a maximum of $\pot$ on the North Pole of $S^2$ and scaling its value to unity. We identified in $\cyl_+$ a number of special regions characterized by different symmetry groups for $\pot$ and ordering of its maxima relative to the \emph{orienting} maximum at the North Pole. For sake of illustration, each of these regions could be further divided in two subregions, which we distinguish by a $^+$ or a $^-$ superscript indicating where the maxima of $\pot$ that supplement the orienting maximum have a larger or a smaller value than the latter.

The axis $\axis$ of the cylinder $\cyl_+$ is characterized by a potential $\pot$ enjoying the $D_{3h}$ symmetry, with $4$ maxima, $3$ of which are equal and either exceed the fourth, orienting maximum (in $\axis^+$) or fall short of it (in $\axis^-$). A special point of $\axis$  separates $\axis^+$ and $\axis^-$: this is the only point with tetrahedral symmetry $T_d$, which, despite its singularity, has given its name to a whole class of bent-core liquid crystal phases.\footnote{Which thus are presumably more complicated than anticipated.} 

On a disk $\disk$, which is the base of $\cyl_+$, $\pot$ has the $D_{2h}$ symmetry and possesses $3$ maxima. $\disk$ can be separated into an inner disk $\disk_1$ and an outer annulus $\disk_2$. The center $\centre$ of $\disk$ has a special nature: there the potential $\pot$ is axially symmetric. The symmetry group is $D_{\infty h}$ and the (primary) maximum at the North Pole is accompanied by a full circle of (secondary) maxima in the southern hemisphere of $S^2$. 

Away from all these special loci is the generic bulk $\bulk$, where $\pot$ has either $4$ (generically unequal) maxima (in $\bulk_4$) or $3$ (unequal) maxima (in $\bulk_3$), separated by a  surface $\mathcal{S}$ in parameter space,  called the separatrix. Both $\bulk_3$ and $\bulk_4$ (where, if necessary, we could distinguish the variants $\bulk_3^\pm$ and $\bulk_4^\pm$) enjoy three planes of symmetry, collectively denoted as $\plane$. Correspondingly, $\pot$ has on $S^2$ a reflection symmetry across a plane through the poles. Therefore, when the parameters fall in particular in $\bulk_4\cap\plane$, two secondary maxima of $\pot$ are equal.

Much like the decomposition of $\disk$ into the union of $\disk_1$ and $\disk_2$, the separatrix $\mathcal{S}$ can be further decomposed into the union of two surfaces, an inner $\mathcal{S}_1$ and an outer $\mathcal{S}_2$, where $\Phi$ has 10 and 12 critical points, respectively. The boundary $\mathcal{L}_1$ of $\mathcal{S}_1$ is a line of cusps in parameter space where $\Phi$ has 8 critical points, whereas the outer boundary $\mathcal{L}_2$ of $\mathcal{S}_2$ is a line where $\Phi$ has 10 critical points.

All secondary maxima of $\pot$ have a remarkable, universal feature, irrespective of the choice of parameters: they fall in the southern hemisphere of $S^2$, when the North Pole is marked by the primary, orienting maximum.

We trust that all the qualitative features of the octupolar potential described in this paper would prompt a better understanding of the many physical instances where an octupolar tensor is at play.

\begin{acknowledgements}
E.G.V. acknowledges the kind hospitality of the Oxford Centre
for Nonlinear PDE, where part of  this work was done while
he was visiting the Mathematical Institute at the University of
Oxford.
\end{acknowledgements}
%

\appendix
\section{The tetrahedron group}
\label{app:tetra}

In this Appendix we give further detail -- beyond those mentioned
in Sect.~\ref{sec:tetra} -- on the tetrahedron group. We will work
on a concrete realization of it in three-dimensional space; the
points identifying  the tetrahedron will be
$$ \( 0 , \frac{2 \sqrt{2}}{3} , -\frac{1}{3} \)  ,  \
\( \sqrt{\frac{2}{3}} , -\frac{\sqrt{2}}{3} ,  -\frac{1}{3} \)  ,\
  \( -\sqrt{\frac{2}{3}} , -\frac{\sqrt{2}}{3} ,  -\frac{1}{3}
\)  , \  \( 0 , 0 , 1 \)  . $$ In angular coordinates $(\vth_{1},\vth_{2})$, these
are
$$ \( \theta_0 , - \frac\pi2\)  ,\   \(\theta_0 , - \frac\pi6 \)  , \  \(\theta_0
, \frac{5 \pi}{6} \)  , \  \(\frac\pi2 , *\)  , $$ where the symbol $*$ means that  in this
case $\vth_2$ is not determined, and
$$ \theta_0 =-
\arcsin \( \frac13 \) \doteq-0.34.$$

The tetrahedron group $T_d \ss O(3)$ is made of $12$ matrices of
determinant $+1$, associated to rotations of an angle $2 \pi / 3$
and multiples around each of the four axes of the tetrahedron,  denoted as $T_d^+ \ss
SO(3)$; and other $12$ matrices of determinant $-1$, collectively denoted as $T_d^-$,  among which are
those associated to reflections through planes containing axes of
the tetrahedron.

We now give the twelve matrices in $T_d^+$; these are:
$$ \begin{array}{ll}
M_1 = \pmatrix{ 1 & 0 & 0\cr 0 & 1 & 0 \cr
 0 & 0 & 1 \cr}, &
M_2 = \pmatrix{-\frac{1}{2} &
 \frac{\sqrt{3}}{2} & 0\cr -\frac{\sqrt{3}}{2} &
   -\frac{1}{2} & 0\cr 0 & 0 & 1\cr} , \\
M_3 = \pmatrix{-\frac{1}{2} & -\frac{\sqrt{3}}{2}&0 \cr
  \frac{\sqrt{3}}{2} & - \frac{1}{2} & 0 \cr 0 & 0 & 1 \cr}, &
M_4 = \pmatrix{ \frac{1}{2} & \frac{\sqrt{3}}{2} & 0 \cr
\frac{1}{2 \sqrt{3}} & -\frac{1}{6} &
 \frac{2 \sqrt{2}}{3}\cr
  \sqrt{\frac{2}{3}} & - \frac{\sqrt{2}}{3} &
  -\frac{1}{3} \cr}, \\
M_5 = \pmatrix{ \frac{1}{2} & \frac{1}{2 \sqrt{3}} &
\sqrt{\frac{2}{3} } \cr \frac{\sqrt{3}}{2} &
 - \frac{1}{6} & -\frac{\sqrt{2}}{3} \cr
  0 & \frac{2 \sqrt{2}}{3} & -\frac{1}{3} \cr}, &
M_6 = \pmatrix{ \frac{1}{2} & -\frac{\sqrt{3}}{2} & 0 \cr
 -\frac{1}{2 \sqrt{3}} & -\frac{1}{6} &
  \frac{2 \sqrt{2}}{3} \cr
  -\sqrt{\frac{2}{3}}&  -\frac{\sqrt{2}}{3} &
   -\frac{1}{3} \cr} ,\\
M_7 = \pmatrix{ \frac{1}{2} &
 - \frac{1}{2 \sqrt{3}} & -\sqrt{\frac{2}{3}} \cr
 -\frac{\sqrt{3}}{2} & -\frac{1}{6} &
  -\frac{\sqrt{2}}{3} \cr
 0 & \frac{2 \sqrt{2}}{3} & -\frac{1}{3} \cr}, &
M_8 = \pmatrix{ - \frac{1}{2} &
 \frac{1}{2 \sqrt{3}} & \sqrt{\frac{2}{3}} \cr
 -\frac{1}{2 \sqrt{3}} & \frac{5}{6} &
  -\frac{\sqrt{2}}{3} \cr
 - \sqrt{\frac{2}{3}} &  - \frac{\sqrt{2}}{3} &
 -\frac{1}{3} \cr} ,\\
M_9 = \pmatrix{ -\frac{1}{2} &
 -\frac{1}{2 \sqrt{3}} &
  -\sqrt{\frac{2}{3}} \cr
 \frac{1}{2 \sqrt{3}} & \frac{5}{6} &
  -\frac{\sqrt{2}}{3} \cr
 \sqrt{\frac{2}{3}} & - \frac{\sqrt{2}}{3} &
  -\frac{1}{3} \cr}, &
M_{10} = \pmatrix{0 & \frac{1}{\sqrt{3}} & -\sqrt{\frac{2}{3}} \cr
 \frac{1}{\sqrt{3}} & - \frac{2}{3} &
  -\frac{\sqrt{2}}{3} \cr
 -\sqrt{\frac{2}{3}} & -\frac{\sqrt{2}}{3} &
  -\frac{1}{3} \cr} ,\\
M_{11} = \pmatrix{0 & -\frac{1}{\sqrt{3}} & \sqrt{\frac{2}{3}} \cr
 -\frac{1}{\sqrt{3}} &  -\frac{2}{3} &
  -\frac{\sqrt{2}}{3} \cr
 \sqrt{\frac{2}{3}} & - \frac{\sqrt{2}}{3} &
 - \frac{1}{3}\cr}, &
M_{12} = \pmatrix{ -1 & 0 & 0 \cr
 0 & \frac{1}{3} & \frac{2 \sqrt{2}}{3} \cr
 0 & \frac{2 \sqrt{2}}{3} &
 -\frac{1}{3} \cr}.
\end{array} $$

The multiplication table for these matrices is the following:
$$ P_{11} =\left(
\begin{array}{llllllllllll}
 1 & 2 & 3 & 4 & 5 & 6 & 7 & 8 & 9 & 10 & 11 & 12 \\
 2 & 3 & 1 & 11 & 7 & 8 & 12 & 10 & 4 & 6 & 9 & 5 \\
 3 & 1 & 2 & 9 & 12 & 10 & 5 & 6 & 11 & 8 & 4 & 7 \\
 4 & 12 & 6 & 5 & 1 & 11 & 9 & 2 & 10 & 7 & 3 & 8 \\
 5 & 8 & 11 & 1 & 4 & 3 & 10 & 12 & 7 & 9 & 6 & 2 \\
 6 & 4 & 12 & 10 & 8 & 7 & 1 & 11 & 3 & 2 & 5 & 9 \\
 7 & 10 & 9 & 2 & 11 & 1 & 6 & 5 & 12 & 4 & 8 & 3 \\
 8 & 11 & 5 & 6 & 10 & 12 & 2 & 9 & 1 & 3 & 7 & 4 \\
 9 & 7 & 10 & 12 & 3 & 4 & 11 & 1 & 8 & 5 & 2 & 6 \\
 10 & 9 & 7 & 8 & 6 & 5 & 3 & 4 & 2 & 1 & 12 & 11 \\
 11 & 5 & 8 & 7 & 2 & 9 & 4 & 3 & 6 & 12 & 1 & 10 \\
 12 & 6 & 4 & 3 & 9 & 2 & 8 & 7 & 5 & 11 & 10 & 1
\end{array}
\right). $$

These generate several subgroups; in particular -- apart from the
trivial ones consisting of $M_1$ alone and of the full group
$T_d^+$ -- we have four subgroups of order three,
$$ G_1 = \{ M_1, M_2, M_3 \}  ,  \
G_2 = \{M_1, M_4 , M_5 \}  ,  \
G_3 = \{ M_1 , M_6 , M_7 \}  ,  \
G_4 = \{ M_1, M_8 , M_9 \}  ; $$
three groups of order two,
$$ G_5 = \{ M_1 , M_{10} \}  ,  \
G_6 = \{ M_1 , M_{11} \}  ,  \
G_7 = \{ M_1 , M_{12} \}  ; $$ and one group of order four,
$$ G_8 = \{ M_1 , M_{10} , M_{11} , M_{12} \}  . $$
The latter is the only nontrivial normal subgroup, and also the
only one acting freely.

We can also easily determine the subspaces $F_k$ admitting each of
these $G_k$ as symmetry subgroups; in particular,
$$ F_1 = (0,0,z)  ,  \
 F_2 = (\sqrt{6} z , \sqrt{2} z , z )  ,  \
 F_3 = (- \sqrt{6} z , \sqrt{2} z , z )  ,  \
 F_4 = (0 , - 2 \sqrt{2} z , z )  ; $$
the subgroups $G_k$, $k=1,2,3,4$ act as rotations (by an angle $2
\pi / 3$) around these axes, which are just the axes of the
tetrahedron. Moreover,
$$ F_5 = \( - \sqrt{\frac32} z , - \sqrt{\frac12} z , z \)  ,  \
 F_6 = \( \sqrt{\frac32} z , - \sqrt{\frac12} z , z \)  ,  \
 F_7 = \( 0 , \sqrt{2} z , z \)  ; $$ these subgroups $G_k$, $k=5,6,7$, represent
rotations by $\pi$ around the given axes $F_k$. Note that $F_8 =
\{ (0,0,0) \}$, and correspondingly $G_8$ represents combined
$\pi$ rotations around the $F_5,F_6,F_7$ axes.

We can give as well the twelve matrices in $T_d^-$; these are:
$$ \begin{array}{ll} M_1^- =  \pmatrix{
 -1 & 0 & 0 \cr 0 & 1 & 0 \cr 0 & 0 & 1\cr}, &
  M_2^- = \pmatrix{\frac{1}{2} & -\frac{\sqrt{3}}{2} & 0 \cr
 -\frac{\sqrt{3}}{2} & -\frac{1}{2},0\cr 0 & 0 & 1\cr}, \\
 M_3^- = \pmatrix{\frac{1}{2} & \frac{\sqrt{3}}{2} & 0\cr
  \frac{\sqrt{3}}{2} & -\frac{1}{2},0\cr  0 & 0 & 1 \cr} , &
 M_4^- =  \pmatrix{ -\frac{1}{2} & -\frac{\sqrt{3}}{2} & 0\cr
  \frac{1}{2 \sqrt{3}} & -\frac{1}{6} &
 \frac{2 \sqrt{2}}{3} \cr
  \sqrt{\frac{2}{3}} & -\frac{\sqrt{2}}{3} &  -\frac{1}{3} \cr}, \\
 M_5^- = \pmatrix{-\frac{1}{2} &
  -\frac{1}{2 \sqrt{3}} & -\sqrt{\frac{2}{3}} \cr
 \frac{\sqrt{3}}{2} & -\frac{1}{6} &
  -\frac{\sqrt{2}}{3} \cr
  0 & \frac{2\sqrt{2}}{3} & -\frac{1}{3}\cr}, &
 M_6^- = \pmatrix{-\frac{1}{2} & \frac{\sqrt{3}}{2} & 0\cr
 -\frac{1}{2 \sqrt{3}} & -\frac{1}{6} &
  \frac{2 \sqrt{2}}{3} \cr
 -\sqrt{\frac{2}{3}} & -\frac{\sqrt{2}}{3} & -\frac{1}{3} \cr}, \\
M_7^- = \pmatrix{ -\frac{1}{2}&\frac{1}{2\sqrt{3}} &
\sqrt{\frac{2}{3}} \cr
 -\frac{\sqrt{3}}{2} & -\frac{1}{6} &
  -\frac{\sqrt{2}}{3} \cr
  0,\frac{2 \sqrt{2}}{3} & -\frac{1}{3}\cr}, &
 M_8^- = \pmatrix{ \frac{1}{2} &
 -\frac{1}{2 \sqrt{3}} & -\sqrt{\frac{2}{3}}\cr
 -\frac{1}{2 \sqrt{3}} & \frac{5}{6} & -\frac{\sqrt{2}}{3}\cr
 -\sqrt{\frac{2}{3}} & -\frac{\sqrt{2}}{3} &
   -\frac{1}{3}\cr}, \\
 M_9^- = \pmatrix{ \frac{1}{2} & \frac{1}{2 \sqrt{3}} &
 \sqrt{\frac{2}{3}} \cr
  \frac{1}{2 \sqrt{3}} & \frac{5}{6} & -\frac{\sqrt{2}}{3} \cr
 \sqrt{\frac{2}{3}} & -\frac{\sqrt{2}}{3} &
   -\frac{1}{3}\cr}, &
 M_{10}^- = \pmatrix{
 0 & -\frac{1}{\sqrt{3}} & \sqrt{\frac{2}{3} } \cr
 \frac{1}{\sqrt{3}} & -\frac{2}{3} & -\frac{\sqrt{2}}{3} \cr
 -\sqrt{\frac{2}{3}} & -\frac{\sqrt{2}}{3} & -\frac{1}{3}\cr}, \\
 M_{11}^- = \pmatrix{ 0 & \frac{1}{\sqrt{3}} & -\sqrt{\frac{2}{3}}
 \cr
   -\frac{1}{\sqrt{3}} & -\frac{2}{3} &
   -\frac{\sqrt{2}}{3} \cr
 \sqrt{\frac{2}{3}} & -\frac{\sqrt{2}}{3} &
   -\frac{1}{3}\cr}, &
 M_{12}^- = \pmatrix{ 1 & 0 & 0 \cr
  0 & \frac{1}{3} & \frac{2\sqrt{2}}{3} \cr
  0 & \frac{2 \sqrt{2}}{3} & -\frac{1}{3} \cr}.
\end{array} $$
Once $M_1^-$ has been defined, they are built by
$$ M_k^- =M_1^-  M_k  . $$
It is obvious that $M_1^-$ represents a reflection (across the
$(y,z)$ plane), so that the matrices $M_k^-$ represent the combination
of rotations and reflections.

If we write $M_{k+12} = M_k^-$, the full multiplication table is
given in block form by
$$ P =\pmatrix{P_{11} & P_{12} \cr P_{21} & P_{22} \cr}, $$
where $P_{11}$ has been given above, and the other blocks are:
$$ P_{12} =\left(
\begin{array}{llllllllllll}
13 & 14 & 15 & 16 &
   17 & 18 & 19 & 20 & 21 & 22 & 23 & 24 \\
15 & 13 & 14 & 21 &
   24 & 22 & 17 & 18 & 23 & 20 & 16 & 19 \\
14 & 15 & 13 & 23 &
   19 & 20 & 24 & 22 & 16 & 18 & 21 & 17 \\
18 & 16 & 24 & 22 &
   20 & 19 & 13 & 23 & 15 & 14 & 17 & 21 \\
19 & 22 & 21 & 14 &
   23 & 13 & 18 & 17 & 24 & 16 & 20 & 15 \\
16 & 24 & 18 & 17 &
   13 & 23 & 21 & 14 & 22 & 19 & 15 & 20 \\
17 & 20 & 23 & 13 &
   16 & 15 & 22 & 24 & 19 & 21 & 18 & 14 \\
21 & 19 & 22 & 24 &
   15 & 16 & 23 & 13 & 20 & 17 & 14 & 18 \\
20 & 23 & 17 & 18 &
   22 & 24 & 14 & 21 & 13 & 15 & 19 & 16 \\
23 & 17 & 20 & 19 &
   14 & 21 & 16 & 15 & 18 & 24 & 13 & 22 \\
22 & 21 & 19 & 20 &
   18 & 17 & 15 & 16 & 14 & 13 & 24 & 23 \\
24 & 18 & 16 & 15 &
   21 & 14 & 20 & 19 & 17 & 23 & 22 & 13
\end{array}
\right), $$
$$ P_{21} =\left(
\begin{array}{llllllllllll}
 13 & 14 & 15 & 16 & 17 & 18 & 19 & 20 & 21 & 22 & 23 & 24 \\
 14 & 15 & 13 & 23 & 19 & 20 & 24 & 22 & 16 & 18 & 21 & 17 \\
 15 & 13 & 14 & 21 & 24 & 22 & 17 & 18 & 23 & 20 & 16 & 19 \\
 16 & 24 & 18 & 17 & 13 & 23 & 21 & 14 & 22 & 19 & 15 & 20 \\
 17 & 20 & 23 & 13 & 16 & 15 & 22 & 24 & 19 & 21 & 18 & 14 \\
 18 & 16 & 24 & 22 & 20 & 19 & 13 & 23 & 15 & 14 & 17 & 21 \\
 19 & 22 & 21 & 14 & 23 & 13 & 18 & 17 & 24 & 16 & 20 & 15 \\
 20 & 23 & 17 & 18 & 22 & 24 & 14 & 21 & 13 & 15 & 19 & 16 \\
 21 & 19 & 22 & 24 & 15 & 16 & 23 & 13 & 20 & 17 & 14 & 18 \\
 22 & 21 & 19 & 20 & 18 & 17 & 15 & 16 & 14 & 13 & 24 & 23 \\
 23 & 17 & 20 & 19 & 14 & 21 & 16 & 15 & 18 & 24 & 13 & 22 \\
 24 & 18 & 16 & 15 & 21 & 14 & 20 & 19 & 17 & 23 & 22 & 13
\end{array}
\right), $$
$$ P_{22} =\left(
\begin{array}{llllllllllll}
1 & 2 & 3 & 4 & 5 & 6 & 7 & 8 & 9 & 10 & 11 & 12 \\
3 & 1 & 2 & 9 & 12 & 10 & 5 & 6 & 11 & 8 & 4 & 7 \\
2 & 3 & 1 & 11 & 7 & 8 & 12 & 10 & 4 & 6 & 9 & 5 \\
6 & 4 & 12 & 10 & 8 & 7 & 1 & 11 & 3 & 2 & 5 & 9 \\
7 & 10 & 9 & 2 & 11 & 1 & 6 & 5 & 12 & 4 & 8 & 3 \\
4 & 12 & 6 & 5 & 1 & 11 & 9 & 2 & 10 & 7 & 3 & 8 \\
5 & 8 & 11 & 1 & 4 & 3 & 10 & 12 & 7 & 9 & 6 & 2 \\
9 & 7 & 10 & 12 & 3 & 4 & 11 & 1 & 8 & 5 & 2 & 6 \\
8 & 11 & 5 & 6 & 10 & 12 & 2 & 9 & 1 & 3 & 7 & 4 \\
11 & 5 & 8 & 7 & 2 & 9 & 4 & 3 & 6 & 12 & 1 & 10 \\
10 & 9 & 7 & 8 & 6 & 5 & 3 & 4 & 2 & 1 & 12 & 11 \\
12 & 6 & 4 & 3 & 9 & 2 & 8 & 7 & 5 & 11 & 10 & 1
\end{array}
\right) .$$

We now have subgroups involving elements of both $T_d^+$ and
$T_d^-$; in particular, using again the 1 through 24 numeration
and denoting $M_k$ directly by $k$, we have the following
subgroups; those of types $\mathcal{G}_a$ through $\mathcal{G}_c$
extending those seen above, while those of type $\mathcal{G}_d$
involve no other element of $T_d^+$ but the identity:
\begin{eqnarray*}
& \mathcal{G}_a: & (1,2,3,13,14,15)  ,  (1,2,3,16,21,23)  ,  (1,2,3,17,19,24)  ,  (1,2,3,18,20,22)  ; \\
& & (1,4,5,13,18,19)  ,  (1,4,5,14,16,22)  ,  (1,4,5,15,21,24)  ,  (1,4,5,17,20,23)  ; \\
& & (1,8,9,13,20,21)  ,  (1,8,9,14,19,23)  ,  (1,8,9,15,17,22)  ,  (1,8,9,16,18,24)  ; \\
& \mathcal{G}_b: & (1,10,13,23)  ,  (1,10,14,17)  ,  (1,10,15,20)  ,  (1,10,16,19)  ,  (1,10,18,21)  ,  (1,10,22,24)  ; \\
& & (1,11,13,22)  ,  (1,11,14,21)  ,  (1,11,15,19)  ,  (1,11,16,20)  ,  (1,11,17,18)  ,  (1,11,23,24)  ; \\
& & (1,12,13,24)  ,  (1,12,14,18)  ,  (1,12,15,16)  ,  (1,12,17,21)  ,  (1,12,19,20)  ,  (1,12,22,23)  ; \\
& \mathcal{G}_c: & (1,10,11,12,13,22,23,24)  ,  (1,10,11,12,14,17,18,21)  ,  (1,10,11,12,15,16,19,20)  ; \\
& \mathcal{G}_d : & (1,13)  ,  (1,14)  ,  (1,15)  ,  (1,20)  ,  (1,21)  , (1,24)  . \end{eqnarray*}

Groups of type $\mathcal{G}_a$ contain rotations by $2 \pi/3$
around an axis and reflections through a plane containing that
axis, hence they are of type $D_{3h}$, and only the first one leaves
the distinguished point at the North Pole untouched; those of type
$\mathcal{G}_b$ contain rotations by $\pi$ around an axis and
reflections through a plane containing that axis, hence are of
type $D_{2h}$, but none of these leaves the distinguished point at
the North Pole untouched; those of type $\mathcal{G}_c$ combine
rotations and reflections through different axes, and none of them
preserves the North Pole; and those of type $\mathcal{G}_d$
consist just of reflections in a plane, hence are of type $D_h$;
the first three preserve the North Pole.

%

\end{document}